\documentclass[twoside,11pt,parskip=half,a4paper]{scrartcl}
\usepackage[utf8]{inputenc}
\usepackage[T1]{fontenc}
\usepackage{hyperref}
\usepackage{a4wide}
\usepackage{array,amsmath,amssymb,makeidx,alltt,mathtools,colordvi}
\usepackage{soul}
\usepackage[dvipsnames]{xcolor}
\usepackage{graphicx}
\usepackage{booktabs}
\usepackage{tabularx}
\usepackage{multirow}
\usepackage[capitalize]{cleveref}
\usepackage{subcaption}
\usepackage[separate-uncertainty,multi-part-units=single,list-units=single]{siunitx}
\usepackage{xspace}
\usepackage{makecell}
\usepackage[citestyle=numeric-comp,bibstyle=HBnumeric,sorting=none]{biblatex}
\usepackage{multicol}

\usepackage{smartdiagram}
\usesmartdiagramlibrary{additions}

\renewcommand{\arraystretch}{1.2}

\newcommand{\ie}{i.e.\ }
\newcommand{\eg}{e.g.\ }
\newcommand{\BR}[1]{\mathrm{BR}(#1)}
\newcommand{\CL}[1]{\ensuremath{\SI{#1}{\percent}~\text{C.L.}}}
\newcommand{\Code}[1]{\texttt{\detokenize{#1}}}
\newcommand{\pvalue}{$p$-value\xspace}

\newcommand{\eqcomma}{\,\text{,}}
\newcommand{\eqdot}{\,\text{.}}

\newcommand{\pvec}[1]{\vec{#1}\mkern2mu\vphantom{#1}} %

\newcommand{\HB}{\texttt{HiggsBounds}\xspace}
\newcommand{\HS}{\texttt{HiggsSignals}\xspace}
\newcommand{\HSv}[1]{\texttt{HiggsSignals-#1}}
\newcommand{\HBv}[1]{\texttt{HiggsBounds-#1}}
\newcommand{\HScurrent}{\HSv{2.6.0}\xspace}
\DeclareSIUnit{\fb}{fb}

\graphicspath{{figs/}}

\begin{document}

\thispagestyle{empty}

\mbox{}\hfill\texttt{ BONN-TH-2020-09},
\texttt{DESY 20-228},
\texttt{IFT-UAM/CSIC-20-081},
\texttt{LU TP 20-53}

\def\thefootnote{\fnsymbol{footnote}}

\begin{center}
    \Large\boldmath\textbf{
        \HSv{2}: Probing new physics with precision\\
        Higgs measurements in
        the LHC 13 TeV era}
    \unboldmath
\end{center}
\vspace{-0.5cm}
\begin{center}
    Philip Bechtle$^{1}$, Sven Heinemeyer$^{2,3,4}$, Tobias Klingl$^1$, \\[.5em]
    Tim Stefaniak$^{5}$, Georg Weiglein$^{5}$ and Jonas Wittbrodt$^{6}$\footnote{Electronic addresses:
        \href{mailto:bechtle@physik.uni-bonn.de}{bechtle@physik.uni-bonn.de},
        \href{mailto:sven.heinemeyer@cern.ch}{sven.heinemeyer@cern.ch},\\
        \href{mailto:klingl@physik.uni-bonn.de}{klingl@physik.uni-bonn.de},
        \href{mailto:tim.stefaniak@desy.de}{tim.stefaniak@desy.de},
        \href{mailto:jonas.wittbrodt@thep.lu.se}{jonas.wittbrodt@thep.lu.se},
        \href{mailto:georg.weiglein@desy.de}{georg.weiglein@desy.de}} \\[0.4cm]
    \textsl{\small
        $^1$Physikalisches Institut der Universit\"at Bonn,
        Nu{\ss}allee 12, D-53115 Bonn, Germany\\[0.1cm]
        $^2$Campus of International Excellence UAM+CSIC, Cantoblanco, E--28049 Madrid, Spain\\[0.1cm]
        $^3$Instituto de F\'\i{}sica Te\'orica, (UAM/CSIC), Universidad
        Aut\'onoma de Madrid,\\ Cantoblanco, E-28049 Madrid, Spain
        \\[0.1cm]
        $^4$Instituto de F\'\i{}sica de Cantabria (CSIC-UC), E-39005 Santander,
        Spain\\[0.1cm]
        $^5$ Deutsches Elektronen-Synchrotron DESY,
        Notkestra{\ss}e 85, D-22607 Hamburg, Germany\\[0.1cm]
        $^6$Department of Astronomy and Theoretical Physics,
        Lund University,
        Sölvegatan~14A, 223~62~Lund,
        Sweden\\[1mm]
    }
\end{center}
\vspace{0.2cm}

\renewcommand{\thefootnote}{\arabic{footnote}}
\setcounter{footnote}{0}

\begin{abstract}
    The program \HS\ confronts the predictions of models with arbitrary Higgs
    sectors with the available Higgs signal rate and mass measurements,
    resulting in a likelihood estimate. A new version of the program, \HSv{2},
    is presented that contains various improvements in its functionality and
    applicability. In particular, the new features comprise improvements in the
    theoretical input framework and the handling of possible complexities of
    beyond-the-SM Higgs sectors, as well as the incorporation of experimental
    results in the form of Simplified Template Cross Section (STXS)
    measurements. The new functionalities are explained, and a thorough
    discussion of the possible statistical interpretations of the \HS\ results
    is provided. The performance of \HS\ is illustrated for some example
    analyses. In this context the importance of public information on certain
    experimental details like efficiencies and uncertainty correlations is
    pointed out. \HS\ is continuously updated to the latest experimental results
    and can be obtained at \url{https://gitlab.com/higgsbounds/higgssignals}.
\end{abstract}

\clearpage

\tableofcontents

\clearpage

\section{Introduction}%
\label{sec:intro}

Elucidating the mechanism that controls electroweak symmetry breaking (EWSB) is
one of the main goals of the LHC\@. The spectacular discovery of a Higgs boson
with a mass around \SI{125}{\GeV} by the ATLAS and CMS
experiments~\cite{Chatrchyan:2012xdj,Aad:2012tfa} marks a milestone of an effort
that has been ongoing for almost half a century and has opened a new era of
particle physics. In the eight years since the discovery of
the new particle at the LHC, its mass has been measured with a few-per-mil accuracy, $M_H^\text{obs} = \SI{125.09 +- 0.24}{\GeV}$~\cite{Aad:2015zhl}.\footnote{This is the latest ATLAS and CMS combined result. More recent data is in agreement with
    this value.} The measured properties are, within current experimental and theoretical
uncertainties, in agreement with the predictions of the Standard Model
(SM)~\cite{Khachatryan:2016vau}. Together with the limits on beyond-the-SM (BSM) particles that were obtained at the LHC with center-of-mass energies of up
to \SI{13}{\TeV} and elsewhere, the requirement that the particle spectrum should include an essentially SM-like Higgs boson
at about 125~GeV imposes important constraints on the parameter space of
possible extensions of the SM\@.

In order to test the predictions of BSM models with arbitrary
Higgs sectors consistently against all the available experimental data,
on the one hand the compatibility with existing BSM Higgs-boson searches
has to be checked. This can be done with the public tool
\HB~\cite{Bechtle:2008jh,Bechtle:2011sb,Bechtle:2013wla,Bechtle:2015pma,Bechtle:2020pkv}.
On the other hand, any BSM model should furthermore be tested
against the measured mass and rates of the observed scalar state. Confronting the predictions of an arbitrary
Higgs sector with the observed Higgs signal\footnote{Here and in the following,
    \emph{Higgs signal} refers to any measurement in analyses at the LHC
    or the Tevatron that can be  associated with the observed state at
    $\SI{125}{\GeV}$, regardless of the statistical significance of each
    individual measurement.} (and potentially with other, future,
signals of additional Higgs states) is the purpose of the public computer program
\HS~\cite{Bechtle:2013xfa}. Here we present the new version
\HSv{2} and update the description of the program
w.r.t.\ version~\texttt{1.0} as presented previously
in Ref.~\cite{Bechtle:2013xfa}.

\HS evaluates a $\chi^2$ measure to provide a quantitative answer to the
statistical question of how compatible the Higgs data
(in particular, measured signal rates and masses) is with the model predictions. The $\chi^2$
evaluation can be performed with various methods~\cite{Bechtle:2013xfa}. Among those the
\textit{peak-centered} $\chi^2$ method (see \cref{sec:newHS}) is the default
method implemented in \HS. In this $\chi^2$ method the (neutral) Higgs signal
rates and masses predicted by the model are tested against the various signal
rate measurements published by the experimental collaborations for a fixed
hypothetical Higgs mass. This hypothetical Higgs mass is typically motivated by
the signal peak seen in the channels with high mass resolution, \ie the searches
for $H\to\gamma\gamma$ and $H\to ZZ^{(*)}\to 4\ell$. In this way, the
model is tested \textit{at the observed peak's mass position}.

With theoretical input provided by the user in form of Higgs masses, production
cross sections, and decay rates in the same format as used in \HB, the code \HS
evaluates the corresponding $\chi^2$ to test the compatibility of any BSM model
with the full set of experimental Higgs-boson data. The experimental data from
LHC (and --- if needed --- Tevatron) Higgs analyses is provided with the program, so
there is no need for the user to include these values manually. However, it is
possible for the user to modify or add to the dataset at will.

The usefulness of a public tool such as \HS\footnote{Other programs that
    perform likelihood calculations from Higgs measurements are
    \texttt{Lilith}~\cite{Bernon:2015hsa,Kraml:2019sis} and
    \texttt{HEPfit}~\cite{deBlas:2019okz}.} has become apparent in the past
years, given the intense work by theorists to use the latest Higgs
measurements as constraints on the SM and theories for new physics. The
$\chi^2$ output of \HS\ can directly be used as input to global fits. Some
example applications can be found in Refs.~\cite{Bechtle:2014ewa,Bechtle:2015nua,Bechtle:2016kui,Bagnaschi:2017tru,Bagnaschi:2018zwg,Bahl:2020kwe,Bahl:2020wee,Kvellestad:2019vxm}.
Performance tests of the \HS\ implementation of experimental results
within selected benchmark models have occasionally  revealed shortcomings in
the presentation of the publicly available experimental data. These
have been reported to the experimental collaborations, which --- in a joint
effort between experiment and theory --- have often led to an improved
usability of the experimental results.

We begin \cref{sec:newHS} with a short overview of the relevant changes made to
the shared \texttt{Higgs\-Bounds} and \HS theoretical input framework. We
then discuss the calculation of the $\chi^2$ measure in \HS including a review
of the peak-centered $\chi^2$ method. We also describe in this context the
$\chi^2$ contributions arising from the LHC Run-1 ATLAS and CMS Higgs
combination and the newly added simplified template cross section (STXS)
observables. Furthermore, we discuss the handling of potential issues that can
arise in BSM Higgs sectors, in particular effects from non-SM-like kinematical
properties of the Higgs candidate, theoretical uncertainties of the mass
predictions, and the possibility of overlapping signals from multiple Higgs
bosons. We provide a short overview of the technical user operation instructions
in \cref{sec:user} and refer to the online documentation~\cite{HSDoc} for
details. We then discuss the options for exploiting the $\chi^2$ result returned
by \HS for several different applications in BSM model testing. In
\cref{sec:performance} we present detailed performance test of the \HS
implementations for several analyses in comparison with official results from
the experimental collaborations. In this context we point out the importance of
sub-channel and correlation information that should be provided by the
experimental collaborations in order to allow an accurate reinterpretation of
their results. We close the section with a list of recommendations for the
publication of Higgs signal measurements. We conclude in \cref{sec:summary}. The
Appendix includes details on the implementation and the format of STXS
observables as well as a listing of all currently included experimental
measurements.

\section{\HSv{2}: Basic concepts and new developments}%
\label{sec:newHS}

\subsection{Extension of the theoretical input framework}

A detailed description of the extensions and modifications of the \HB input
framework for the theory predictions is given in Ref.~\cite{Bechtle:2020pkv}. In
this section we only highlight the relevant changes affecting
\texttt{HiggsSignals}.

With the enlarged experimental program of precision Higgs rate measurements at
the LHC during Run 2 more of the sub-dominant Higgs-boson production and decay
modes came under consideration. While \HSv{1} included only the five main LHC
Higgs-boson production modes (single Higgs production, vector boson fusion
(VBF), $W$- and $Z$-boson associated Higgs production, top-quark pair associated
Higgs production), the code now also accounts for processes with smaller cross
sections. The upper part of \cref{Tab:pdidentifier} lists all Higgs processes
currently implemented in \HSv{2}. Furthermore, the input of some of the
inclusive production processes is supplemented by input for their dominant
exclusive subprocesses: First, the gluon fusion production and $b\bar{b}$
associated Higgs production processes are now treated separately, while in
\HSv{1} they were always taken together as inclusive single Higgs production.
This is required in order to properly implement experimental analyses that
exhibit sensitivity to the $b$-jet multiplicity. Second, the quark-initiated and
gluon-initiated $Z$-boson associated Higgs production processes are now treated
separately. Further additions to the available production modes in the \HS\
input are the single top-quark associated Higgs production channel in the $t$-
and $s$-channel and Higgs  production in association with a single top quark and
a $W$ boson. The lower part of \cref{Tab:pdidentifier} lists all implemented
Higgs-boson decay modes.

\begin{table}[p]
    \centering
    \begin{tabular}{ll}
        \toprule
        identifier $p$ & production mode                                                             \\
        \cmidrule(lr){1-1}\cmidrule(lr){2-2}
        0              & no production mode                                                          \\
        1              & single Higgs production, $pp \to \phi$                                      \\
        2              & vector boson fusion, $pp \to qq \phi$                                       \\
        3              & $W$-boson associated Higgs production, $pp \to W^\pm\phi$                   \\
        4              & $Z$-boson associated Higgs production, $pp \to Z\phi$                       \\
        5              & $t\bar{t}$ associated Higgs production, $pp \to t\bar{t}\phi$               \\
        6              & gluon fusion Higgs production, $gg\to \phi$                                 \\
        7              & $b\bar{b}$ associated Higgs production, $gg\to  b\bar{b}\phi$               \\
        8              & single top associated Higgs production ($t$-channel), $pp\to t\phi$         \\
        9              & single top associated Higgs production ($s$-channel), $pp\to t\phi$         \\
        10             & quark-initiated $Z$-boson associated Higgs production, $q\bar{q} \to Z\phi$ \\
        11             & gluon-initiated $Z$-boson associated Higgs production, $gg \to Z\phi$       \\
        12             & single top and $W$-boson associated Higgs production, $gb \to tW^\pm \phi$  \\
        \midrule
        identifier $d$ & decay mode                                                                  \\
        \cmidrule(lr){1-1}\cmidrule(lr){2-2}
        0              & no decay mode                                                               \\
        1              & $\phi \to \gamma\gamma$                                                     \\
        2              & $\phi \to W^+W^-$                                                           \\
        3              & $\phi \to ZZ $                                                              \\
        4              & $\phi \to \tau^+\tau^-$                                                     \\
        5              & $\phi \to b\bar{b}$                                                         \\
        6              & $\phi \to Z\gamma$                                                          \\
        7              & $\phi \to c\bar{c}$                                                         \\
        8              & $\phi \to \mu^+\mu^-$                                                       \\
        9              & $\phi \to gg$                                                               \\
        10             & $\phi \to s\bar{s}$                                                         \\
        11             & $\phi \to t\bar{t}$                                                         \\
        \bottomrule
    \end{tabular}
    \caption{Production and decay mode identifiers, $p$ and $d$, for the various
    Higgs production and decay processes included in \HSv{2}.}%
    \label{Tab:pdidentifier}
\end{table}

A Higgs signal \emph{channel} composed of one production mode and one decay mode is
specified by the \emph{channel identifier (ID)} \texttt{c}, given by the string
construct \texttt{c = "p.d"}, where \texttt{p} (\texttt{d}) is the production
(decay) mode identifier given in \cref{Tab:pdidentifier}.\footnote{In \HSv{1}
    the channel ID was given by a two-digit integer composed of \texttt{p} and
    \texttt{d} as first and second digit, respectively. With the larger amount of
    production and decay modes this is no longer possible. For compatibility,
    two-digit channel IDs without a period in between are still supported in
    \HSv{2}.} In rare cases the collider process is specified only as a production
or decay mode. In such cases, the unspecified mode has the identifier $0$. Some examples are given in \cref{tab:chanEx}.

\begin{table}
    \centering
    \begin{tabular}{ll}
        \toprule
        channel ID (\texttt{c}) & collider process                           \\
        \midrule
        \texttt{1.1}            & $pp \to \phi$, $\phi \to \gamma\gamma$     \\
        \texttt{7.10}           & $gg \to b\bar{b}\phi$, $\phi \to s\bar{s}$ \\
        \texttt{11.0}           & $gg \to Z \phi$                            \\
        \texttt{0.2}            & $\phi \to W^+W^-$                          \\
        \bottomrule
    \end{tabular}
    \caption{Examples for processes encoded by the \HS\ channel IDs.}\label{tab:chanEx}
\end{table}

Per default, \HB and \HS employ the narrow width approximation, \ie we assume
that the production cross section and the branching ratio can be factorized, so
that the signal rates can simply be calculated from the user input of the cross
sections and branching ratios. However, within \HBv{5} (and thus also within
\HSv{2}) it is also possible to set the rate directly for each signal channel
via dedicated \texttt{Fortran} subroutines (see Ref.~\cite{Bechtle:2020pkv}),
which allows treating cases where the narrow width approximation is not
applicable (as e.g.~relevant in the case of a hypothetical or future signal in
the high mass range).

\subsection{The \texorpdfstring{$\chi^2$}{chi-squared} calculation and individual contributions}%
\label{Sec:chi2_overview}

In the previous version, \HSv{1}, three different run modes were supported:
\textsl{(1)} the \emph{peak-centered $\chi^2$ method} which uses the signal
strength measurements, $\hat{\mu}$, performed for one specific Higgs boson mass
value, $\hat{m}$ (\eg at $\hat{m} = \SI{125}{\GeV}$); \textsl{(2)} the
mass-centered $\chi^2$ method which uses the signal strength measured as a
function of a hypothesized Higgs boson mass as experimental input, thus enabling
a $\chi^2$-test using the signal strength measured at the model-predicted Higgs
boson mass, $m$; \textsl{(3)} a combination of both methods. Furthermore, an
additional, separate $\chi^2$ contribution arises from the comparison of the
mass measurements with the model-predicted Higgs boson mass.

However, soon after the Higgs boson signal was established at a Higgs mass
$\simeq \SI{125}{\GeV}$, the LHC experiments ceased to provide signal strength
measurements for an extended Higgs boson mass interval. Without updated
experimental input, the mass-centered $\chi^2$ method quickly became irrelevant,
and the peak-centered $\chi^2$ method became the main run mode of \HS. Hence,
the current version \HSv{2} only uses the peak-centered $\chi^2$ method while
the other run modes have been removed. Yet, different contributions to the
total $\chi^2$  arise according to the experimental input the user chooses to
apply. The different types of $\chi^2$ contributions are listed in \cref{Tab:chi2contributions}.

\begin{table}[tb]
    \centering
    \begin{tabularx}{\textwidth}{lllX}
        \toprule
        Total $\chi^2$            & rate contribution               & mass contribution             & Experimental input                                   \\
        \midrule
        $\chi^2_\text{peak}$      & $\chi^2_{\text{peak},\mu}$      & $\chi^2_{\text{peak},m}$      & user-selected ``peak observables''                   \\
        $\chi^2_\text{LHC~Run-1}$ & $\chi^2_{\text{LHC~Run-1},\mu}$ & $\chi^2_{\text{LHC~Run-1},m}$ & LHC Run-1 ATLAS and CMS combined measurements        \\
        $\chi^2_{\text{STXS}}$    & $\chi^2_{\text{STXS},\mu}$      & $\chi^2_{\text{STXS},m}$      & Simplified Template Cross Section (STXS) observables \\
        \bottomrule
    \end{tabularx}
    \caption{Individual $\chi^2$ contributions in \HSv{2}. The total
        $\chi^2$ (\emph{left column}) is defined as the sum of the $\chi^2$
        contribution from the rate measurements (\emph{second column}) and the
        $\chi^2$ contribution from the mass measurements (\emph{third
            column}).}%
    \label{Tab:chi2contributions}
\end{table}

The $\chi^2$ calculation using the so-called peak observables (\emph{first row}
in \cref{Tab:chi2contributions}) closely follows the peak-centered $\chi^2$
method employed in \HSv{1}. A brief review and a description of relevant
modifications will be given in \cref{Sec:chi2peak}. The experimental input to
this calculation is specified by the user by selecting an ``observable set''
which may contain measurements performed by the Tevatron experiments, as well as
by the LHC experiments at center-of-mass energies of \SIlist{7;8;13}{\TeV}.
Predefined observable sets are provided with the \HS program, but the users can
also create their own observable sets with selected observables. We stress, however, that the user needs to avoid a statistical overlap with other observables contributing to the total $\chi^2$ (\emph{see below}) when compiling
an observable set that differs from the ones that are predefined in \HS.

After the completed LHC runs at \SIlist{7;8}{\TeV} the two experiments ATLAS and
CMS released combined measurements of the signal
rates~\cite{Khachatryan:2016vau} and the Higgs boson mass~\cite{Aad:2015zhl}.
The signal rates were presented as unfolded inclusive measurements in 20 pure
channels, \ie one production process combined with one decay process (see Tab.~8
and Fig.~7 of Ref.~\cite{Khachatryan:2016vau}), accompanied with a $20\times20$
correlation matrix. The Higgs boson mass was determined to
$\hat{m}_\text{LHC~Run-1} = \SI[parse-numbers=false]{125.09 \pm 0.21
        (\text{stat.}) \pm 0.11(\text{syst.})}{\GeV}$~\cite{Aad:2015zhl}. These
experimental results give rise to a separate $\chi^2$ contribution in \HSv{2}
(\emph{second row} in \cref{Tab:chi2contributions}); more details will be given
in \cref{Sec:chi2LHCRun1}.

Another experimental input format that became available during Run-2 of the LHC
are the simplified template cross section (STXS)
measurements~\cite{deFlorian:2016spz} (\emph{third row} in
\cref{Tab:chi2contributions}).
\texttt{Higgs\-Signals-2} features a new \texttt{Fortran}
module to handle this input. While the treatment of these measurements in \HS is
similar to the usual peak observables, the new module allows for a more
versatile handling and additional features, as will be discussed in
\cref{sec:STXS}. In the long run, the \HS\ framework handling the STXS observables will
supersede the peak-centered $\chi^2$ method also when the conventional (inclusive) signal strength measurements are used. In the meantime, STXS and peak observables can be used in parallel within one ``observable set'', each type giving rise to a separate $\chi^2$ value which can be added if there is no statistical overlap in the corresponding measurements.
As the Higgs mass measurements are always
implemented in association with a signal rate measurement of the relevant
experimental channel within the \HS program, the new STXS observables can also
handle accompanying mass measurements.

\subsection{The peak-centered \texorpdfstring{$\chi^2$}{chi-squared} method in \HSv{2}}%
\label{Sec:chi2peak}

Within a given BSM model, the Higgs boson signal strength $\mu$ predicted for a specific bin or category of an experimental Higgs analysis is defined as
\begin{equation}
    \mu = \frac{\sum_i \epsilon_i {\left[ \sigma \times \mathrm{BR}\right]}_i}
    {\sum_j \epsilon_\text{SM,j} {\left[ \sigma_\text{SM} \times \mathrm{BR}_\text{SM}\right]}_j}\eqcomma
    \label{Eq:mu_pred}
\end{equation}
where the sums run over all contributing \emph{channels} $i$, which by
definition consist of one Higgs production and one decay mode (see above). The
numerator contains the experimentally observable signal rate predicted by the
model, where $\epsilon_i$ is the experimental \emph{signal
efficiency}\footnote{Experimentally, one distinguishes between the efficiency
(related to the detector performance and object reconstruction) and signal
acceptance $\mathcal{A}$ (\ie the analysis-specific fraction of signal events
passing the full signal selection). Here, the \emph{signal efficiency}
$\epsilon$ refers to the product of the experimental efficiency and signal
acceptance.}$^,$\footnote{\label{footnotezeta} Without loss of generality, the
normalization of the signal rate to the SM expectation in \cref{Eq:mu_pred}
allows us to redefine all $\epsilon_i$ and $\epsilon_{\text{SM},i}$ such that
$\epsilon_{\text{SM},1} \equiv 1$. In this way, the $\epsilon_i$ and
$\epsilon_{\text{SM},i}$ describe the \emph{relative} model and SM signal
efficiency, respectively, of the channel $i$ with respect to the first ($i=1$)
channel in the SM.} and ${\left[\sigma \times \mathrm{BR} \right]}_i$ is the
model-predicted signal rate (i.e., in the narrow width approximation, the
production cross section, $\sigma$, times the branching ratio, $\mathrm{BR}$)
for channel $i$. The denominator contains the corresponding quantities for a SM
Higgs boson. Hence, the signal strength $\mu$ is a SM-normalized signal rate.
The signal efficiencies in the model, $\epsilon_i$, can be different than those
in the SM, $\epsilon_{\text{SM},i}$, if the Higgs candidate has different
kinematical properties, e.g., arising from higher-dimensional operators. Per
default, we assume $\epsilon_i = \epsilon_{\text{SM},i}$, but the user can
directly set $\epsilon_{i}$ if desired, see \cref{ssec:efficiencies} for further
details.

On the experimental side, in an analysis of the Higgs signal at $\SI{125}{\GeV}$, the measurement of the signal strength $\mu$ in a specific bin or category is performed by assuming SM Higgs properties for the observed particle, $\epsilon_i =
    \epsilon_{i,\text{SM}}$. The signal strength is determined by rescaling the SM-predicted signal rate for all involved
channels $i$ by a universal factor $\tilde{\mu}$,
\begin{equation}
    {[\sigma \times \mathrm{BR}]}_i = \tilde{\mu} \cdot {\left[\sigma_{\text{SM}} \times \mathrm{BR}_{\text{SM}}\right]}_i\eqdot
\end{equation}
The scale factor $\tilde\mu$ that best fits the observation is the central
value of the measurement, which we shall denote as $\hat{\mu} =
    \tilde{\mu}|_\text{best-fit}$. Furthermore, the fitting procedure provides the
upper and lower $\CL{68}$ uncertainties, $\Delta \hat{\mu}_\mathrm{up}$ and $\Delta
    \hat{\mu}_\mathrm{low}$, respectively.

In some analyses the measurements in experimental bins or categories are unfolded onto pure production and decay channels by fitting the corresponding signal strengths to all measurements simultaneously. This results in signal strength measurements of pure channels, however, with often significant correlations induced by the unfolding process. These types of measurements are special cases of the above definition, \cref{Eq:mu_pred} (without a sum over $i$), and can be treated analogously in this method. However, it is then very important that a correlation matrix is provided by the experiments and properly included in the $\chi^2$ calculation in \HS.\footnote{Even in the case of signal strength measurements in experimental bins or categories there are correlated systematic uncertainties which can be incorporated in \HS.} More details and discussions will be given in the following subsections and in \cref{sec:performance}.

As it will be useful in the following, we
decompose $\mu$, \cref{Eq:mu_pred}, as
\begin{equation}
    \mu = \sum_i \zeta_i \,\omega_{i}^\text{SM}\, c_i\eqcomma
    \label{Eq:mu_decomposed}
\end{equation}
with \emph{relative efficiency modifiers} $\zeta_i \equiv
    \epsilon_i/\epsilon_{\text{SM},i}$, as well as the \emph{SM channel weights},
$\omega_{\text{SM},i}$, and \emph{individual channel signal strengths}, $c_i$,
given by
\begin{align}
    \omega_{i}^\text{SM} & = \frac{\epsilon_{\text{SM},i} {\left[ \sigma_{\text{SM}} \times \mathrm{BR}_{\text{SM}}\right]}_i}{\sum_j \epsilon_{\text{SM},j} {\left[ \sigma_{\text{SM}} \times \mathrm{BR}_{\text{SM}}\right]}_j}\eqcomma \label{Eq:SMweight} \\
    c_i                  & = \frac{{\left[\sigma \times \mathrm{BR}\right]}_i}{{\left[\sigma_\text{SM} \times \mathrm{BR}_\text{SM}\right]}_i}\eqdot
\end{align}

As explained in the \HSv{1} documentation~\cite{Bechtle:2013xfa} the  aim of the
peak-centered $\chi^2$ method is to perform a $\chi^2$ test for the hypothesis
that a \emph{local excess, ``signal'' (or ``peak observable''), in the observed
    data at a specified mass is generated by the model}. The experimental input for
this method are the signal strength measurements performed at a \emph{specified
    mass}, $\hat{m}$, as well as measurements of the Higgs boson mass from the
$\gamma\gamma$ and $ZZ^{(*)}\to 4\ell$ final states. Of course, ideally, the
mass measurements and the specified mass of the signal strength measurements
should at least approximately coincide (assuming the signals originate from the
same Higgs boson). In this way, each signal strength measurement $\hat{\mu}$ is
intertwined with a specific Higgs mass value, which plays either the role of
just a reference point or an actual measurement (that will contribute to the
$\chi^2$ test). Thus, a \emph{peak observable} $\hat{p}$ is defined as one
$\hat{\mu}$ measurement at a mass value $\hat{m}$ (with mass resolution $\Delta
    \hat{m}$). If $\hat{m} \pm \Delta\hat{m}$ is furthermore treated as a
measurement, we call it a \emph{mass-sensitive peak
    observable}.\footnote{Observables handled via the new \HSv{2} module for STXS
    measurements are treated analogously, \ie among STXS observables we again
    distinguish between rate measurements at a hypothesized mass value and
    measurements of both the rate and mass (see \cref{sec:STXS}). However, none of
    the STXS observables implemented in the current version \HScurrent\ includes a
    mass measurement.}

\subsubsection{The \texorpdfstring{$\chi^2$}{chi-squared} calculation from the signal rates}

We now briefly review the $\chi^2$ evaluation in the peak-centered $\chi^2$
method of \HS. As already mentioned in \cref{Sec:chi2_overview},
\cref{Tab:chi2contributions}, the total $\chi^2$ value obtained in this method
is composed of a $\chi^2$ part from the signal strength observables and a
$\chi^2$ part from the Higgs mass observables, $\chi^2_\text{peak} =
    \chi^2_{\text{peak},\mu} + \chi^2_{\text{peak},m}$. For $N$ peak observables,
the signal strength part is given by
\begin{equation}
    \chi^2_{\text{peak},\mu} = {(\boldsymbol{\hat\mu} - \boldsymbol{\mu})}^T \boldsymbol{C_\mu}^{-1}  (\boldsymbol{\hat\mu} - \boldsymbol{\mu})\eqcomma
\end{equation}
where $\boldsymbol{\hat\mu}$ and $\boldsymbol{\mu}$ are $N$-dimensional vectors
of the measured and predicted signal strength, respectively.
The signal strength covariance matrix  $\boldsymbol{C_\mu}$ describes the signal
rate uncertainties and incorporates correlations of the major uncertainties
between the peak observables using publicly available information from the
experimental analyses and theory predictions. For all peak observables, we
include the correlation of the luminosity uncertainty (in $\%$ of the measured
$\mu$) for each experiment and center-of-mass energy, as well as the
correlations of theoretical rate uncertainties. For the latter, we use the
predictions from the LHC Higgs cross section work group (LHC HXSWG) Yellow
report 4~\cite{deFlorian:2016spz} for the parametric and theoretical rate
uncertainties of each production and decay process. Assuming a SM Higgs boson
with mass \SI{125.09}{\GeV}, we construct \emph{relative covariance matrices}
for the production cross sections and branching ratios of the processes listed
in \cref{Tab:pdidentifier}, denoted as $\boldsymbol{C}^\text{SM}_\sigma$ and
$\boldsymbol{C}^\text{SM}_\mathrm{BR}$, respectively.\footnote{We neglect
    correlations of theoretical/parametric uncertainties between production and
    decay rates. These correlations cannot be unambiguously reconstructed from the
    information given in Ref.~\cite{deFlorian:2016spz}. Furthermore, these
    correlations are expected to be subdominant, at least for the main experimental
    channels. See also Ref.~\cite{Arbey:2016kqi} for a discussion.} For instance,
correlations are induced by common error sources, \eg the uncertainty in the
strong coupling constant $\alpha_s$, or --- for the branching fractions ---
through the division of the partial decay widths by the total decay width. As
the theoretical rate uncertainties in the tested model can be different, we
define analogous matrices, $\boldsymbol{C}^\text{model}_\sigma$ and
$\boldsymbol{C}^\text{model}_\mathrm{BR}$, for the tested model. Per default,
these are assumed to be the same as in the SM, but can be changed by the
user.\footnote{Instructions for how these matrices can be evaluated and
    incorporated in \HS are given in the online documentation~\cite{HSDoc}.}

In the construction of the signal strength covariance matrix
$\boldsymbol{C_\mu}$ the theoretical rate uncertainties enter as
\begin{equation}
    {(\boldsymbol{C_\mu})}_{\alpha\beta} = \left( \sum_{a=1}^{k_\alpha} \sum_{b=1}^{k_\beta}  \left[ {(\boldsymbol{C_\sigma}^\text{model})}_{p(a)p(b)} + {(\boldsymbol{C_\mathrm{BR}}^\text{model})}_{d(a)d(b)} \right] \cdot \omega_{\alpha,a}^\text{model} \omega_{\beta,b}^\text{model} \right) \mu_\alpha \mu_\beta\eqcomma
\end{equation}
where the notation is as follows (see also Ref.~\cite{Bechtle:2013xfa}): The
index $a$ ($b$) runs over the $k_\alpha$ ($k_\beta$) contributing channels of
the peak observable $\hat{p}_\alpha$ ($\hat{p}_\beta$). The index mappings
$p(x)$ and $d(x)$ project onto the production and decay mode identifier
(cf.~\cref{Tab:pdidentifier}), respectively, for the channel $x$. The
\emph{model channel weights} $\omega_{\alpha,a}^\text{model}$ for the peak
observable $\hat{p}_\alpha$ are defined in analogy to \cref{Eq:SMweight} as
\begin{equation}
    \omega_{i}^\text{model} = \frac{\epsilon_{i} {\left[ \sigma \times \mathrm{BR}\right]}_i}{\sum_j \epsilon_{j} {\left[ \sigma \times \mathrm{BR} \right]}_j}\eqcomma
\end{equation}
i.e.~the relative contribution of the channel $i$ to the total signal rate, as
\emph{predicted by the model}. Through the multiplication with the channel
weight and the signal strength modifier, the \emph{relative} squared theory
uncertainties encoded in $\boldsymbol{C}^\text{model}_\sigma$ and
$\boldsymbol{C}^\text{model}_\mathrm{BR}$ are converted into \emph{absolute}
squared signal strength uncertainties, as required for the covariance matrix.
The theoretical rate uncertainties in the SM, which are typically already
included in the signal strength measurement and therefore need to be subtracted
from the $\hat\mu$ uncertainty beforehand (see Ref.~\cite{Bechtle:2013xfa} for
more details), are evaluated in a similar way using
$\boldsymbol{C}^\text{SM}_\sigma$ and $\boldsymbol{C}^\text{SM}_\mathrm{BR}$.

In addition to the luminosity and theoretical rate uncertainties, other
correlations between the peak observables can be included in \HS if the relevant
information is available. This can be done by providing a corresponding correlation matrix within the observable set which is then loaded into \HS\footnote{Such
    correlation matrices are loaded automatically with the observable set during the
    initialization of the program. Each matrix should be given as three column data
    (index1, index2, correlation coefficient) in a separate file with extension
    \texttt{.corr} in the observable set directory.}. Ideally such matrices are
published by the experiment along with the measurements. Indeed, as
we will discuss in \cref{sec:performance}, including the correlation matrix in
general drastically improves the accuracy of the $\chi^2$ test, in particular if the $\mu$ measurements correspond to pure channels obtained after an unfolding process.

\subsubsection{The \texorpdfstring{$\chi^2$}{chi-squared} calculation from the
    Higgs mass}%
\label{ssec:chisqpeakmass}

In the calculation of the $\chi^2$ contribution from the Higgs boson mass
measurements, $\chi^2_{\text{peak},m}$, \HS\ allows three different choices to
model the probability density function (pdf) of the Higgs boson mass: (\emph{1})
as a uniform (box) distribution; (\emph{2}) as a Gaussian, or (\emph{3}) as a
box with Gaussian tails. For the Gaussian pdf, the theoretical mass uncertainty
is treated as fully correlated among mass sensitive peak observables to which
the same Higgs boson has been assigned.

If a Higgs boson $h_i$ with mass $m_i$ and theoretical mass uncertainty $\Delta
    m_i$ is assigned to a mass-sensitive peak observable $\hat{p}_\alpha$ with
mass measurement $\hat{m}_\alpha \pm \Delta\hat{m}_\alpha$, its $\chi^2$
contribution is given by\footnote{We suppress in the following the
    subscript '$\text{peak},m$' for the individual $\chi^2$ contributions from
    the peak observables.}
\begin{equation}
    \chi^2_{\alpha} = \begin{cases}
        0      & \mbox{for}~|m_i - \hat{m}_\alpha| \le \Delta m_i + \Lambda \Delta \hat{m}_\alpha \\
        \infty & \mbox{otherwise}\eqcomma                                                         \\
    \end{cases}
    \label{eq:chi2m_box}
\end{equation}
for a uniform (box) pdf, and
\begin{align}
    \chi^2_{\alpha} =
    \begin{cases}
        0                                                              & \text{for } |m_i - \hat{m}_\alpha | \le \Delta m_i, \\
        {(m_i -\Delta m_i - \hat{m}_\alpha)}^2/{(\Delta \hat{m}_\alpha)}^2 & \text{for } m_i - \Delta m_i > \hat{m}_\alpha,      \\
        {(m_i +\Delta m_i - \hat{m}_\alpha)}^2/{(\Delta \hat{m}_\alpha)}^2 & \text{for } m_i + \Delta m_i < \hat{m}_\alpha,      \\
    \end{cases}
    \label{eq:chi2m_boxgaussian}
\end{align}
for a box-shaped pdf with Gaussian tails. In both cases, the total $\chi^2$
contribution from the Higgs mass measurements is given by
$\chi^2_{\text{peak},m} = \sum_\alpha^M \chi^2_{\alpha}$, with $\alpha$ running
over the $M$ mass-sensitive peak observables. The parameter $\Lambda$ appearing
in \cref{eq:chi2m_box} is the Higgs assignment range which will be discussed in
detail in \cref{sec:overlappingHiggs}. Note that the uniform (box) pdf is a
rather poor description of the Higgs mass pdf (in particular, of the
experimental uncertainty) and is included mostly for illustrative purposes. In
the (default) case of a Gaussian pdf, the total $\chi^2$ contribution from the
Higgs mass reads
\begin{align}
    \chi^2_{\text{peak},m} = {(\boldsymbol{\hat{m}} - \boldsymbol{m}_i )}^T \boldsymbol{C_m}^{-1} (\boldsymbol{\hat{m}} - \boldsymbol{m}_i ).
    \label{eq:chi2m_gaussian}
\end{align}
Here, $\boldsymbol{\hat{m}}$ and $\boldsymbol{m}$ are $M$-dimensional vectors
and contain the measured and predicted Higgs mass for the mass-sensitive peak
observables, respectively. The diagonal elements of the Higgs mass covariance
matrix, $\boldsymbol{C_m}$, contain the squared experimental uncertainty,
${(\Delta \hat{m}_{\alpha})}^2$, while the squared theory mass uncertainty,
${(\Delta m_i)}^2$, enters all matrix elements ${(\boldsymbol{C_m})}_{\alpha\beta}$
(including the diagonal) where the Higgs boson $h_i$ is assigned to both peak
observables $\hat{p}_\alpha$ and $\hat{p}_\beta$.

\subsection{The \texorpdfstring{$\chi^2$}{chi-squared} contribution from the LHC Run-1 combination}%
\label{Sec:chi2LHCRun1}

In Ref.~\cite{Khachatryan:2016vau} the ATLAS and CMS collaborations have
published a combined analysis of their datasets at
$\sqrt{s}=\text{\SIlist{7;8}{\TeV}}$ center-of-mass energy. In its most general
form, the results were presented as signal rate measurements for the five
dominant production modes ($pp\to H$, VBF, $WH$, $ZH$, $t\bar{t}H$) times the
five dominant decay modes ($H\to \gamma\gamma, ZZ, WW, \tau^+\tau^-, b\bar{b}$).
Five channels --- for which the sensitivity for a measurement had not been
reached in Run-1 --- were omitted  resulting in a total of 20 measured
channels. The analysis was performed for a Higgs mass of
$m_H=\SI{125.09}{\GeV}$. The correlations of all theoretical uncertainties
(assuming the SM Higgs boson) and experimental systematic uncertainties were
provided in the form of a $20\times 20$ correlation matrix. In addition to the signal rate
measurements, the Higgs mass was determined to $m_H =
    \SI[parse-numbers=false]{125.09 \pm 0.21 (\text{stat.}) \pm 0.11
        (\text{syst.})}{\GeV}$ from the combined ATLAS and CMS Run-1
data~\cite{Aad:2015zhl}.

\HS incorporates the 20 Run-1 signal rate measurements (including their
correlations) via a dedicated run routine (see online documentation~\cite{HSDoc}). Furthermore, a
$\chi^2$ contribution from the Higgs mass measurement, taken to be \SI{125.09 +-
    0.24}{\GeV} in \HS, arises, if a Higgs boson in the tested model is assigned to
these measurements. The resulting $\chi^2$ contributions from the LHC Run-1
signal rates and mass measurements can then be added to other $\chi^2$ values
obtained by the \HS runs (\eg using $\SI{13}{\TeV}$ results).

As such, the ATLAS and CMS combined analysis represents the benchmark of
$\sqrt{s}=\text{\SI{7}{\TeV}}$ and $\text{\SI{8}{\TeV}}$ results. For the
validation of the \HS
methodology, reproducing these results from the signal strength measurements of
the individual analyses at $7$ and $\SI{8}{\TeV}$ provides an excellent cross-check, as will be shown in
\cref{sec:performance:Run1combined}.

\subsection{Simplified Template Cross Section (STXS) measurements}%
\label{sec:STXS}

During Run~2 of the LHC the experimental collaborations started to employ the
STXS framework~\cite{deFlorian:2016spz} for the presentation of Higgs rate
measurements. These measurements comprise a different type of input observables
in \HS, resulting in an additional $\chi^2$ contribution (see
\cref{Tab:chi2contributions}) that is evaluated in a new module. Although
the basic information given for STXS measurements is very similar to the
conventional peak (or ``$\mu$'') measurements, we choose to treat them
separately in order to provide and enable a more flexible handling and
additional features for the STXS observables. It should be noted that ---
although we denote the new module and the observables as STXS --- any signal
rate measurement can be handled via this framework, regardless whether this
measurement fulfills the official definitions of an STXS bin. For instance,
inclusive measurements or fiducial differential cross section measurements can
be treated, as long as the necessary information is provided. Furthermore, if
the peak-centered $\chi^2$ method and the STXS method are simultaneously used,
each measurement can only be implemented either as peak or STXS-observable in
order to avoid double-counting.

The STXS framework was designed in order to (\emph{i}) provide
commonly-defined exclusive phase-space regions for Higgs production processes facilitating the incorporation of information on differential cross sections,
(\emph{ii}) maximize the measurement's sensitivity to the underlying physical
process(es) while minimizing its dependence on theory assumptions, (\emph{iii})
isolate possible BSM contributions by defining bins that have an enhanced
sensitivity to BSM effects. By addressing these targets, the STXS framework
has the goal to simplify the procedure of BSM model testing against the experimental
data.

The defined exclusive regions of phase space, called ``bins'' for
simplicity, are specific to the different Higgs production modes. With
increasing amounts of data, measurements of differential distributions of the
various Higgs processes become possible. Therefore, STXS bins have been defined
for three stages (``stage~0'', ``stage~1'', ``stage~2'', and substages thereof) to allow a transition
from more inclusive to more differential measurements, see Ref.~\cite{deFlorian:2016spz}
for details. This transition can be performed independently for each production
mode. In order to maximize the sensitivity of the current data, various decay
modes can be combined in the determination of the STXS bins, assuming the SM
Higgs boson as a kinematic template.

Within \HS each experimental measurement of an STXS bin enters the $\chi^2$
calculation as an individual observable. It should be noted, however, that many of the STXS bins (in particular at Stage~0) do not represent ``pure'' Higgs signal channels but instead are comprised of different production channels (similar to the conventional ``peak'' observables, see
\cref{Sec:chi2peak}). This is because the STXS definitions are driven by the
particle level objects produced in association with the Higgs boson, whereas the signal
channels within \HS are distinguished by the topology and Higgs coupling
dependence. For instance, the two distinct production channels of
vector boson fusion, $pp\to q\bar{q} h$, and Higgs-strahlung with a hadronically
decaying vector boson, $pp\to Vh$ (with $V\to q\bar{q}$), are considered
together in one Stage-0 STXS bin, as they lead to the same final state
particles. Only at a later stage the STXS framework aims to separate these
processes into exclusive bins by employing dedicated cuts. Therefore, \HS in
general treats the STXS observables as multi-channel measurements, and --- if no
further information is given --- assumes that the combined processes have
similar signal efficiency (using the SM Higgs boson as a kinematic template).
Furthermore, the STXS measurements are implemented in \HS preferably as
\emph{absolute} rates, along with corresponding rates for the SM prediction as
an important reference value (see below). This avoids the complication or peak observables that the implemented SM-normalized rates (i.e.~signal strengths $\mu$) have to be adjusted in case an updated SM prediction for the signal rate becomes available.

The STXS framework provides a smooth transition from more inclusive to more
exclusive, \ie differential, Higgs rate measurements, as more data is collected.
In \HS however, the standard user input --- handled via the \HB framework ---
are the predictions for the \emph{inclusive} (SM-normalized) cross sections and
decay rates. In order to enable tests of non-trivial new physics effects on
{differential distributions of Higgs boson processes}, the user can specify
independent predictions individually for every STXS observable. This is done by
providing \emph{rate modification factors} $r_i^j$ for each STXS bin, defined as
\begin{align}
    r_i^j \equiv \frac{\sigma_i^j (\text{STXS-bin}) / \sigma_i^j (\text{incl.})}{\sigma_{\text{SM}, i} (\text{STXS~bin}) / \sigma_{\text{SM}, i} (\text{incl.})},
    \label{eq:STXS_ratemodifier1}
\end{align}
where the index $i$ labels the signal channels of the STXS~bin, and the index
$j$ labels the neutral Higgs bosons in the model. The prediction for the
exclusive STXS~bin rate (inclusive rate) for channel $i$ and Higgs boson $h_j$
in the model is denoted $\sigma_i^j(\text{STXS-bin})$
($\sigma_i^j(\text{incl.})$), whereas the corresponding SM predictions are given
by $\sigma_{\text{SM}, i}(\text{STXS-bin})$ ($\sigma_{\text{SM}, i}(\text{incl.})$). Using the
inclusive signal strength modifier from the conventional input, $\mu_i^j =
    \sigma_i^j (\text{incl.})/\sigma_{\text{SM}, i} (\text{incl.})$, the model-predicted
signal rate for the STXS~bin is internally calculated as
\begin{equation}
    \sigma_i^j (\text{STXS-bin}) = \mu_i^j \, r_i^j \, \sigma_{\text{\text{SM}},i} (\text{STXS-bin}).\label{eq:STXS_rate}
\end{equation}
This quantity then enters the $\chi^2$ test. Note that the knowledge of the SM
prediction for the STXS-bin is essential here and is taken from the experimental analysis along with the corresponding
measurement. The rate modification factors can be set using the subroutine
\Code{STXS::assign_modification_factor_to_STXS}  (see online documentation~\cite{HSDoc}). Per default, $r_i^i = 1$, i.e.~the SM Higgs boson is used as a kinematic template. Note that the quantities $\sigma_{\text{SM}, i} (\text{STXS~bin})$ and $\sigma_{\text{SM}, i} (\text{incl.})$ in \cref{eq:STXS_ratemodifier1} can in principle be extracted from \HS, i.e.~only the model-specific numerator of \cref{eq:STXS_ratemodifier1} needs to be evaluated externally. However, we recommend to also evaluate the denominator (i.e., the SM predictions) of \cref{eq:STXS_ratemodifier1} with the same computing setup in order to provide predictions for both the model and SM at the same level of accuracy.

The STXS functionality in \HS\ and the associated $\chi^2$ test can be
employed with measurements of individual cross sections, or branching ratios, or
cross section times branching ratios, or even ratios of branching ratios,
either given in absolute numbers, or in terms of a SM-normalized quantity. In
particular, as mentioned above, any peak observable could in principle be implemented as an STXS
observable (despite of generally not being consistent with the definition of
STXS bins). More information on the technical implementation of STXS
observables in \HS can be found in \cref{app:STXS}.

\subsection{Handling the potential complexity of BSM Higgs sectors}

When testing the Higgs sector predictions of a BSM theory with Higgs
measurements some issues that affect the calculation of the predicted signal
strength $\mu$, \cref{Eq:mu_pred}, warrant special consideration.

\subsubsection{Non-trivial signal efficiencies \texorpdfstring{$\epsilon_i \ne \epsilon_{i,\text{SM}}$}{}}%
\label{ssec:efficiencies}
Per default, \HS assumes $\epsilon_i \equiv \epsilon_{i,\text{SM}}$ for all
experimental analyses and channels. If significant modifications of the signal
efficiencies $\epsilon_i$ with respect to the SM expectation
$\epsilon_{i,\text{SM}}$ are anticipated in the model, these should be
evaluated \emph{externally} by the user for every relevant experimental analysis
and every contributing channel $i$, for instance, by running a Monte-Carlo
simulation of the BSM Higgs signal and processing it through the experimental
analysis. In addition, the SM Higgs boson signal should also be simulated and
processed for comparison, such that \emph{relative efficiency modifiers}
$\zeta_i \equiv \epsilon_i/\epsilon_{i,\text{SM}}$ can be determined. These can
then be fed into \HS in order to account for the modified signal efficiencies in
the $\chi^2$ test (through the subroutine
\Code{assign_modelefficiencies_to_peak}, see also \cref{sec:user}). The usage of
the relative efficiency modifiers $\zeta_i$ has the advantage that uncertainties
related to the recasting method largely cancel, as they typically affect both
the model's Higgs signal and the SM Higgs signal in a similar way. For the
STXS observables these effects are captured by the usage of rate modification factors,
$r^j_i$, described in \cref{sec:STXS}. For a recent study of the CP properties of the Higgs--top-quark interaction in which these features have been employed see Ref.~\cite{Bahl:2020wee}.

\subsubsection{Different predicted and observed mass with theory
    uncertainties}\label{ssec:diffmasses} If the Higgs signal candidate's mass $m$
is a free parameter of the model, it can be chosen to exactly match the observed
Higgs mass $\hat{m}$, for which the signal strength measurements have been
performed. However, this is not always possible for all observables
simultaneously, as often subsets of signal strength measurements have been
performed at slightly different hypothesized Higgs masses by ATLAS and CMS\@. If
$m \ne \hat{m}$, the mass dependence should be taken into account in the signal
strength calculation, \cref{Eq:mu_pred}, by evaluating the model predicted
signal rate at the predicted mass $m$, while evaluating the SM signal rate at
$\hat{m}$.

In contrast, if the Higgs signal candidate's mass $m$ is not a free parameter,
but instead a model-prediction with a non-negligible theoretical uncertainty
$\Delta m$, there are two reasonable choices: First, one could neglect the
uncertainty of the mass prediction and do the same as above by taking
into account the mass
dependence of the model-predicted signal rate in \cref{Eq:mu_pred};
second, one could factor out the mass dependence of the rates and
calculate both the model- and SM-predicted signal rates at the predicted mass
value $m$.\footnote{Here, an implicit assumption is made that the
    relative variation of the Higgs boson candidate's cross sections and branching
    ratios with its mass are the same in the model and the SM\@.} In \HSv{2} the default option is a combination of both: If the
difference between observed and predicted mass is less than the theoretical
uncertainty, $|m - \hat{m}| \le \Delta m$, \HS evaluates both the numerator and
denominator in \cref{Eq:mu_pred} at $m$. Otherwise, if $m > \hat{m} + \Delta m$
or $m < \hat{m} - \Delta m$, it takes into account the mass dependence by
evaluating the numerator at $m$ and the denominator at $\hat{m} + \Delta m$ or
$\hat{m} - \Delta m$, respectively. In this way, the mass value chosen for the
reference SM rate is considered to be the nearest mass value to $m$ allowed by
the theoretical mass uncertainty. Furthermore, this choice leads to a smooth
transition between the different mass regimes. This behavior can be controlled
through the \Code{setup_rate_normalization} subroutine, and the theoretical
mass uncertainties $\Delta m$ can be given as optional input by the user, see the
online documentation~\cite{HSDoc} for details.

\subsubsection{Multiple overlapping Higgs-boson candidates}%
\label{sec:overlappingHiggs}
A prime application of \HS are BSM models with an extended Higgs sector, \ie
scenarios where more than one neutral Higgs boson can potentially contribute to
the observed Higgs signal. \HS therefore employs a dedicated algorithm --- the
\emph{assignment procedure} --- in order to determine which Higgs bosons are
contributing in the signal strength evaluation, \cref{Eq:mu_pred}, and thus
enter the $\chi^2$ test against the observed signal. If a Higgs boson is
\emph{assigned} to an observable, then its signal rate enters
\cref{Eq:mu_pred}. If no Higgs boson is assigned to an observable, the
$\chi^2$ contribution from the signal rate measurement is evaluated for zero
predicted signal rate, $\mu=0$, which typically results in a large $\chi^2$
penalty. The assignment procedure of a Higgs boson $h_i$ to an observable
$\hat{p}_\alpha$ depends on the experimental mass
resolution, $\Delta \hat{m}_\alpha$, as well as on the theoretical
Higgs mass uncertainty, $\Delta m_i$. As a general rule, if
\begin{equation}
    |m_i - \hat{m}_\alpha| \le \begin{cases}
        \Lambda \sqrt{{(\Delta m_i)}^2 + {(\Delta \hat{m}_\alpha)}^2} & \text{for a Gaussian Higgs mass pdf,} \\
        \Delta m_i + \Lambda \Delta \hat{m}_\alpha                & \text{otherwise,}
    \end{cases}\label{Eq:assignment}
\end{equation}
then the Higgs boson $h_i$ is assigned to the observable $\hat{p}_\alpha$. Here,
$\Lambda$ is a control parameter called \emph{assignment range}, with default
value $\Lambda = 1$. A few exceptions to this rule exist in the case where the
observable $\hat{p}_\alpha$ contains a mass measurement that enters the $\chi^2$
contribution from the Higgs mass, or in the case that several peak observables
are collected in an \emph{assignment group}~\cite{Bechtle:2013xfa}. In
particular, for the former case, a Higgs boson assignment is also possible even
if \cref{Eq:assignment} is not fulfilled. This happens when the total $\chi^2$
contribution from the peak observable in the case of assignment is lower than
the $\chi^2$ contribution in the case where the Higgs boson is not assigned (see
Ref.~\cite{Bechtle:2013xfa} for details). Two different assignment range
parameters are introduced for the two kinds of observables: $\Lambda$ for all
observables (peak, STXS and Run-1) that do not have an associated mass
measurement, and $\Lambda_m$ for mass sensitive observables (which are currently
available in the $\gamma \gamma$ and $ZZ\to 4\ell$ channels). This provides the
user with more flexibility to specify the mass range that is regarded as
phenomenologically viable using the \Code{setup_assignmentrange} and
\Code{setup_assignmentrange_massobs} subroutine, respectively (see online
documentation~\cite{HSDoc} for details). The default choice is $\Lambda = 1$ and
$\Lambda_m = 2$.

\begin{figure}[t]
    \includegraphics[width = 0.5\textwidth]{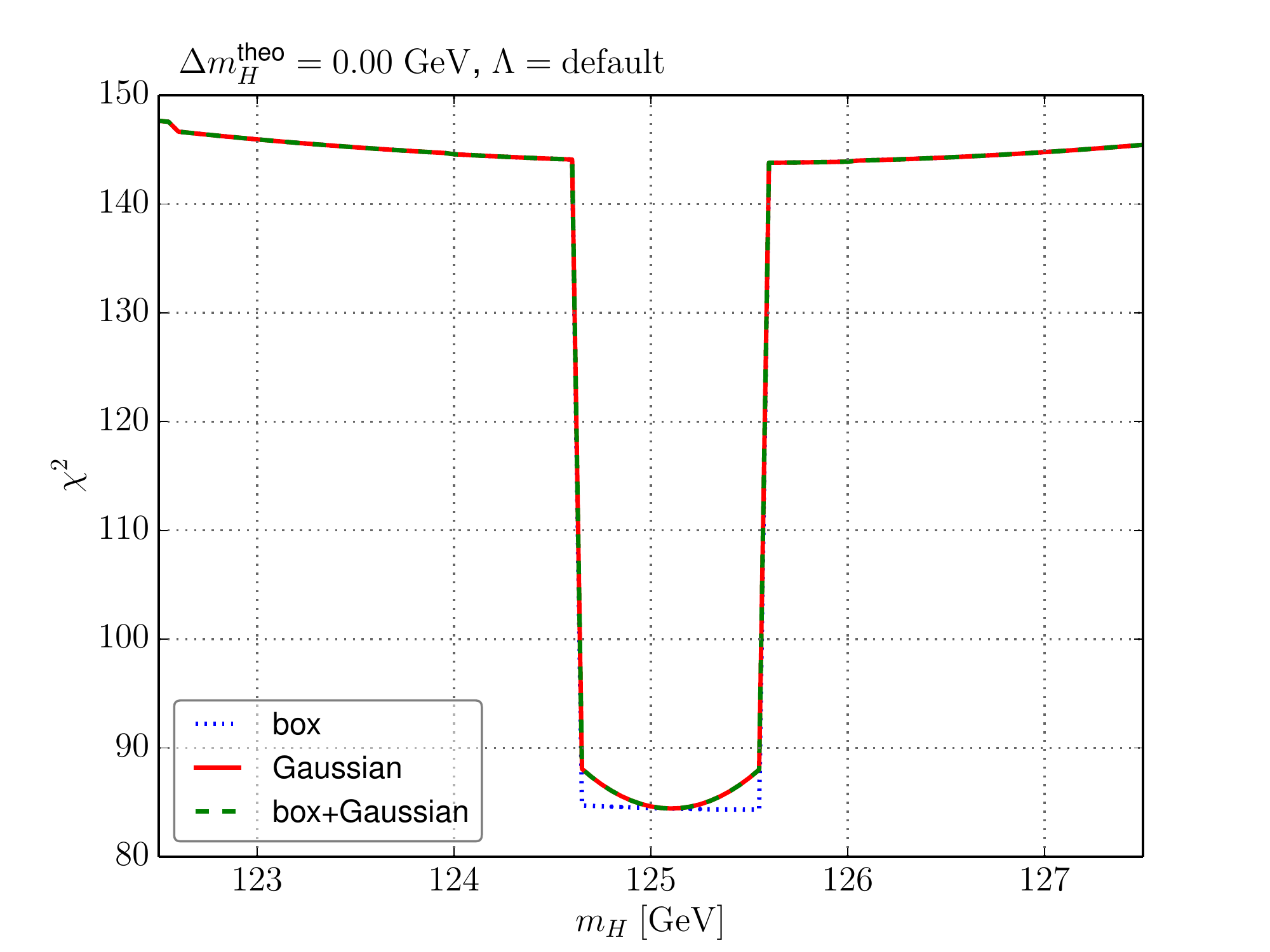}\hfill
    \includegraphics[width = 0.5\textwidth]{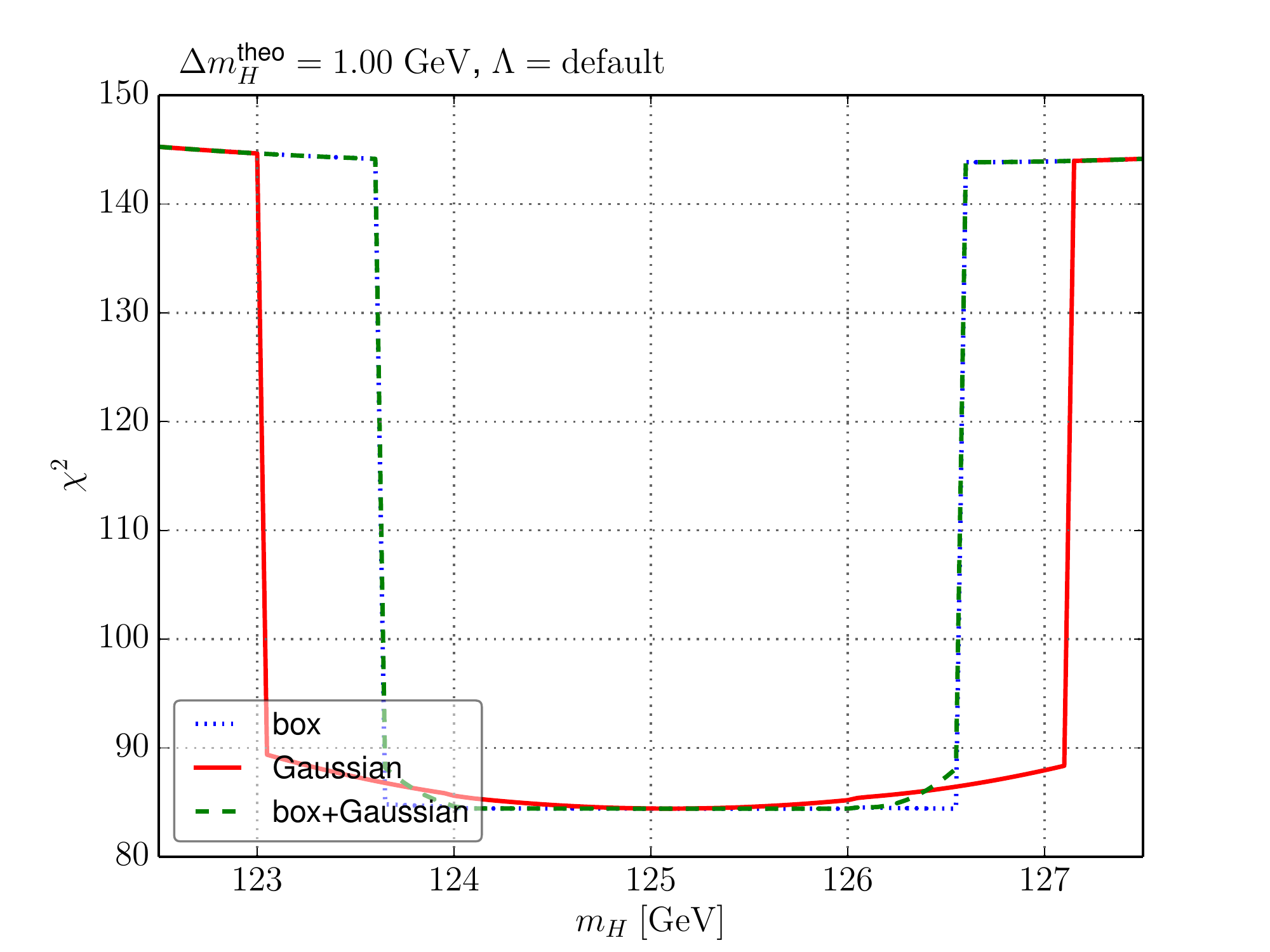}
    \includegraphics[width = 0.5\textwidth]{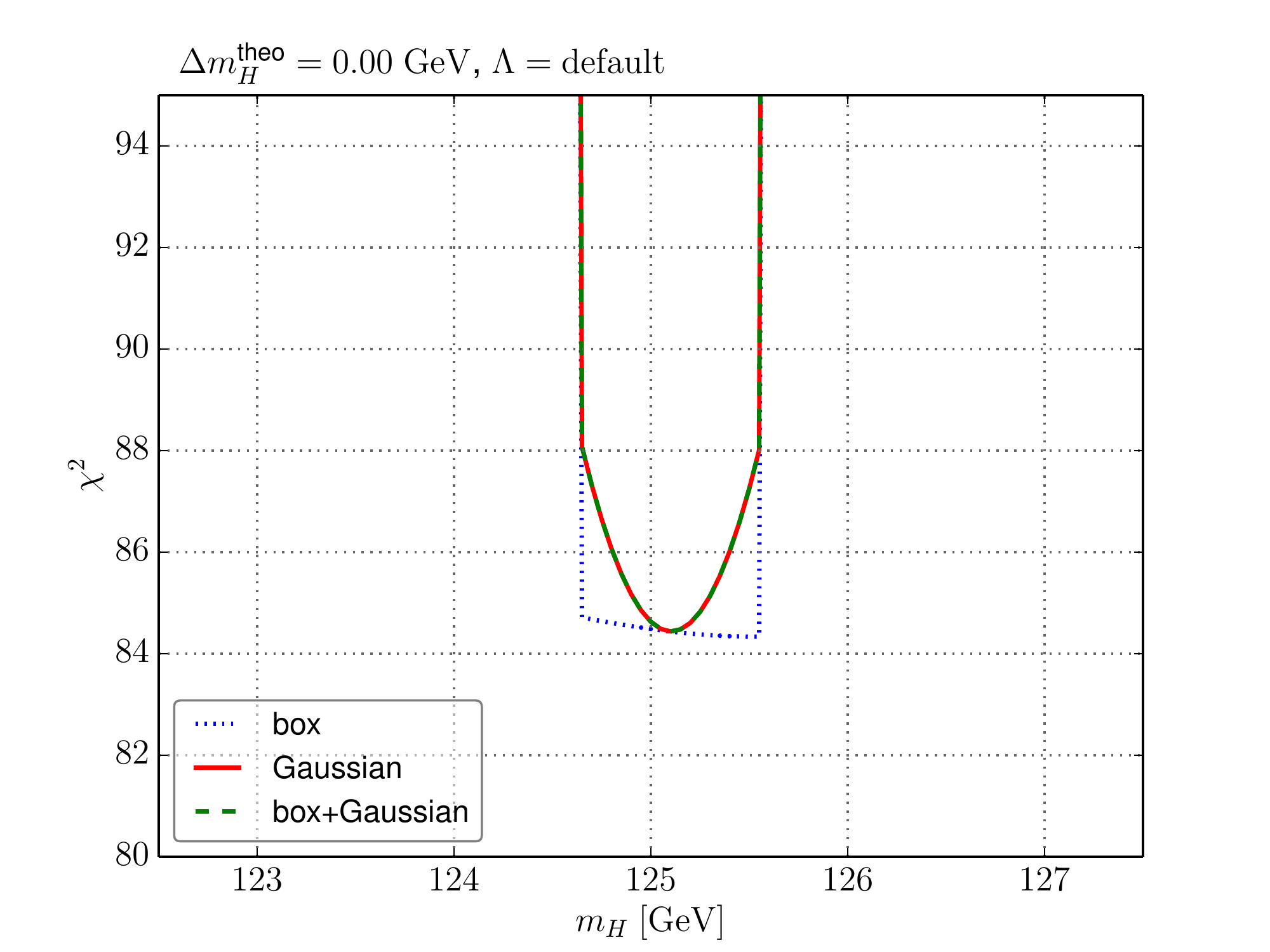}\hfill
    \includegraphics[width = 0.5\textwidth]{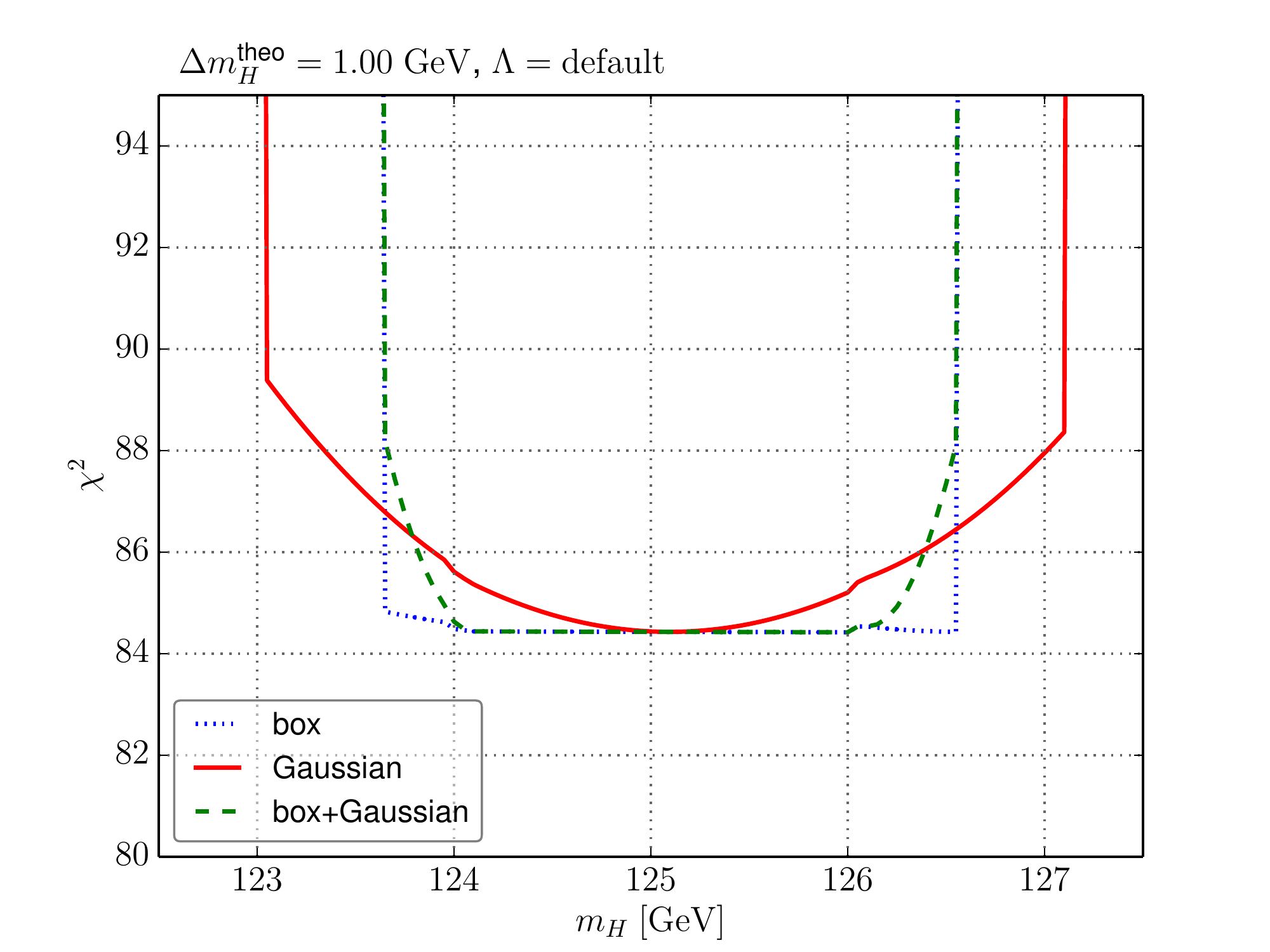}
    \caption{Total $\chi^2$ as a function of the Higgs mass, $m_H$, for the case
        of no theoretical mass uncertainty (\emph{left panels}) and a
        theoretical mass uncertainty of \SI{1}{\GeV} (\emph{right panels}). The
        Higgs couplings are set to the SM prediction, and the assignment range
        parameters ($\Lambda, \Lambda_m$) are set to their default values
        (\emph{see text}). The \emph{top panels} show the $\chi^2$ function in
        the full mass interval, whereas the \emph{bottom panels} show a zoom to
        the region of the lowest $\chi^2$ values. }\label{Fig:assignmentexample}
\end{figure}

Before we move on to cases with multiple Higgs bosons, we illustrate the
assignment procedure with a simple example. \Cref{Fig:assignmentexample}
displays the total $\chi^2$ as a function of the mass of a Higgs boson that has
couplings that are identical to those of the SM Higgs boson, for the three
possible choices of the Higgs mass pdf (box, Gaussian, box+Gaussian). The left
panels (\emph{right panels}) assume zero (a \SI{1}{\GeV}) theoretical mass uncertainty.
The step-like shape originates from the Higgs assignment procedure, where at
every step the assignment of the Higgs boson to some observables changes. For
instance, in the top left panel, the $\chi^2$ distribution for all mass pdfs
features a jump at $m =\SI{124.61}{\GeV}$ and at \SI{125.57}{\GeV}. Here, the
assignment to the LHC Run-1 observables changes due to the implemented Higgs
mass measurement of $\hat{m} = \SI{125.09 +- 0.24}{\GeV}$ and the default
setting $\Lambda_m = 2$. If $\Delta m \ne 0$ the mass interval in which the
Higgs boson is assigned to the LHC Run-1 observables is larger. Outside this
range, the $\chi^2$ rises steeply, rendering these mass values as highly
disfavored. Two more features are illustrated in this example. First, the
differences between the three Higgs mass pdf choices,
\cref{eq:chi2m_box,eq:chi2m_boxgaussian,eq:chi2m_gaussian}, becomes apparent, in
particular in the case of non-zero theoretical mass uncertainty. Second, the
plots display the treatment of the mass dependence of the signal rate
prediction, as discussed above, resulting in an asymmetric $\chi^2$ shape around
the minimum for all three Higgs mass pdf choices. The way such plots can be
obtained from \HS is illustrated in the \Code{HS_mass} example program.

We now continue with the case of multiple neutral Higgs bosons. If $N>1$ Higgs
bosons are assigned to an observable $\hat{p}_\alpha$, \cref{Eq:mu_pred} is
generalized to $\mu_\alpha = \sum_j \mu_{\alpha,j}$, where $j$ runs over all $N$
assigned Higgs bosons $h_j$.\footnote{The observable index $\alpha$ is made
    explicit here, whereas in \cref{Eq:mu_pred} it was
    omitted.}\textsuperscript{,}\footnote{Note that the procedure described in
    \cref{ssec:diffmasses} is applied to the $\mu_{\alpha,j}$ calculation, \ie for
    each assigned Higgs boson $h_j$ individually.} If the observable contains a mass
measurement, $\hat{m}_\alpha \pm \Delta \hat{m}_\alpha$, the predicted mass
value and its uncertainty are determined from the masses of the assigned Higgs
bosons, $m_j$, and their uncertainties, $\Delta m_j$, through a signal-strength
weighted average,
\begin{equation}
    \overline{m}_\alpha = \frac{\sum_j \mu_{\alpha,j} m_j}{\sum_{j} \mu_{\alpha,j}}, \qquad \overline{\Delta m_\alpha} = \frac{\sum_j \mu_{\alpha,j} \Delta m_j}{\sum_{j} \mu_{\alpha,j}},
    \label{Eq:massaverage}
\end{equation}
respectively. These averaged quantities in \cref{Eq:massaverage} then enter the
$\chi^2$ test against the Higgs mass measurement. In other words, we assume that
multiple Higgs bosons overlapping within the experimental mass resolution
(according to \cref{Eq:assignment}) would show up as a single signal peak
approximately located at $\overline{m}$. In case of a Gaussian pdf the
theoretical mass uncertainties are still treated as fully correlated
uncertainties among the observables $\hat{p}_\alpha$ and $\hat{p}_\beta$, if the
same Higgs boson $h_j$ has been assigned (denoted by the symbol `$\bowtie$'
below). The diagonal and off-diagonal entries in the covariance matrix are then
given by
\begin{align}
    \text{cov}_{\alpha\alpha} & = {(\Delta \hat{m}_\alpha)}^2 + {(\overline{\Delta m_\alpha})}^2,                                                                                                   \\
    \text{cov}_{\alpha\beta}  & = \sum_{j~\bowtie~\alpha,\beta} \frac{\mu_{\alpha,j}}{\mu_{\alpha}} \frac{\mu_{\beta,j}}{\mu_{\beta}} \overline{\Delta m}_\alpha \overline{\Delta m}_\beta \, ,
    \label{Eq:Cov_offdiag}
\end{align}
respectively, where the relative signal strength contributions
$\mu_{\alpha,j}/\mu_{\alpha}$ of the assigned Higgs boson act as
uncertainty weights in \cref{Eq:Cov_offdiag}.

The described procedure is well-motivated and accurate in the case that the
individual signals cannot even partially be resolved by the experimental
analysis, or in other words, the predicted mass differences between the assigned
Higgs bosons $h_i$ and $h_j$ are small compared to the experimental mass
resolution, $|m_i- m_j| \ll \Delta \hat{m}_\alpha$ (for all $i,j$). However, the
procedure becomes inaccurate if $|m_i - m_j| \gtrsim \Delta \hat{m}_\alpha$,
i.e.~when the individual signals are at least partially resolved. In order
to accommodate this limitation, we introduce another $\chi^2$ contribution for
the mass separation of the assigned Higgs bosons with respect to the mass
average, which for the three mass pdf choices reads as
\begin{align}
    \chi^2_{\text{ sep},\alpha} & = \sum_{j} \Theta(\mu_{\alpha,j} - c_\mu \mu_\alpha) \cdot \begin{cases}
        0      & \text{if }| m_{j} - \overline{m}_\alpha | \le \Delta m_j + \Delta \hat{m}_\alpha\eqcomma \\
        \infty & \text{else,}\end{cases}                                                 &  & \text{(box pdf)} \label{Eq:chi2sep_box} \\[.2em]
    \chi^2_{\text{sep},\alpha}  & = \sum_{j} \frac{\mu_{\alpha,j}}{\mu_\alpha} \frac{{(m_j - \overline{m}_\alpha)}^2}{{(\Delta m_j)}^2 + {(\Delta \hat{m}_\alpha)}^2}\eqcomma &  & \text{(Gaussian pdf)}                   \\[.2em]
    \chi^2_{\mathrm{sep},\alpha}     & = \sum_{j}  \frac{\mu_{\alpha,j}}{\mu_\alpha} \cdot \begin{cases}
        \frac{{(m_j + \Delta m_j - \overline{m}_\alpha)}^2}{{(\Delta \hat{m}_\alpha)}^2} & \text{if }m_j + \Delta m_j < \overline{m}_\alpha\eqcomma \\
        \frac{{(m_j - \Delta m_j - \overline{m}_\alpha)}^2}{{(\Delta \hat{m}_\alpha)}^2} & \text{if }m_j - \Delta m_j >\overline{m}_\alpha\eqcomma  \\
        0                                                                            & \text{else,}
    \end{cases}                                                        &  & \text{(box+Gaussian)}
\end{align}
where $\Theta$ is the Heaviside function. In order to avoid artificial penalties
in the box-pdf case, \cref{Eq:chi2sep_box}, only {$h_j$ that contribute at
least a fraction of $c_\mu$ (with a default value of \SI{1}{\percent}) to the
total signal strength $\mu_\alpha$ are assigned to the peak observable
$\hat{p}_\alpha$.

As a result, the total $\chi^2$ value is composed of the signal rate part,
$\chi^2_\mu$, the averaged Higgs mass part, $\chi^2_{\overline{m}}$, and the
Higgs mass separation part, $\chi^2_{\text{sep}}$. Note that the above formulas
include the ``standard'' case with only one assigned Higgs boson. Moreover, due
to the signal strength weighted averaging, a (technically) assigned Higgs boson
with zero signal strength does not contribute to the $\chi^2$.

\begin{figure}
    \centering
    \includegraphics[width = 0.55\textwidth]{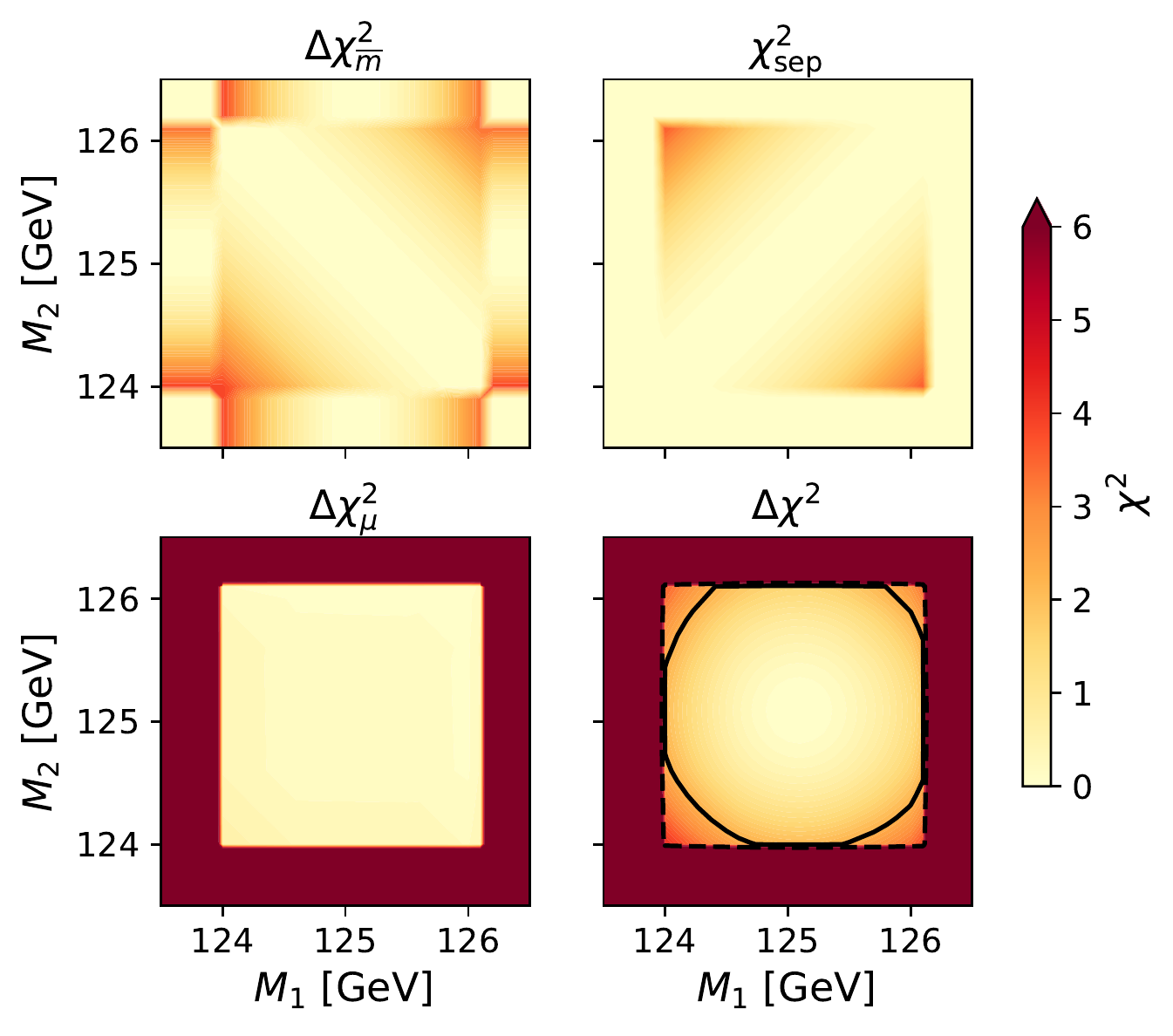}
    \caption{Individual $\chi^2$ contributions from the averaged mass,
    $\chi^2_{\bar{m}}$, the mass separation, $\chi^2_\text{sep}$, the signal
    rates, $\chi^2_\mu$, and the resulting total $\chi^2$ for a toy example with
    two Higgs bosons of masses $M_{1,2}$, theoretical mass uncertainties $\Delta
        M_{1,2}=\SI{0.5}{\GeV}$, and universal signal strengths
    $\mu_{\alpha;1,2}=0.5$. A Gaussian pdf is used, and all $\Lambda$ are at
    their default values.}\label{Fig:2Higgses}
\end{figure}

A simple example for the total $\chi^2$ and its individual contributions is
shown in \cref{Fig:2Higgses} for a toy example with two Higgs bosons $H_{1,2}$
with masses $M_{1,2}$, theoretical mass uncertainties $\Delta M_{1,2} =
    \SI{0.5}{\GeV}$ and universal signal strengths $\mu_{\alpha;j} = 0.5$ (for all
$\alpha$ and $j=1,2$). The current default observable set is used as
experimental input, i.e., the only mass measurement included is $\SI{125.09 \pm
        0.24}{\GeV}$ from the LHC Run-1 combination. The $\chi^2$ contribution from the
averaged mass $\bar{m}$ (\emph{top left panel}) shows four distinct regions. In
the corner squares neither of $H_{1,2}$ is assigned, and the $\chi^2_{\bar{m}}$
is zero. In the central square both scalars are assigned, and $\chi^2_{\bar{m}}$
has a flat direction along the diagonal of constant $\bar{m}$. In the four
rectangular regions at the sides only one of the scalars is assigned, and the
$\chi^2_{\bar{m}}$ profile corresponds to the one-particle case discussed in
\cref{Fig:assignmentexample}. The flat direction along constant $\bar{m}$ is
resolved by the $\chi^2_\text{sep}$ contribution (\emph{top right panel}). This
penalizes large mass splittings and is only non-zero in the region where both
scalars are assigned. The $\chi^2_\mu$ contribution from the signal rates
(\emph{bottom left panel}) is minimal and almost constant,
$\Delta\chi^2\approx0$, as long as both scalars are assigned since their rates
sum to the SM rates. As soon as one of the scalars drops from assignment, the
rates are only half the SM rates, or zero if neither of $H_{1,2}$ is assigned.
In both cases $\Delta\chi^2_\mu$ incurs a very large penalty. The resulting
total $\Delta\chi^2$ is shown in the \emph{bottom right panel} with the \CL{68}
(\CL{95}) contour indicated by a black solid (dashed) line. The resulting
profile has no flat direction and favors $M_1\approx M_2 \approx \hat m_\alpha$.
\Cref{Fig:2Higgses} was generated with the example program \Code{HS_2Higgses}
that is provided with the \HS package.

\section{User operating instructions}\label{sec:user}
\HSv{2} incorporates a number of modernizations to the
source code with respect to \HSv{1}. Furthermore, the project has moved the \HSv{2} development to the \textsf{GitLab}
repository
\begin{center}
    \url{https://gitlab.com/higgssbounds/higgssignals}
\end{center}
where the source code is available. Additionally, we are now using CMake as
build system such that \HS is now compiled by
\begin{verbatim}
    mkdir build && cd build
    cmake ..
    make
\end{verbatim}

This will compile the \HS library, the main executable, as well as a number of
example programs that illustrate different use cases. \HS depends on the \HB
library which has to be available on the system and will be automatically
found and used by CMake. More detailed information on building and linking
\HS can be found in the \texttt{Readme} on the above-mentioned web-page.

\subsection{\texttt{Fortran} subroutines}
\label{ssec:subroutines}

The \texttt{Fortran} subroutine interface provides access to all of the
functionality in \HS. Up-to-date, detailed documentation for the subroutines is
available online at
\begin{center}
    \url{https://higgsbounds.gitlab.io/higgssignals}.
\end{center}
The most important interface change in \HSv{2} stems from the separation of
the observable set into three parts --- the peak observables, the STXS
observables, and the LHC Run-1 combination --- as discussed in the previous
section. These can be separately accessed through the subroutines
\Code{run_HiggsSignals} (for the \SI{13}{\TeV} LHC peak observables),
\Code{run_HiggsSignals_STXS}, and
\Code{run_HiggsSignals_LHC_Run1_combination} that return the corresponding
$\chi^2$ values defined in \cref{Tab:chi2contributions}, the number of
observables, and the \pvalue. In most applications, a combined result
including all of these observable sets is required. In this case the
subroutine \Code{run_HiggsSignals_full} can be used or the three subroutines
can be run separately and the resulting $\chi^2$ values can be
summed.

\HSv{2} also includes a \texttt{C} interface to all of the \texttt{Fortran}
subroutines to make them more accessible from \texttt{C} or \texttt{C++} codes.
This interface automatically handles the necessary type conversions and is
included in the online documentation.

\subsection{Command-line version}
Compiling \HS generates a main executable that can be run as
\begin{verbatim}
    ./HiggsSignals <expdata> <pdf> <whichinput> <nHzero> <nHplus> <prefix>
\end{verbatim}

where the arguments specify the following: \Code{expdata} selects which
experimental dataset of peak and STXS observables is used. The default value is
\Code{latestresults} which refers to the latest available experimental results
included in your version of \HS. Alternatively, any of the folders in
\Code{data/Expt_tables/} can be named, and the observables in that folder will
be used. The parameter \Code{<pdf>} specifies the probability density function
as discussed in \cref{ssec:chisqpeakmass}. The input method is selected by
\Code{<whichinput>}, while \Code{<nHzero>} and \Code{<nHplus>} specify the
number of neutral and charged Higgs bosons in the model, respectively. These
parameters are passed to the \HB\ framework in order to handle the theoretical
input, see the \HB manual for further details~\cite{Bechtle:2020pkv}. The
\Code{<prefix>} path specifies the longest common path to the input data files.

\HS can use files in the SLHA format for input and output. The input blocks are
described in detail in the \HB manual~\cite{Bechtle:2020pkv}. In SLHA format \HS
produces an output block called \Code{HiggsSignalsResults} containing the
resulting $\chi^2$ values.

\subsection{Interpretations of the \texorpdfstring{$\chi^2$}{chi-squared}
    for model testing}%
\label{sec:user:interpretation}

The main application of the \HS~$\chi^2$ result is the statistical
discrimination between models on the basis of Higgs boson signal
observables (mass and rates, partly in kinematical bins). This is
achieved by statistically testing either a particular hypothesis or
two hypotheses against each other. Three typical applications can
be distinguished:

\begin{enumerate}
    \item[1)] Within a given model with model parameters $\vec{p} = p_i$
          ($i=1,2,\dots,N$) one may wish to determine the parameter regions that
          are preferred at a specific confidence level (C.L.) by the
          observation. Or, as a related issue, one may consider the question
          which parameter regions are excluded at a specific C.L., given the
          fact that the model contains other regions that are in better
          agreement with the data. This can be answered by calculating two-sided
          confidence intervals (C.I.) for the parameters $\vec{p}$. These
          C.I.~are typically presented in one or two dimensions (with the
          remaining parameters profiled or marginalized), but can in general be
          of higher dimension $D \le N$. Examples for typical applications of
          this kind can be found in
          Refs.~\cite{Bechtle:2015gvr,Bahl:2020kwe,Bechtle:2016kui,Bechtle:2015pma,Bahl:2020wee,Bagnaschi:2017tru}.
          These \emph{global fit} studies often combine the \HS~$\chi^2$ with
          other relevant constraints and observables. This hypothesis test will
          be described in \cref{sec:user:interpretation:fit};

    \item[2)] If one is interested \emph{only} in the lower \emph{or} upper
          boundary of the C.I.\ of the model parameter(s), the calculation of
          one-sided confidence intervals can be appropriate. This is called
          \emph{limit setting}. It may be of use if the parameter(s) represent
          so-far unobserved phenomena. For instance, one could ask the following
          questions: How large can the branching ratio for Higgs boson decays
          into invisible final states be? What is the maximally allowed value of
          the CP-violating phase of the Higgs-top-quark interaction? How low can
          the masses of the additional MSSM Higgs bosons be? How large can a
          Higgs doublet-singlet mixing be? While a \emph{lower} or \emph{upper}
          limit on a single parameter can be derived unambiguously, this is no
          longer true in a higher dimensional parameter space. However, in such
          cases a suitable mapping onto a one-dimensional parameter (e.g., a
          common signal strength modifier) may still be found to enable the
          derivation of a one-sided C.I.. Details on this hypothesis test are
          given in \cref{sec:user:interpretation:limit}.

\end{enumerate}

In both applications the calculation of the two-sided or one-sided C.I.~is a
\emph{hypothesis test} based on a likelihood ratio (LR), which quantifies the
(dis-)agreement of two competing statistical models based on the ratio of their
likelihoods. Two categories need to be distinguished in the hypothesis test:
First, which models are actually compared, and second, for which model
comparison is the choice of LR optimal? For reasons beyond the scope of this
discussion, these two categories do not necessarily coincide. In the analyses of
the LHC experiments, both for limit setting and for fitting, the LR is
constructed such that one of these models is typically determined by maximizing
the likelihood over the entire parameter space, defining the best-fit (BF)
scenario, and the other represents a specific model parameter point under study.
If the likelihood of the alternative hypothesis (\ie the non-BF model point
under study) is significantly lower than the likelihood of the null hypothesis
(\ie the BF point), the alternative hypothesis can be rejected (see
\cref{sec:user:interpretation:fit,sec:user:interpretation:limit} for details).
The Neyman-Pearson lemma~\cite{Neyman1992} shows that the LR test is the most
powerful test among all possible statistical tests in this case.

The full likelihood of the model parameters $\vec{p}$ given the
observed data $\vec{x}$, denoted as $\mathcal{L}(\vec{p}|\vec{x})$, as
evaluated by the LHC experiments, is not publicly available. Based on
the publicly available information on the measurements the \HS
$\chi^2$ approximates the full log-likelihood,
$\chi^2(\vec{p}|\vec{x}) \approx
    -2\ln\mathcal{L}(\vec{p}|\vec{x})$. The log-likelihood ratio (LLR) can
therefore be constructed as
\begin{equation}
    t(\vec{p}) = -2\ln\frac{\mathcal{L}(\vec{p}|\vec{x})}{\mathcal{L}(\hat{\vec{p}}|\vec{x})}
    =-2\left(\ln\mathcal{L}(\vec{p}|\vec{x}) - \ln\mathcal{L}(\hat{\vec{p}}|\vec{x})\right)
    \approx\chi^2(\vec{p}|\vec{x})-\chi^2 (\hat{\vec{p}}|\vec{x})
    \equiv\Delta\chi^2\eqdot
\end{equation}
Here, $\hat{\vec{p}}$ denotes the parameter point at which the
$\chi^2$ is minimized (or the likelihood
$\mathcal{L}(\vec{p}|\vec{x})$ is maximized), the ``BF point''. The likelihood ratio is
also denoted as a test statistics $t(\vec{p})$. In the presence of
additional nuisance parameters $\vec{\theta}$, corresponding to all
possible systematic or parametric uncertainties, the test statistics
generalizes to
\begin{equation}
    t(\vec{p}) =  -2\ln
    \frac{\mathcal{L}(\vec{p},\hat{\hat{\vec{\theta}}}|\vec{x})}{\mathcal{L}(\hat{\vec{p}},\hat{\vec{\theta}}|\vec{x})}.
    \label{eqn:teststat}
\end{equation}
In the numerator the nuisance parameters are optimized for each
tested parameter point $\vec{p}$ in the nuisance parameter space, with
the optimum denoted by $\hat{\hat{\vec{\theta}}}$, while in the
denominator $\vec{p}$ and $\vec{\theta}$ are optimized simultaneously
to find the global likelihood maximum at the point $\hat{\vec{p}}$
and~$\hat{\vec{\theta}}$.

In a few cases, likelihood distributions are given for single observables or
model parameters by the LHC experiments (see
e.g.~Ref.~\cite{Sirunyan:2018koj,Sirunyan:2020sum} for one-dimensional
likelihood functions of $\kappa$ parameters, the Higgs decay rate to
invisible final states, the total decay width, and Higgs CP-sensitive
parameters). However, this is not generally the case for many Higgs
signal rate measurements, which are usually only presented in terms of
central values and $1\sigma$ uncertainties, ideally accompanied by a
correlation matrix. Based on this information, the $\chi^2$ calculated
by \HS{} is the closest possible approximation to the full likelihood
for many observables.

\begin{enumerate}
    \item[3)] The third application is the goodness-of-fit test which is
          different to the above choices in that it \emph{typically} tests only
          one statistical model, i.e.~it does not compare two different model
          parameter points, but is evaluated for a single parameter point.
          Typically, this test is performed for specific scenarios, e.g., the BF
          scenario that has been identified in a preceding global model fit
          (case 1). Certain caveats apply, however, if the user wants to exclude
          a model purely on grounds of a goodness-of-fit test (see below).

          An approximation to the hypothesis test using the LR (as discussed
          above) can be constructed based on the goodness-of-fit test. This
          simpler implementation can be used to compare, for instance, the BF
          scenarios of different models. In particular, with a so-called
          $F$-test, one can determine whether model A leads to a significantly
          better description of the observation than model B, despite the fact
          that model A has more free parameters than model B. The $F$ test is
          recommended to analyze which of the models under study delivers a
          better fit, but it is not recommended as a means to exclude models or
          model parameter ranges. More details on goodness-of-fit test are given
          in \cref{sec:user:interpretation:goodness}.
\end{enumerate}

\HS can be used for all three applications. The most common hypothesis test
employed in phenomenological studies with \HS\ is certainly the model fit
(case~1). In the following sections we elaborate on the three choices and
discuss how in each case the $\chi^2$ value calculated by \HS is used in
practice. A decision chart for choosing the appropriate statistical treatment
for a variety of common applications is given in \cref{fig:decisionchart}. The
chart also includes the possible attempt to exclude the SM hypothesis in favor
of an alternative model (``discovery mode''). Here, the user must be aware of
some important caveats in such an interpretation before making any far-reaching
claims. These will be discussed in \cref{sec:user:interpretation:nogoes}.

\begin{figure}
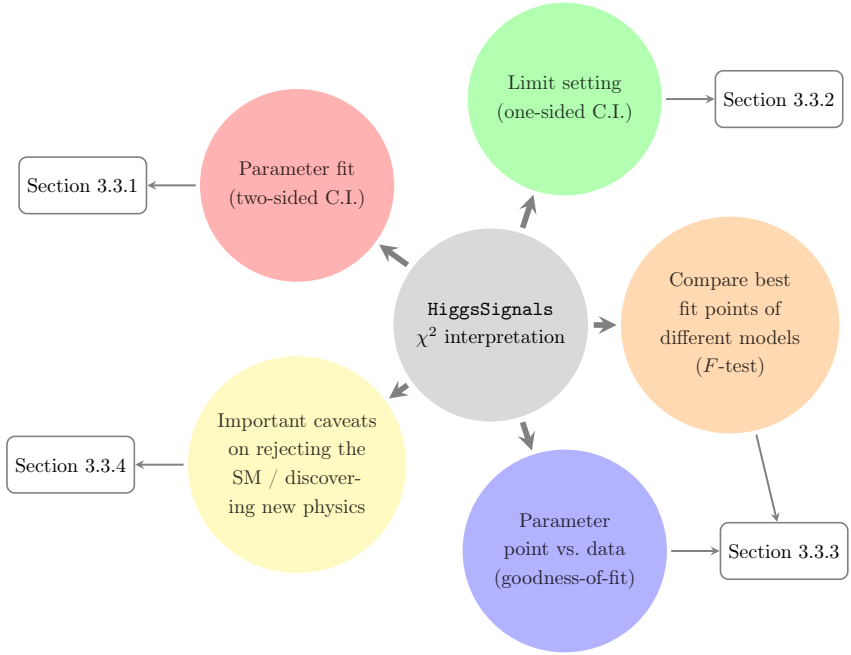

    \centering
    \scalebox{0.75}{
        \small
        \smartdiagramset{
            set color list={green!60, red!60, yellow!60, blue!60, orange!60},
            bubble center node font = \small,
            distance planet-text = .2cm,
            uniform connection color=true,
            planet text width=2.8cm,
            planet font=\small,
            distance planet-satellite=4.2cm,
            satellite text width=3cm,
            /tikz/connection planet satellite/.append style={->},
            additions={
                    additional item offset=0.85cm,
                    additional item border color=gray,
                    additional connections disabled=false,
                    additional item text width = 2cm,
                    additional arrow color=gray,
                    additional arrow tip=stealth,
                    additional arrow line width=1pt,
                    additional arrow style=<-,
                }
        }
        \smartdiagramadd[constellation diagram]{%
            \texttt{HiggsSignals} $\chi^2$ interpretation,
            Limit setting (one-sided C.I.),
            Parameter fit (two-sided C.I.),
            Important caveats on rejecting the SM~/ discovering~new~physics,
            Parameter point vs.\ data (goodness-of-fit),
            Compare best fit points of different models\\ ($F$-test)%
        }{right of satellite2/Section 3.3.2,
            left of satellite3/Section 3.3.1,
            right of satellite5/Section 3.3.3,
            left of satellite4/Section 3.3.4
        }
        \smartdiagramconnect{->}{satellite6/additional-module3}%

    }
    \caption{Decision chart for the statistical interpretation of the \HS\
        $\chi^2$ result.}%
    \label{fig:decisionchart}
\end{figure}

\subsubsection{Usage in a fit}%
\label{sec:user:interpretation:fit}

\begin{figure}
    \centering
    \includegraphics[width=0.3\textwidth]{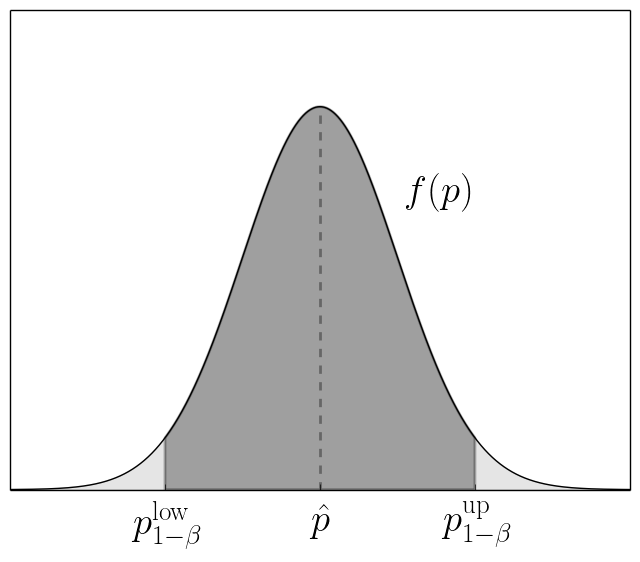}\hspace{0.1cm}
    \includegraphics[width=0.3\textwidth]{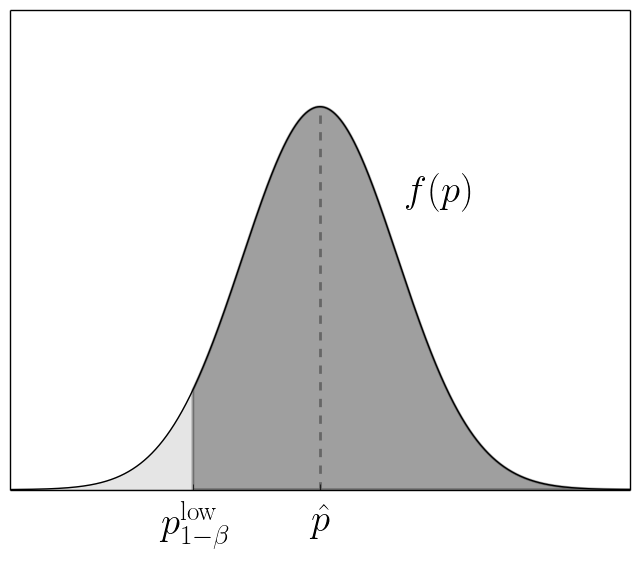}\hspace{0.1cm}
    \includegraphics[width=0.3\textwidth]{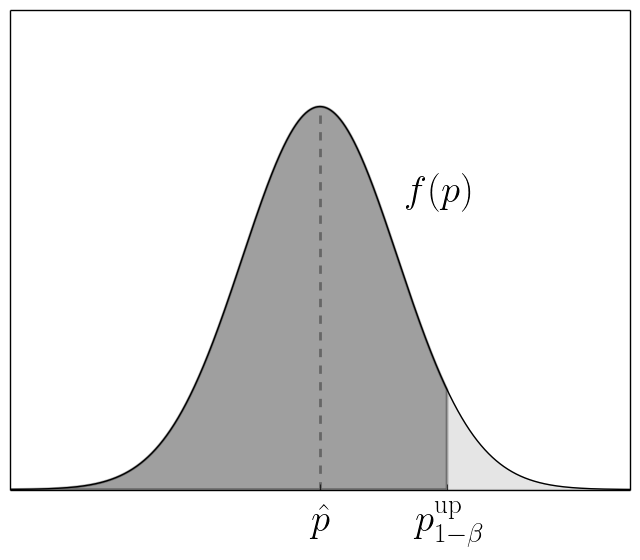}
    \caption{Illustration of a two-sided confidence interval (\emph{left
            panel}), a one-sided confidence interval with a lower bound
        (\emph{middle panel}) and a one-sided confidence interval with
        an upper bound (\emph{right panel}), for a
        given probability density function $f(p)$ of a parameter $p$.
    }\label{fig:CIs}
\end{figure}

When fitting the parameters of a given model to the observation one employs the
calculation of two-sided confidence intervals, as one is generally
interested in the lower \emph{and} upper boundaries of the allowed parameter
intervals. This is illustrated in the left panel of \cref{fig:CIs}. The plot
shows a one-dimensional probability density function $f(p)$ of a parameter
$p$, the best-fit value   $\hat{p}$ and the lower and upper boundaries of
the C.I.{} of probability $1-\beta$, denoted $p_{1-\beta}^\text{low}$ and
$p_{1-\beta}^\text{up}$, respectively.\footnote{For illustration the pdf is
    Gaussian-shaped, but can in general have an arbitrary shape as long as it is
    properly normalized, $\int f(p)dp = 1$.} The dark-shaded area (\ie the
integral over the pdf $f(p)$) corresponds to the probability of $1-\beta$.
For instance, for a $1\sigma$ C.I.\ in a one-dimensional parameter $p$, we
choose $\beta\approx 0.32$, and $p_{0.68}^\text{low}$ ($p_{0.68}^\text{up}$)
is the corresponding lower (upper) boundary of the \SI{68}{\percent} C.I..\footnote{The
    other two panels of \cref{fig:CIs} are described in
    \cref{sec:user:interpretation:limit}.}

It is well-known that in the so-called Gaussian limit --- where near
the best fit point the uncertainties of the observables are approximately independent of
the parameters, the relation between parameters and observables are
approximately linear, and the
uncertainties are approximately Gaussian ---
for one-dimensional parameter spaces the two-sided confidence
interval corresponding to a $1\sigma$ ($2\sigma$) C.L.
is given by the ensemble of model parameter points with
$\Delta\chi^2(\vec{p}) \le 1~(4)$ above the best fit point at
$\Delta\chi^2(\hat{\vec{p}})=0$. In two-dimensional parameter spaces,
the $1\sigma$ and $2\sigma$ confidence regions are found at
$\Delta\chi^2(\vec{p})\le 2.30$ and $\Delta\chi^2(\vec{p})\le 6.18$,
respectively. For these values, it does not matter how many additional
parameters of interest or nuisance parameters are free in the fit, as
long as all other parameters apart from those for which the C.I.
is determined are profiled (marginalized) in the frequentist
(bayesian) interpretation. An overview of the $\Delta\chi^2$ values for
different C.L.\ and different values of the number of parameters under
study (also called degrees of freedom $\nu$) are given in
\cref{tab:chi2values}. While there is no logical necessity to restrict
the number of parameters for whose common variation a C.I. is
given to one or two, it is uncommon to determine the C.I. for
more than two parameters at a time, as this becomes difficult to
visualize.
If it is important to convey information about the relation
between many different parameters, one usually provides a covariance matrix
based on linear correlation factors and on the $1\sigma$ one-dimensional
uncertainties. These confidence intervals and covariances based on profiling can
be established numerically using tools like MIGRAD in MINUIT~\cite{James:1994vla} or
by using methods like Markov Chain Monte Carlo techniques or other numerical
optimizers.

Caution is required for cases where the requirements for the
Gaussian limit are not reached, \ie where the $\chi^2$ profile around
the minimum is not parabolical. This situation appears in many
examples of model fits. It is common practice to use the
$\Delta\chi^2$ ranges given here also in such a situation, albeit it
cannot be guaranteed that they still correspond to the quoted
frequentist coverage of \eg \SI{68}{\percent} for the one-dimensional
$1\sigma$ C.I. Furthermore, linear correlations might not describe the
interdependence of the parameters properly. We emphasize that the
user should treat these cases with caution and discuss possible
implications, or instead use a Toy-Monte-Carlo based
technique. Ref.~\cite{Bechtle:2015nua} gives such an example for determining the parameter
range for the actual C.L. instead of using the profiling
technique.

\begin{table}
    \centering
    \sisetup{table-format=2.2,round-precision=2,round-mode=places}
    \begin{tabular}{lc SSSSSS}
        \toprule
                     &             & \multicolumn{6}{c}{$\nu$}                                         \\
        C.L.         & (s.d.)      & 1                         & 2     & 3     & 4     & 5     & 6     \\
        \cmidrule(lr){1-2}\cmidrule(lr){3-8}
        $68.27\%$    & ($1\sigma$) & 1.                        & 2.30  & 3.53  & 4.72  & 5.89  & 7.04  \\
        $95\%$       &             & 3.84                      & 5.99  & 7.81  & 9.49  & 11.07 & 12.59 \\
        $95.45\%$    & ($2\sigma$) & 4.                        & 6.18  & 8.02  & 9.72  & 11.31 & 12.85 \\
        $99\%$       &             & 6.63                      & 9.21  & 11.34 & 13.28 & 15.09 & 16.81 \\
        $99.73\%$    & ($3\sigma$) & 9.                        & 11.83 & 14.16 & 16.25 & 18.21 & 20.06 \\
        $99.99994\%$ & ($5\sigma$) & 25.                       & 28.74 & 31.81 & 34.55 & 37.09 & 39.49 \\
        \bottomrule
    \end{tabular}
    \caption{$\Delta \chi^2$ as function of confidence level (C.L.)
        and degree of freedom ($\nu$). The values for $\nu>2$ are
        rarely used and only given for
        completeness.}\label{tab:chi2values}
\end{table}

It is strongly recommended to use the likelihood-ratio based fitting
technique to find the uncertainty range of the model parameters. We
\emph{discourage} from calculating the (approximate) goodness-of-fit
for each model parameter point and then combine
those parameter points with acceptable goodness-of-fit to form an
``allowed region''. The reason for this recommendation is
explained in \cref{sec:user:interpretation:goodness}.

\subsubsection{Usage in limit setting}%
\label{sec:user:interpretation:limit}

The main difference between limit setting and fitting is the choice of
a one-sided confidence interval for the former. As discussed
above, limit setting and fitting are subclasses of hypothesis
testing, which can be used for both discoveries or exclusions. For
discoveries, hypothesis testing addresses the question ``How unlikely
is the observed pattern of data if the \emph{null hypothesis} is
true?'' where the null hypothesis is typically the assumption of the
SM without any additional signal, $s=0$. In contrast limit setting
is a hypothesis test where exclusions of an \emph{alternative
    hypothesis} at a pre-defined C.L. are sought. Exclusions are
related to the question ``How unlikely is the data if the
alternative hypothesis is true?''. It is practically often
implemented using a one-dimensional parametrization of the strength
of the model predictions. In the \HS{} case, each model prediction
of observables $\vec{x}(\vec{p})$ can be parametrized as
\begin{equation}
    \vec{x}(s,\vec{p}) \equiv
    \vec{x}_{\text{SM}}+s\,\left(\vec{x}(\vec{p})-\vec{x}_{\text{SM}}\right).
    \label{eq:deviationrescaling}
\end{equation}
Here, we introduced $s$ as a global rescaling parameter of the model-predicted
deviations in the observables $\vec{x}$ from the SM prediction, which we further
abbreviate as $\vec{d}(\vec{p}) \equiv \vec{x}(\vec{p})-\vec{x}_{\text{SM}}$.
The true model prediction at $\vec{p}$ is retained when the rescaling parameter
--- also called strength of BSM effects --- is $s=1$. Under a change of
parameters $\vec{p}\to \pvec{p}'$ two possible changes of $\vec{d}$ need to be
distinguished: First, the case where the parameter transformation leads to a
different \emph{pattern of deviations} in the observables $\vec{x}$ from the SM
prediction, \ie
\begin{align}
    \vec{d}(\vec{p})  \xrightarrow{\vec{p}\to\pvec{p}'} \pvec{d}'(\pvec{p}') \ne \alpha\,\vec{d}(\vec{p}),
    \label{eq:deviations_patternchanges}
\end{align}
with a scalar constant $\alpha$. Second, the case where the parameter
transformation leaves the \emph{pattern of deviations} invariant,
\begin{align}
    \vec{d}(\vec{p})  \xrightarrow{\vec{p}\to{\pvec{p}}'} \pvec{d}'(\pvec{p}') = \alpha\,\vec{d}(\vec{p}).
\end{align}
Here the scalar multiplicative factor $\alpha$ can simply be absorbed into
the transformed strength of deviations,
\begin{align}
    s \pvec{d} \xrightarrow{\vec{p}\to\pvec{p}'} s' \pvec{d}' = s \alpha \, \pvec{d},
    \label{eq:deviations_patterninvariant}
\end{align}
\ie only the strength of BSM effects $s$ changes to $s' = \alpha s$ under the change of parameters.

Using these definitions, limits on model parameter points can be derived from
measurements of the Higgs mass and rates in \HS by employing a limit setting
procedure on $s$. Depending on the two possible transformation properties under
a change of parameters described above, two statistical tests are often
performed in experimental and phenomenological analyses in the absence of a
significant deviation of the data from the SM\@:
\begin{enumerate}
    \item Is the alternative model with parameters $\vec{p}$ excluded at a given
          C.L.? In this case, the C.L. for $s=1$ needs to be calculated. If the
          C.L. falls below a pre-defined cut, the model point $\vec{p}$ is
          regarded as excluded. This is the recommended application for all
          models where the parameters of interest \emph{cannot} be mapped onto a
          single rescaling parameter $s$ that only affects the strength of the
          model-predicted deviation from the SM, \ie for the first case of
          parameter transformation properties,
          \cref{eq:deviations_patternchanges}. This is typically the case in
          relatively complicated parameter spaces, \eg in the ($M_A,
              \tan\beta$) parameter plane of the 2HDM or MSSM\@.
    \item What is the maximum signal strength $s_{\text{up}}$ at which the
          pattern of deviations from the SM defined by $\vec{x}(\vec{p})$ is
          allowed? In this case, $s$ is varied until the set of predictions of
          observables $\vec{x}(s_{\text{up}},\vec{p})$ is found which is
          excluded exactly at the pre-defined C.L.. This is the recommended
          application for models where a (sub)set of parameters leaves --- at
          least to a sufficiently accurate approximation --- the shape of the
          pattern of deviations invariant and only affects the strength $s$
          according to \cref{eq:deviations_patterninvariant}. This applies, for
          instance, when constraining an additional Higgs branching ratio into
          an invisible final state, or a singlet-doublet-mixing angle in a Higgs
          singlet extension model. In this case, only the parameters which
          \emph{do} affect the shape of the pattern of deviations need to be
          tested separately, while the (sub)set of parameters only affecting the
          strength can directly be tested against $s_\text{up}$. For
          illustration we discuss an explicit example in \cref{fig:BRinvLim}
          below.
\end{enumerate}

In the following, we will describe the two tests on the C.L.\ of $s=1$ and of
the determination of $s_{\text{up}}$.

The C.L.\ is chosen by defining the desired \emph{power} of the test, $\beta$.
In particle physics, for exclusions typically $\beta=0.05$ is used, such that
the required C.L. is $1-\beta = \SI{95}{\percent}$. In order to find the observed C.L., the
test statistics $t$ is chosen in complete analogy to \cref{eqn:teststat}, but
this time parametrized as a function of the signal strength $s$ for a fixed
parameter point $\vec{p}$. Such a test can then be done for each choice of fixed
$\vec{p}$. The commonly used choice of test statistics $t$ in LHC related
analyses is based on
\begin{equation}
    t(s) =
    -2\ln\lambda(s)=-2\ln\frac{\mathcal{L}(s,\hat{\hat{\vec{\theta}}}|\vec{x})}{\mathcal{L}(\hat{s},\hat{\vec{\theta}}|\vec{x})}\approx \chi^2(s,\hat{\vec{\theta}})-\chi^2(\hat{s}, \hat{\hat{\vec{\theta}}})\approx \chi^2(s)-\chi^2(\hat{s}),
    \label{eqn:teststat2}
\end{equation}
which optimally distinguishes the alternative hypothesis $s$ from the best-fit
hypothesis $\hat{s}$, under consideration of all possible uncertainties
(systematic, parametric etc.) using the vector of nuisance parameters
$\vec{\theta}$. The same convention for the definition of $\hat{\vec{\theta}}$
and $\hat{\hat{\vec{\theta}}}$ as in \cref{eqn:teststat} applies. The second
approximation, omitting the nuisance parameters $\theta$, relates to the fact
that the nuisance parameters are not fitted explicitly in \HS{}, since the
experimental systematic errors are not reported separately by their source in
the experimental input to \HS{}. Hence, correlated and uncorrelated systematic
errors are only treated in the covariance matrix entering the $\chi^2$
calculation. This similarity in the choice of test statistics between limit
setting and fitting exemplifies the convergence of the consistent statistical
treatment of one-sided confidence limits and two-sided confidence limits in the
LHC era. This approach goes back to the ideas discussed in
Ref.~\cite{Feldman:1997qc}, but was not used in high energy physics at earlier
major experiments. The discussion of the likelihood ratio based hypothesis test
in this section follows the discussion in Ref.~\cite{Cowan:2010js}.

A detailed discussion of model exclusions based on model-independent likelihood
values, as employed in \HB, has been presented in
Ref.~\cite{Bechtle:2020pkv}.\footnote{In \HB, likelihoods are parametrized for a
    few search limits, e.g.~from the LEP experiments~\cite{Schael:2006cr} and for
    searches for additional Higgs bosons decaying into $\tau^+\tau^-$ final states
    at the LHC~\cite{Bechtle:2013wla,Bechtle:2015pma}.} The difference between the
\HB\ and \HS\ approach lies on the one hand in the approximation to the
likelihood, which in \HB{} is based on published observed and expected
likelihood values, while in \HS{} the likelihood ratio is approximated from the
$\Delta\chi^2$. On the other hand, the approaches differ in the parametrization
of the model, where physical parameters are applied for \HB, and the signal
strength $s$ is used in the \HS{} example discussed here. In the following, we
will focus on the aspect of limit setting as we strongly discourage from using
\HS for an attempted discovery. See \cref{sec:user:interpretation:goodness} for
detailed considerations on this issue.

When setting a limit on $s$, \cref{eqn:teststat2} is not used directly. Instead, several separate cases are considered: First, in order not to hold an overshoot of the data above the signal hypothesis against the signal, and still allowing $s<0$, the test statistics from \cref{eqn:teststat2} is further modified in the following way: \begin{equation}
    q_s^{\chi^2}=\begin{cases}
        t(s)\approx \Delta\chi^2(s) = \chi^2(s)-\chi^2(\hat{s}) & \text{for }\hat{s}\leq s, \\
        0                                                       & \text{for }\hat{s}>s,
    \end{cases}
    \label{eqn:teststat3}
\end{equation}
where we have introduced the notation $q_s^{\chi^2}$ to distinguish this
modified test statistics from the test statistics $t(\vec{p})$ employed in the
fits described in \cref{sec:user:interpretation:fit} as given in
\cref{eqn:teststat}. The superscript $^{\chi^2}$ refers to the $\chi^2$
approximation of the LR\@. The difference between the two test statistics
further lies in the case separation of \cref{eqn:teststat3}: for $\hat{s} > s$,
$q_s^{\chi^2} = 0$ is used in order to prevent the signal hypothesis $s$ from
being excluded if an even larger signal $\hat s$ would actually better describe
the experimental results.

Often, $s>0$ is required from physical considerations\footnote{In the example
    given here, where $s$ parametrizes the relative strength of the deviation
    from the SM prediction as given by the parameter point $\vec{p}$, negative values of $s$
    usually have no physical meaning.}, in which
case a negative best-fit-value $\hat{s}$ should be interpreted as
``The no-signal hypothesis fits data the best''. Then,
\begin{equation}
    q^{\chi^2}_s=\begin{cases}
        \chi^2(s) - \chi^2(0) & \text{for }\hat{s}<0          \\
        \Delta\chi^2(s)       & \text{for }0\leq\hat{s}\leq s \\
        0                     & \text{for }\hat{s}>s
    \end{cases}
\end{equation}
is used,\footnote{We do not consider the case of possible discoveries here,
    which would employ a dedicated test statistics $q_0$ for refuting the $s=0$
    hypothesis. See Ref.~\cite{Cowan:2010js} for details.} where $\chi^2(0)$
denotes the $\chi^2$ obtained for the null hypothesis, typically the SM\@.

In order to derive the desired C.L.\ for $s=1$ or find
$s_{\text{up}}$, the calculation of a probability density function
$f(q^{\chi^2}_s|s)$ is needed to compare the range of
possible expected outcomes for $q^{\chi^2}_s$ (the random variable, i.e.\
the quantity which would show a different outcome for each simulated
repetition of all measurements) with the observed
$q^{\chi^2}_s$ in the case of an assumed signal strength $s$. For
calculating the C.L.\ at $s=1$, only the pdfs $f(q^{\chi^2}_s|1)$
and $f(q^{\chi^2}_s|0)$ are required. For the calculation of
$s_{\text{up}}$, $s$ needs to be varied, and eventually pdfs need to
be calculated for several choices of $s$, until the desired C.L.\ at
$1-\beta$ is found at $s_{\text{up}}$.
Details are given in
Ref.~\cite{Cowan:2010js}. In this reference, for the case discussed
here, the correspondence between the observed test statistics $q^{\chi^2}_s$
and the observed C.L.\ of the signal plus background hypothesis
$\mathrm{CL}_{s+b}$\footnote{It should be noted that in the brief discussion presented
    in this paper, every C.L.\ and C.I.\ is described as an observed
    result corresponding to the observed data. In order to obtain the expected
    quantities, the statistical treatment is exactly the same, apart
    from replacing the observed data with the measurements expected
    under the $s=0$ hypothesis.}
is approximately found using
\begin{equation}
    \Phi^{-1}(1-\mathrm{CL}_{s+b})=\sqrt{q^{\chi^2}_s} \eqcomma
    \label{eqn:phitotheminusone}
\end{equation}
where $\Phi$ is the cumulative normal distribution function.

As discussed above, the power $1-\beta$ is typically chosen such that
$\beta=0.05$ holds for a one-sided confidence level which is usually called a
``$2\,\sigma$'' C.L.\footnote{While $\beta=0.05$ is not the exact value
corresponding to a $2\,\sigma$ exclusion, it is so close to it that it has
become a commonly used term.}. This case is exemplified in the right panel of
\cref{fig:CIs}.  The desired upper limit on the allowed signal rate
$s_\text{up}$ (corresponding to $p^\text{up}_{1-\beta}$ in \cref{fig:CIs}) is
adjusted such that its $\mathrm{CL}_{s+b}$ corresponds to the chosen C.L.\ of
$1-\beta=0.95$.  Then, $\Phi^{-1}(0.05)=1.64$ can be used to determine the
\SI{95}{\percent} C.L.\ upper limit
\begin{equation}
    s_\text{up}=\hat{s}+1.64 \, \sigma_s.
    \label{eqn:sup}
\end{equation}
The standard deviation $\sigma_s$ on the signal hypothesis $s$ (as
constrained by the test statistics) can be determined numerically from
the Wald approximation~\cite{citeulike:6130790} using the same
minimization and profiling techniques as in case of a fit (see
\cref{sec:user:interpretation:fit}). The
approximate uncertainty $\sigma_s$ on the parameter $s$ in a simple
one-dimensional fit is
$\sigma_s=|\hat{s}-s(\Delta\chi^2=1)|$, where $s(\Delta\chi^2=1)$ is the
signal strength below or above $\hat{s}$ for which $\Delta\chi^2=1$ is
obtained. This uncertainty connects the
two-sided limit calculated for a fit to the one-sided limit discussed
here. This connection can be shown as follows: In the Wald
approximation, near the optimum the $\Delta\chi^2$ forms a parabola
and hence can be parametrized as
\begin{equation}
    \Delta\chi^2(s) = {\left(\frac{s-\hat{s}}{\sigma_s}\right)}^2.
\end{equation}
Using $1.64=\Phi^{-1}(0.05) = \sqrt{q^{\chi^2}_{s_\text{up}}}=\sqrt{\Delta\chi^2(s_\text{up})}=(s_\text{up}-\hat{s})/\sigma_s$ in the allowed
range of $\hat{s}$, we can solve for $s_\text{up}$ and find
\cref{eqn:sup}. In this case, the model parameter point $\vec{p}$ can
be regarded as excluded if $s_{\text{up}}\le 1$ is found, and as
allowed if $s_{\text{up}}> 1$ is observed.

It is important to note that the above confidence level construction yields
$\mathrm{CL}_{s+b}$, \ie the confidence level of the signal plus background
hypothesis at the signal rate $s_\text{up}$. In most experimental applications
of signal exclusions, $\mathrm{CL}_s=\mathrm{CL}_{s+b}/\mathrm{CL}_b$ is used in
order to avoid accidental exclusion of parameter points for which the search has
no sensitivity~\cite{Read:2002hq}. Based on Ref.~\cite{Cowan:2010js} and the
example in Ref.~\cite{Bechtle:2020pkv}, we find for the notation of the latter
reference $q_s^\text{exp}(s)\equiv {((s-\hat{s})/\sigma_s)}^2$ and
$q_s^\text{obs}(\hat{s})\equiv {((\hat{s}-\hat{s})/\sigma_s)}^2=0$. Here,
$q_s^\text{exp}(s)$ in the notation from Ref.~\cite{Bechtle:2020pkv} is the
square of the separation between the \emph{tested} value $s$ and the best fit
$\hat{s}$ in units of the uncertainty $\sigma_s$, while
$q_s^\text{obs}(\hat{s})$ is the test statistics of the \emph{observed} best fit
point, $\hat{s}$, which in the definition of this section is always
0.\footnote{The equivalence of using the squared relative separation of $s$ from
$\hat{s}$ as an ``expected'' value, which may seem surprising at first, arises
from the fact that only differences between $\chi^2$ values have a meaning in
this interpretation, and not absolute values. This is analogous to the
definition of the likelihood, where only ratios of likelihoods (or differences
of log-likelihoods) have a physical meaning. Fundamentally, $s/\sigma_s$
parametrizes the \emph{expected} distance (in units of $\sigma_s$) of the median
of the pdf of the test statistics of the signal plus background hypothesis from
the median of the pdf of the $s=0$ hypothesis, and $\hat{s}/\sigma_s$ is the
relative \emph{observed} distance from $s=0$. However, we define $q^{\chi^2}_s$
such that $q^{\chi^2}_s(s=\hat{s})=0$. Thus, in contrast to an implementation as
in Ref.~\cite{Bechtle:2020pkv}, where the test statistics $q^{\text{obs}}$ at
the best fit point is not defined to be 0, the $\chi^2$ based input to the limit
calculation applied here is shifted by $\hat{s}/\sigma_s$, and thus $s/\sigma_s
- \hat{s}/\sigma_s$ parametrizes the relative distance of the expected result
from the background-only result, and $\hat{s}/\sigma_s - \hat{s}/\sigma_s=0$
always holds for the observed result.} For the confidence levels, from inverting
\cref{eqn:phitotheminusone} and inserting into Eqs.~(25) and (26) of
Ref.~\cite{Bechtle:2020pkv}, we thus find
\begin{align}
    \mathrm{CL}_{s+b} & = 1 - \Phi\left((s-\hat{s})/\sigma_s\right),  \\
    \mathrm{CL}_b     & = 1 - \Phi\left(-(s-\hat{s})/\sigma_s\right),
    \label{Eq:CL}
\end{align}
where the symmetry stems from the symmetry of the parabolic approximation to the
likelihood. In order to find the limit $s_\text{up}$ for the $\mathrm{CL}_s$
case, the above procedure has to be repeated iteratively by varying $s$ until it
is adjusted to the value $s_\text{up}$ such that $\mathrm{CL}_s=0.05$ (for a
\SI{95}{\percent} confidence level) is found. For testing the exclusion of a
model point at $s=1$, no iteration is required. $\mathrm{CL}_s$ can be
calculated using the equations above, and if $\mathrm{CL}_s<\beta$ is found, the
parameter point $\vec{p}$ in the model can be regarded as excluded.

\begin{figure}
    \centering
    \includegraphics[width=0.6\textwidth]{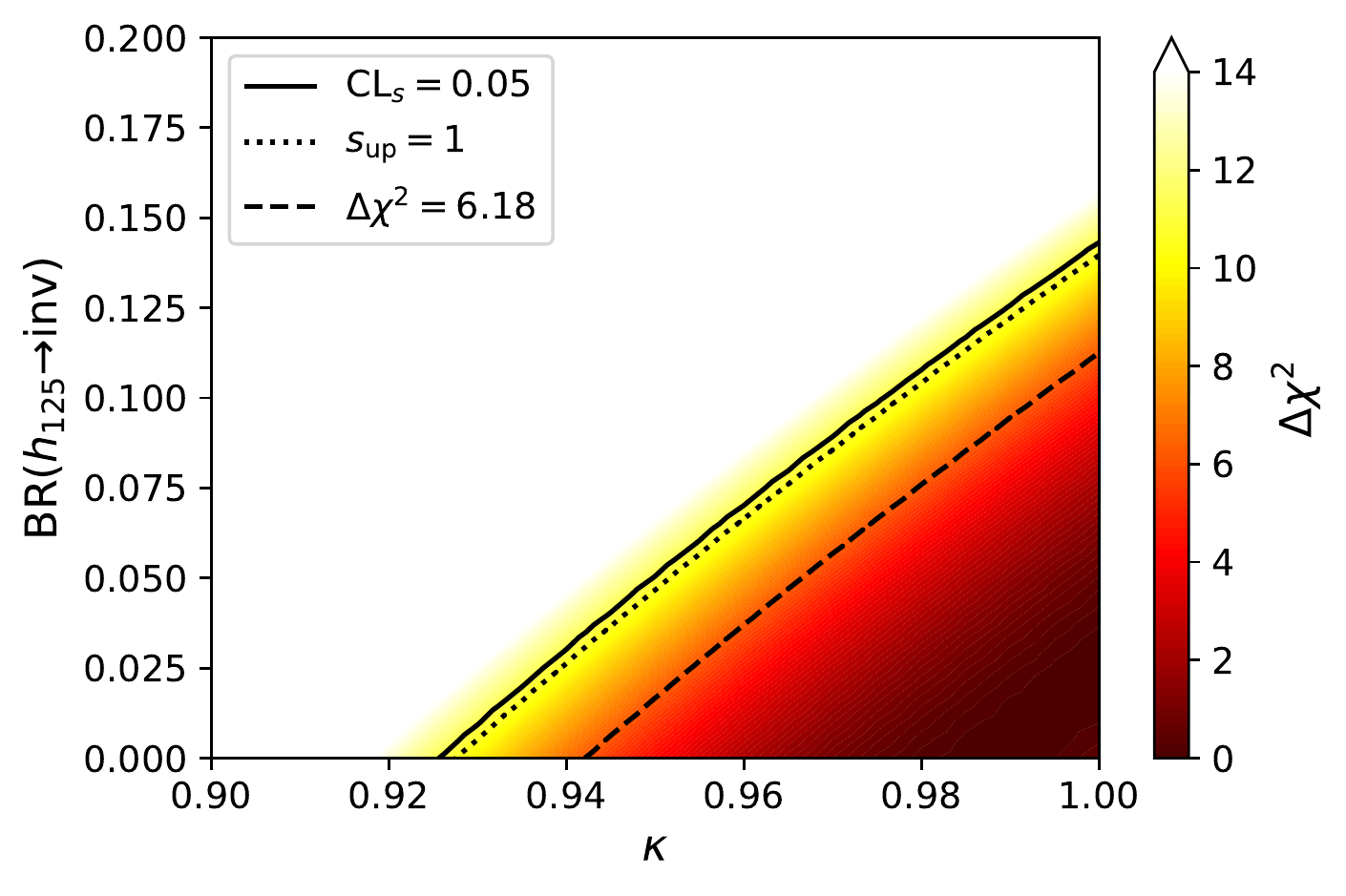}
    \caption{The \HS $\Delta\chi^2$ profile in a toy model with a global
        Higgs-coupling scale factor $\kappa$ and an additional Higgs decay into
        invisible final states with decay rate
        $\text{BR}(h_{125}\to\text{inv})$. The contour lines indicate the
        borders of the \CL{95} allowed regions derived from a fit (dashed) and a
        limit setting procedure using $\text{CL}_{s+b}$ (dotted), which
        corresponds to $s_{\text{up}}= 1$, and $\text{CL}_s$ (solid).}%
    \label{fig:BRinvLim}
\end{figure}

\Cref{fig:BRinvLim} illustrates the difference between a limit setting and a
fitting interpretation in a toy model with a single SM-like Higgs boson at
\SI{125.09}{\GeV} that is modified by a global coupling scale factor
$\kappa\leq1$ and an additional decay into invisible final states with a
branching ratio $\text{BR}(h_{125}\to\text{inv})$. For instance, this toy model
can directly represent parts of the parameter space of models with additional
scalar singlet(s)~\cite{Robens:2015gla,Robens:2019kga}. The figure shows three
different ways to determine the allowed parameter region. The dashed line
indicates the $2\sigma$ favored region in a fit (compare \cref{tab:chi2values}).
The dotted line is the limit derived from $\text{CL}_{s+b}$ using $s_\text{up}$
from \cref{eqn:sup}, and the solid line indicates the \CL{95} limit based on
$\text{CL}_s$. Both of the limit setting approaches lead to a less aggressive
limit than the fit since they use one-sided instead of two-sided confidence
intervals. For this simple example the difference between $\text{CL}_{s+b}$ and
$\text{CL}_s$ is small, but in general the $\text{CL}_s$ based limit should be
used whenever possible. This example is implemented in the \Code{HSLimitEffC}
example code.

\subsubsection{Usage in a goodness-of-fit test}%
\label{sec:user:interpretation:goodness}
The third typical application of the \HS $\chi^2$ is the goodness-of-fit test.
Fundamentally, it aims to determine the consistency of the model compared to the
data only, in contrast to the comparisons between different (parameter)
hypotheses in the above two cases. In practice, there are only very few cases
where a goodness-of-fit test is preferential over a hypothesis test. One example
is a case where the null hypothesis is deprecated by the data in a similar way
as the alternative hypothesis, which contains the null hypothesis within its
parameter range. This seemingly paradox situation can happen for example when
comparing a supersymmetric model to the SM (see \eg
Ref.~\cite{Bechtle:2015nua}). The SM as a null hypothesis is so well established
that a very tight requirement on its rejection is placed --- typically
$5\,\sigma$ significance corresponding to $p_0<2.7\times 10^{-7}$. On the other
hand, the alternative hypothesis shall be rejected at the level $p_s<0.05$. In
this case, if a hypothesis test is performed following
\cref{sec:user:interpretation:limit}, in this hypothetical example the
alternative hypothesis can never be refuted because in the decoupling limit it
would fit the data at least as well as the SM, even though the SM would provide
a bad fit to the data at the level of $2.7\times 10^{-7}<p_0<0.05$. In such a
case a careful goodness-of-fit test can be used to refute a model.

The simplest application of a goodness-of-fit test is to calculate
the probability $P(\chi^2_{\min}|\text{ndf})$ of the observed $\chi^2_{\min}$ under the assumption
that it follows a $\chi^2$ distribution after the
minimization in the fit, as described in
\cref{sec:user:interpretation:fit}, for the given number of degrees of
freedom ($\mathrm{ndf}$). In the construction of the \HS $\chi^2$ it is
assumed that the uncertainties are Gaussian, all correlations are
linear and that the relation between observables and parameters is
approximately linear around the minimum. In this case, the analytic
form of the $\chi^2$ distribution can be used to calculate the
$\chi^2$ probability, and the user can require a certain confidence
level for an
allowed model. Often, the ratio
$\chi^2_{\min}/\mathrm{ndf}$ is used to estimate the goodness-of-fit,
and it is expected that it is around 1 for a model that provides a
good fit to the data. The variance of the $\chi^2$
distribution $\sqrt{V_{\chi^2}}=\sqrt{2\,\mathrm{ndf}}$ can be taken
as a rough guideline for the expected range of deviation from 1 for a
good fit.

However, this popular approach, while not generally invalid, comes with two
caveats. First, in many fits of BSM models, two different parameter regimes are
within the allowed parameter ranges in 1 or 2\,$\sigma$ C.I.: parameter ranges
where the observables effectively do not depend on the parameters (the decoupled
regime) as well as regions where the observables depend strongly on the same
parameters. In such a case, the construction of the analytic $\chi^2$
probability in the Gaussian approximation is invalid because it assumes a purely
linear relation between parameters and observables. The user should handle such
a situation with care, \eg by using toy Monte Carlo simulations to obtain the
true probability density function of the~$\chi^2$.

The second caveat is that there is a notable difference between the
goodness-of-fit and the confidence interval calculation (see
\cref{sec:user:interpretation:fit,sec:user:interpretation:limit}): By
construction, the confidence interval cannot increase (or in other words, the
constraint on the allowed parameter space cannot get looser) if the data is
presented in an increased number of subchannels.\footnote{Assuming that
systematic uncertainties do not grow with smaller subchannel size.} Splitting
the measurement in many individual measurements can only decrease the confidence
interval or leave it unchanged.  However, if the measurements are split up in
many individual results which the model can only vary together, but not
individually, the goodness-of-fit in the average of all statistical tests in an
ensemble of possible experimental outcomes increases if the separate
measurements are statistically consistent with each other. This statistical
increase of the goodness-of-fit on average buries the tension between the
physical variations and the data under the expected statistical variations of
the individual measurements. Therefore, using many experimental subchannels as
implemented in \HS can artificially increase the goodness-of-fit. The
consequences of this effect will also clearly appear in the validation of
the LHC Run~1 combination in \cref{sec:performance:Run1combined}. Note that this
statement about the probable outcome of such a test assumes statistical
consistency amongst all observables. The situation is different if separate
measurements of the same or a similar physical quantity do not agree with each
other at a significant level, and thus are not at all or only marginally
consistent with each other. A notable occurrence of such a case is the
electroweak fit~\cite{Haller:2018nnx}, where different individual
measurements for the forward-backward and left-right asymmetries at the
$Z$~pole are in disagreement with each other while their average agrees with the
SM prediction. In this case, the goodness-of-fit of course increases when
fitting to physically meaningful averages instead of fitting to the separate
measurements of the physical quantity which the SM cannot vary independently. As
a consequence, the result for the goodness-of-fit depends on whether individual
measurements are directly used as input for the fit or whether instead certain
averages of individual measurements are used as input.  For a discussion of both
of these caveats see Ref.~\cite{Bechtle:2015nua}.

Due to the latter caveat we further discourage using the goodness-of-fit on
multiple parameter points in order to find an allowed area. The result would
strongly depend on the structure of the experimental data and not only on the
genuine predictions of a model parameter point. The goodness-of-fit should be
evaluated with care, and the effort needed to do so is typically only warranted
for the best-fit point. If it has an acceptable goodness-of-fit, the preferred
parameter regions are much more robustly determined using the fit procedure
described in \cref{sec:user:interpretation:fit}. Generally, for all the
above reasons, we advise to treat all goodness-of-fit test results as acceptable
as long as each of them lies within the range deemed acceptable by the analyzer,
and we advise against excluding models based on a comparison of the
goodness-of-fit between models. However, while models should not be
\emph{excluded} based on a relative comparison of their goodness-of-fits, it is
a valid question to ask whether models with different numbers of degrees of
freedom provide a better or worse description of the data. This is often of
interest if a more general model with a large number of parameters contains a
more constrained model with less parameters.

In order to achieve this comparison between models of different complexity, a
generalization, called the $F$-test, can be
derived~\cite{10.1093/biomet/40.3-4.318}. In practice, the $F$-test might be
easier to implement than the hypothesis test described in detail in the limit
setting description of \cref{sec:user:interpretation:limit}. However, it is not
as statistically stringent.

The $F$-test is a possibility to quantify how much a model \textbf{B} improves
over the description of the data compared to model \textbf{A} with a different
number of free parameters. This is particularly useful if one of the two models
is a special case of the other model. If \textbf{B} contains \textbf{A}, as in
extensions of the SM, \textbf{B} will always provide a better or equal fit to
the data than \textbf{A}. The $F$-test weighs the higher complexity of
\textbf{B} over \textbf{A} against the improvement in the fit. The test
statistic $f$ is calculated as
\begin{align}
    f = \frac{\chi^2_\textbf{A}}{\nu_\textbf{A}} \left/
    \left(\frac{\chi^2_\textbf{B}}{\nu_\textbf{B}}\right)\right.,
\end{align}
with the number of degrees of freedom $\nu_\textbf{A,B} = n_\text{obs} -
    n_\text{par,\textbf{A,B}}$, where $n_\text{obs}$ denotes the number of
    measurements, and $n_\text{par,\textbf{A,B}}$ corresponds to the number of
    parameters in the two models. The $\chi^2$ values refer to the minimal
    $\chi^2$ value found in the parameter space (\ie the best-fit value). For
    instance, we can consider the SM as the restricted model \textbf{A}, and a
    BSM model that extends the SM as model \textbf{B}. Since for the expectation
    value $E[\chi^2]=\nu$ and thus $E[\chi^2/\nu]=1$ holds if the model truly
    describes the data, and since $E[\chi^2/\nu]>1$ if the model does not
    truly describe the data, $E[f]<1$ is expected if the restricted model
    \textbf{A} describes the data better, and $E[f]>1$ is expected if
    \textbf{B} describes the data better.

The cumulative probability $F$ quantifies the significance of the
$\chi^2$ improvement found in the more general model, while
accounting for the larger number of parameters. It is found by
integrating the probability density function of the test statistic $f$ from zero to the
$f$-value determined by the data via the fit. The null hypothesis, which
is that model~\textbf{B} does not provide a significantly better fit to
the data than model~\textbf{A} (i.e.~the SM), can be rejected, for
instance, at the $68\%$ ($95\%$) confidence level if $F > 0.68~(0.95)$.

\subsubsection{Caveats on rejections of the Null Hypothesis}%
\label{sec:user:interpretation:nogoes}
The decision chart in \cref{fig:decisionchart} also contains the case where a
combination of measurements available in \HS{} could be used to reject the SM
hypothesis in favor of a specific model of New Physics. We urge for a lot of
caution in such an application. For instance, consider the combination of ATLAS
and CMS results via \HS{}. On a purely statistical basis, this test is more
sensitive than a test using experimental results from a single experiment, and
thus it is possible that such a phenomenological analysis might be the first to
claim a rejection of the null hypothesis (\ie the hypothesis that the observed
Higgs boson has precisely the properties predicted by the SM) in favor of a
specific alternative hypothesis (\eg a model where the \SI{125}{\GeV} Higgs
boson has a modified coupling structure with respect to the SM). As long as the
model of New Physics is \emph{not} constructed to specifically fit the observed
deviations of the data from the SM prediction, it is \emph{in principle}
statistically viable to perform such a test. However, if the sensitivity of the
single experiments is not sufficient to reject the null hypothesis, it is very
likely that the exclusion derived in a combination of measurements from both
experiments is statistically marginal, making the numerical outcome strongly
dependent on factors like the treatment of systematics and their correlations.
While these are implemented with care in \HS{}, they contain the relevant
correlations only for the theory uncertainties and the luminosity, and the
published experimental correlations \emph{within} one analysis. Experimental
correlations between different analyses, which could arise \eg from particle
reconstruction effects or common selection cuts are usually not published and
hence cannot be taken into account in \HS{}. Therefore, it is not advisable to
claim a statistically meaningful rejection of the SM based on \HS{} alone. In
addition, as explained in \cref{sec:user:interpretation:goodness}, one
should also consider the custom to place a much more stringent criterion of
$5\,\sigma$ significance, corresponding to $p_0<2.7\times 10^{-7}$, on the C.L.
required for the exclusion of the null hypothesis.

Furthermore, the rejection of the null hypothesis always depends --- to a
varying degree, depending on the definition of the likelihood ratio --- on the
signal hypothesis. This is also true for the profile likelihood technique
employed at the LHC~\cite{Cowan:2010js}, where a discovery is claimed if the
likelihood of the background only (or SM) hypothesis compared to the best fit
hypothesis of an alternative model is sufficiently small. It is conceivable that
a model of New Physics is tested as an alternative hypothesis which was
specifically and intentionally constructed to describe known deviations of the
measurements from the SM\@. Testing such a model against the \emph{same} data as
was used for its construction in order to claim a rejection of the SM is
obviously \emph{not} statistically meaningful.

\section{Performance tests}%
\label{sec:performance}

In this section we discuss the performance of \HS for a few selected
experimental analyses by comparing official and \HS-reproduced results for
various models that parametrize Higgs couplings or certain production rates. We
first present the \HS performance for the ATLAS and CMS Run-1
combination~\cite{Khachatryan:2016vau} of Higgs boson measurements, either using
the officially combined measurements or measurements from the individual Run-1
analyses. We then discuss example results for Run-2 analyses, either using
signal strength ($\mu$) measurements or STXS measurements as input. We conclude
this section with a recommendation for the presentation of future Higgs signal
rate measurements.

\subsection{Reproduction of the  ATLAS and CMS Run-1 combination}%
\label{sec:performance:Run1combined}

ATLAS and CMS published results for the production cross sections and decay
rates of the observed Higgs boson from a combined analysis of the LHC $pp$ collision data
at $\sqrt{s}=\text{\SIlist{7;8}{\TeV}}$~\cite{Khachatryan:2016vau}.
For the reproduction of the results with \HS we use two different experimental inputs:
\begin{enumerate}
    \item The combined ATLAS and CMS results for production cross section,
          $\sigma_i$, times decay branching ratio, $\mathrm{BR}(H\to f)$, of the
          various signal channels together with the
          provided correlation matrix from Ref.~\cite{Khachatryan:2016vau}. This input and the associated $\chi^2$ calculation is described in \cref{Sec:chi2LHCRun1}.
    \item The signal strength measurements published in individual Run-1 analyses by ATLAS~\cite{Aad:2014eha,Aad:2015vsa,ATLAS:2014aga,Aad:2014eva,Aad:2015gra,ATLAS:2015zda} and CMS~\cite{Chatrchyan:2013iaa,CMS:2013xda,Khachatryan:2014ira,Chatrchyan:2014nva,CMS:2014ega,Chatrchyan:2013mxa,Khachatryan:2014qaa,CMS:2013dda,CMS:utj}\@.
\end{enumerate}
We consider several generic parametrizations based on Higgs coupling scale
factors and compare the \HS results for both input methods to the official
experimental fit.

\subsubsection{Parametrization through coupling scale factors}
In the $\kappa$ framework~\cite{LHCHiggsCrossSectionWorkingGroup:2012nn} we parametrize BSM effects through seven
independent Higgs coupling scale factors --- the generation-independent fermionic scale factors for up-type and down-type quarks $\kappa_u$ and $\kappa_d$, respectively, for leptons $\kappa_{\ell}$, as well as the heavy gauge boson scale factors $\kappa_Z$, $\kappa_W$, and the loop-induced gluon and photon scale factors $\kappa_g$ and $\kappa_{\gamma}$.
Since the total width of the Higgs boson cannot be constrained at the LHC with sufficient precision in a model-independent way, an additional assumption is necessary. Possible assumptions to overcome the degeneracy induced by the unknown total width are the following~\cite{Bechtle:2014ewa}:
\begin{enumerate}
    \item There are no BSM decays of the Higgs boson, \ie $\BR{H\to\text{NP}}=0$.
    \item New physics decays are allowed, \ie $\BR{H\to\text{NP}}\geq0$ but the
          scale factors for the Higgs-gauge boson couplings are required to be
          $|\kappa_{W,Z}|\leq 1$. This assumption breaks the degeneracy introduced
          by the unknown total width by limiting the VBF and $VH$ Higgs production
          channels~\cite{Duhrssen:2004uu,Duhrssen:2004cv}.
    \item All additional Higgs decay modes are required to yield an invisible
          final state, $\BR{H\to\mathrm{NP}}=\BR{H\to\mathrm{inv.}}$. Such decay models can then be constrained by direct searches at the LHC which exploit the Higgs boson recoil when produced in association with other objects ($Z$ or $W$ boson, quarks, etc.).
\end{enumerate}
In the following, we only consider the first two scenarios. The second one is
compatible with a wide range of BSM physics. In particular, it is valid for
models that contain only singlet and doublet Higgs fields. By including the
branching fraction into states of new physics
\begin{equation}
    \BR{H\to\mathrm{NP}}=1-\frac{\kappa^{2}_H\Gamma^H_{\text{SM}}}{\Gamma^H}\eqcomma
    \label{eqn:BRHnp}
\end{equation}
as additional free parameter in the fit we have eight free parameters in the
second scenario. In this equation a new modifier, $\kappa_H$, defined as
\begin{equation}
    \kappa^2_H=\sum\limits_j \mathrm{BR}^j_{\text{SM}}\kappa^2_j\eqcomma
    \label{eq:kappaH}
\end{equation}
is introduced to characterize modifications to the sum of the partial widths of
decays of the Higgs boson into SM final states. Here, $\mathrm{BR}^j_{\text{SM}}
= \Gamma_{\text{SM}}^j/\Gamma_{\text{SM}}^H$ is the branching ratio as predicted
for a SM Higgs boson. In both fits, it is assumed that the coupling scale factor
$\kappa_u$ is positive, without loss of generality.

\begin{figure}[t]
    \centering
    \includegraphics[width=0.4\textwidth,height = 0.55\textwidth]{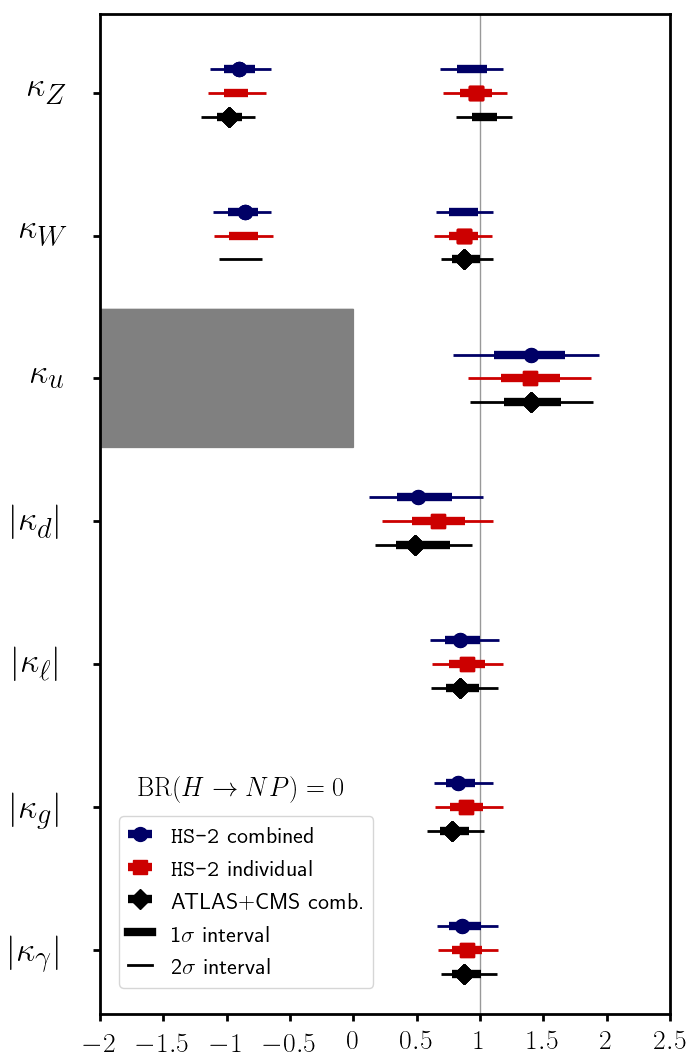}\hspace{1cm}
    \includegraphics[width=0.4\textwidth,height = 0.55\textwidth]{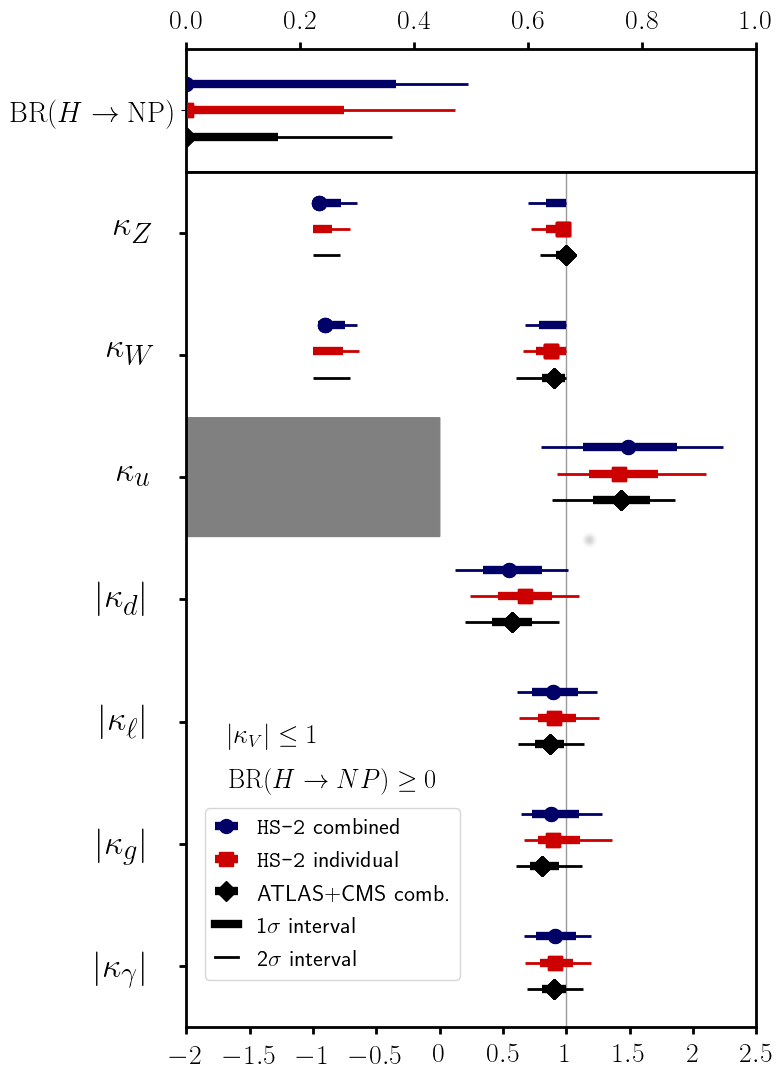}
    \caption{Official and reproduced ATLAS and CMS combined Run-1 results
        in the Higgs coupling scale factor parametrization, assuming no new BSM Higgs
        decay modes (\emph{left panel})  or $\kappa_V \le
            1$ (\emph{right panel}). The \HSv{2} results have been obtained using as
        input either the combined measurements (blue) or measurements
        from individual analyses (red).
        The official results are included in black.
        The gray areas indicate that $\kappa_u>0$ is assumed without loss of generality. Only absolute values are shown for sign-insensitive parameters. The error bars indicate
        the $1\sigma$ (thick lines) and $2\sigma$ (thin lines)
        intervals.}\label{Combined_Zwudlgag}
\end{figure}

\Cref{Combined_Zwudlgag} shows the fit results for the two scenarios. The
\HSv{2} fit results with the ATLAS and CMS combination input and the results
with the individual signal strength input are shown in blue and red,
respectively. The results obtained from \HS with the different input datasets
agree well with each other, and both results are consistent with the official
results shown in black. Differences in the sign of the best fit point in the \HS
results are typically due to $\chi^2$ distributions that are exactly (\eg for
$\kappa_W$ in the $\mathrm{BR}(H\to\mathrm{NP})=0$ case) or almost insensitive
to the sign.

For $\BR{H \to \mathrm{NP}} = 0$ (\cref{Combined_Zwudlgag}, \emph{left panel})
in the case of the individual signal strength input the best-fit point has a
goodness-of-fit of $\chi^2_{\min}/\mathrm{ndf}=52.6/69$ which leads to a \pvalue
compatibility between the data and the predictions of the SM of
\SI{92.87}{\percent}. In the case of the ATLAS and CMS combination input we find
$\chi^2_{\min}/\mathrm{ndf}=14.95/13$ and a \pvalue estimate of
\SI{31.0}{\percent}.  The reason for this large difference is the sensitivity of
the goodness-of-fit to the structure of the experimental data discussed in
\cref{sec:user:interpretation:goodness}. When many separate measurements are
included, the \pvalue mostly measures the agreement between the separate
measurements instead of quantifying the agreement of the model with the
data~\cite{Bechtle:2015nua}. For this reason we refrain from quoting the
goodness-of-fit for the remainder of this section.

\newpage
For the $|\kappa_V| \le 1$ assumption  in \cref{Combined_Zwudlgag} (\emph{right
panel}), the largest differences between the \HSv{2} and the official results
appear in the contributions of BSM decays. The \HS analysis returns
$\mathrm{BR}(H\to\mathrm{NP}) < 0.37$ ($<0.28$) as the \CL{68} region using the
combined (individual) ATLAS and CMS signal strength input. Both are notably
larger than the official results $\mathrm{BR}(H\to\mathrm{NP})<0.16$. We suspect
that the assumption of Gaussian uncertainties is not fully applicable in all
parts of the parameter space and therefore causes the observed discrepancy.

\subsubsection{Parameterization using ratios of coupling scale factors}%
\label{sec:ratios}

We now turn to a Higgs coupling parametrization in terms of ratios of $\kappa$
scale factors. The $gg\to H\to ZZ$ channel serves as reference channel for the
normalization because its overall uncertainties and background contamination are
very small. It is parameterized as a function of
$\kappa_{gZ}=\kappa_g\cdot\kappa_Z/\kappa_H$, with $\kappa_H$ defined
in~\cref{eqn:BRHnp}.\footnote{This definition of $\kappa_{gZ}$ implicitly
assumes the absence of Higgs decays to new physics,
$\mathrm{BR}(H\to\mathrm{NP}) = 0$. However, the fit results can be generalized
to the case $\mathrm{BR}(H\to\mathrm{NP}) \ne 0$ if $\kappa_{gZ}$ is redefined
to $\kappa_{gZ}=\kappa_g\cdot\kappa_Z/\tilde{\kappa}_H$ with $\tilde{\kappa}_H^2
= \kappa_H^2/(1-\mathrm{BR}(H\to\mathrm{NP}))$.} The ratios
$\lambda_{Zg}=\kappa_Z/\kappa_g$ and $\lambda_{tg}=\kappa_t/\kappa_g$ are probed
by taking the ratios of measurements of VBF and $ZH$ production and measurements
of $t\bar{t}H$ production, respectively, to the measured rate for the $gg\to
H\to ZZ$ channel. Similarly, the ratios obtained from gluon fusion production
with the decay modes $H\to WW$, $H\to \tau\tau$, $H\to bb$ and $H\to
\gamma\gamma$ probe the four ratios $\lambda_{WZ}=\kappa_W/\kappa_Z$,
$\lambda_{\gamma Z}=\kappa_\gamma/\kappa_Z$, $\lambda_{\tau
Z}=\kappa_{\tau}/\kappa_Z$ and $\lambda_{bZ}=\kappa_b/\kappa_Z$. Without loss of
generality, $\kappa_Z$ and $\kappa_g$ are assumed to have the same sign,
constraining $\lambda_{Zg}$ and $\kappa_{gZ}$ to be positive.

\begin{figure}[t]
    \centering
    \includegraphics[width=0.4\textwidth,height = 0.55\textwidth]{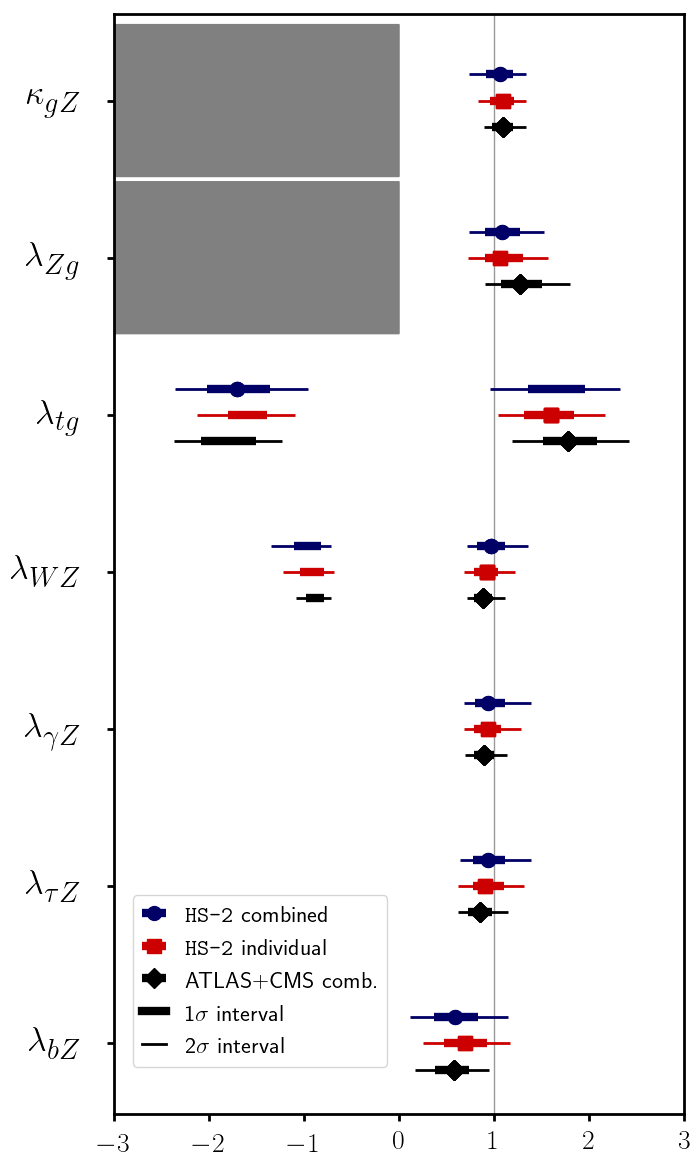}
    \caption{Official and reproduced ATLAS and CMS combined Run-1 results for
        ratios of Higgs coupling scale factors. The results obtained with
        \HSv{2} using the combined ATLAS and CMS data as input and the results
        using the individual signal strength measurements as input are shown in
        blue and red, respectively, while the official results are shown in
        black. The error bars indicate the $1\sigma$ (thick lines) and $2\sigma$
        (thin lines) intervals. The gray areas indicate the parameters that are
        assumed to be positive.}%
    \label{Combined_lambda}
\end{figure}

\Cref{Combined_lambda} compares the official fit results with the ones obtained
from \HSv{2}, displaying the corresponding  best-fit values for the different
$\lambda$ parameters. We find very good agreement between the results obtained
from \HSv{2} and the official results, though the \HS fits tend to have slightly
larger \SI{68}{\percent} and \CL{95} intervals. There are two reasons for these
differences. Firstly, the Gaussian approximation may not be valid for the
experimental uncertainties in all regions of parameter space. Secondly, the
parameterization in terms of ratios should lead to partial cancellations of
common theoretical uncertainties which are not expected to be entirely captured
in the \HS results. While our result with the individual signal strength input
reproduces the positive sign of the best-fit value for $\lambda_{tg}$, the
\HSv{2} fit with the ATLAS and CMS combination input appears to prefer negative
values. However, the $\Delta\chi^2$ between the negative and the positive
best-fit value is tiny and thus there is no clear preference for one particular
sign.

\subsubsection{Parametrization using ratios of cross sections and branching fractions}\label{sec:crossSection}

Using the narrow-width approximation, the signal strength $\mu_i^f$ can be
decomposed into the signal strength for production $\mu_i=\sigma_i
    /\sigma_{i,\text{SM}}$ and the signal strength for the decay,
$\mu^f=\mathrm{BR}^f/\mathrm{BR}_\text{SM}^f$, with the short-hand notation $\text{BR}^f\equiv \BR{H\to f}$.

Choosing the $gg\to H\to ZZ$ channel as reference, the product of the
production cross section $\sigma_i$ and the branching fraction
$\text{BR}^f$ can be expressed as
\begin{equation}
    \sigma_i\cdot\mathrm{BR}^f =\sigma(gg\to H\to ZZ) \cdot\left(\frac{\sigma_{i}}{\sigma_{\mathrm{ggF}}}\right)\cdot \left( \frac{\mathrm{BR}^f}{\mathrm{BR}^{ZZ}}\right).
    \label{eqn:crossSections}
\end{equation}

In accordance with the ATLAS and CMS analysis~\cite{Khachatryan:2016vau}, we
assume that the gluon fusion (ggF) and the $bbH$ production signal strengths are
equal, $\mu_{\mathrm{ggF}}=\mu_{bbH}$, the $H\to Z\gamma$ and $H\to\gamma\gamma$
decay signal strengths are equal, $\mu^{Z\gamma}=\mu^{\gamma\gamma}$, and the
$H\to gg$, $H\to cc$, and $H\to bb$ decay signal strengths are equal,
$\mu^{gg}=\mu^{cc}=\mu^{bb}$.

\begin{figure}[t]
    \centering
    \includegraphics[height = 0.55\textwidth]{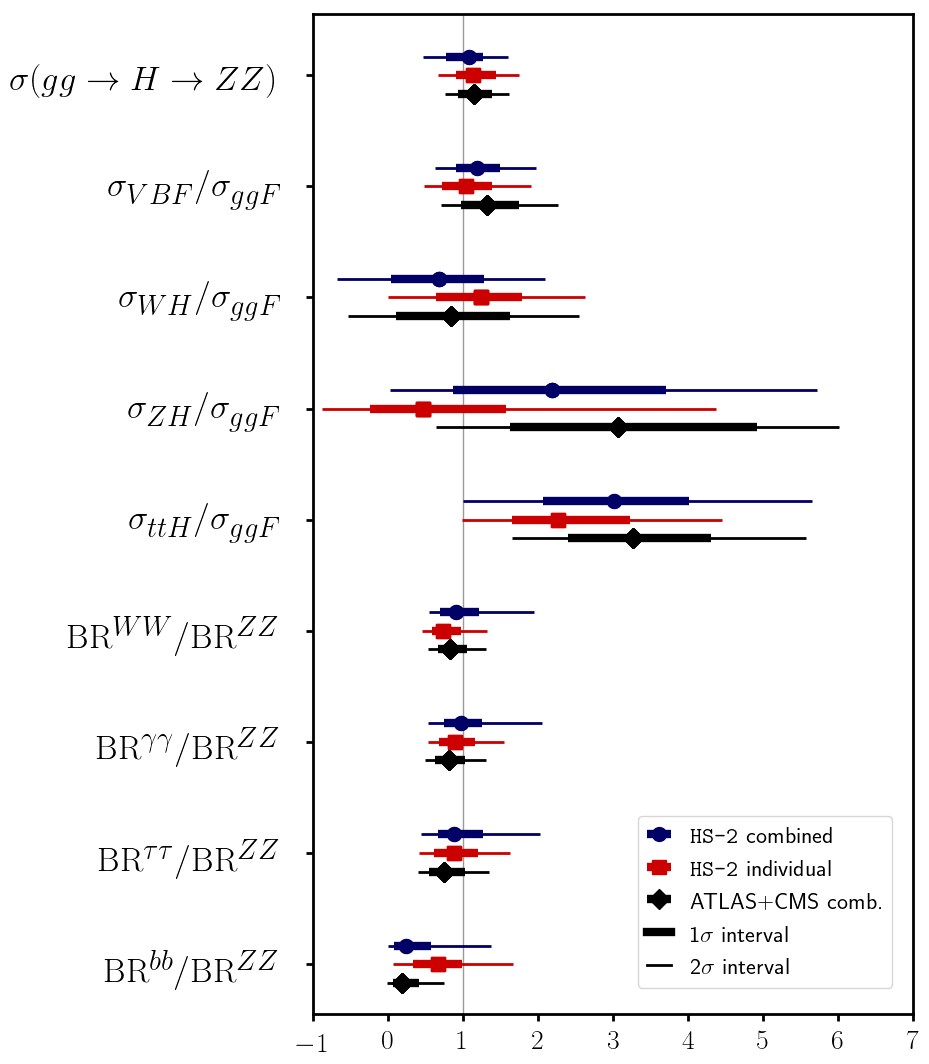}
    \caption{Results obtained with \HSv{2} in comparison with the official fit
            results for the $\sigma (gg\to H\to ZZ)$ cross section and for
            ratios of cross sections and branching fractions. The {\HSv2}
            results using the combined ATLAS and CMS data and the results using
            the individual signal strength measurements as experimental input
            are shown in blue and red, respectively. Also shown are the official
            results from ATLAS and CMS (black). The error bars indicate the
            $1\sigma$ (thick lines) and $2\sigma$ (thin lines) intervals. The
            results are normalised to the respective SM predictions.}%
    \label{Combined_crossSection}
\end{figure}

\Cref{Combined_crossSection} shows the fit results from the combined ATLAS and
CMS analysis and the fit to individual signal strength measurements in blue and
red, respectively. The fit result from the official ATLAS and CMS combination is
shown in black. While the result obtained with {\HSv2} using the combined input
agrees well with the official result, we observe some larger discrepancies for
various parameters when using individual measurements, most strikingly for
$\sigma_{ZH}/\sigma_{\mathrm{ggF}}$ and $\mathrm{BR}^{bb}/\mathrm{BR}^{ZZ}$
which are strongly anti-correlated. Furthermore the central values of
$\sigma_{ttH}/\sigma_{\mathrm{ggF}}$ and
$\sigma_{\text{VBF}}/\sigma_{\mathrm{ggF}}$ are shifted towards the SM
prediction. We mainly relate these features to the fact that some analyses ---
in particular in CMS --- have been improved for the combined result, but no
updated individual measurements have been released. \Cref{fig:only} shows a
comparison between the official ATLAS-only~\cite{Aad:2015gba} (\emph{left
panel}) and CMS-only~\cite{Khachatryan:2014jba} (\emph{right panel}) fit results
and the corresponding results obtained with \HSv{2} when using the individual
signal strengths from ATLAS and CMS\@. \HSv{2} reproduces the official
ATLAS-only fit result very well. However, for the CMS-only fit we observe
similar discrepancies as for \cref{Combined_crossSection}, namely smaller values
for $\sigma_{\mathrm{VBF}}/\sigma_{\mathrm{ggF}}$,
$\sigma_{ZH}/\sigma_{\mathrm{ggF}}$, and $\sigma_{ttH}/\sigma_{\mathrm{ggF}}$ as
well as a larger value for $\mathrm{BR}^{bb}/\mathrm{BR}^{ZZ}$.

\begin{figure}[t]
    \centering
    \includegraphics[height = 0.55\textwidth]{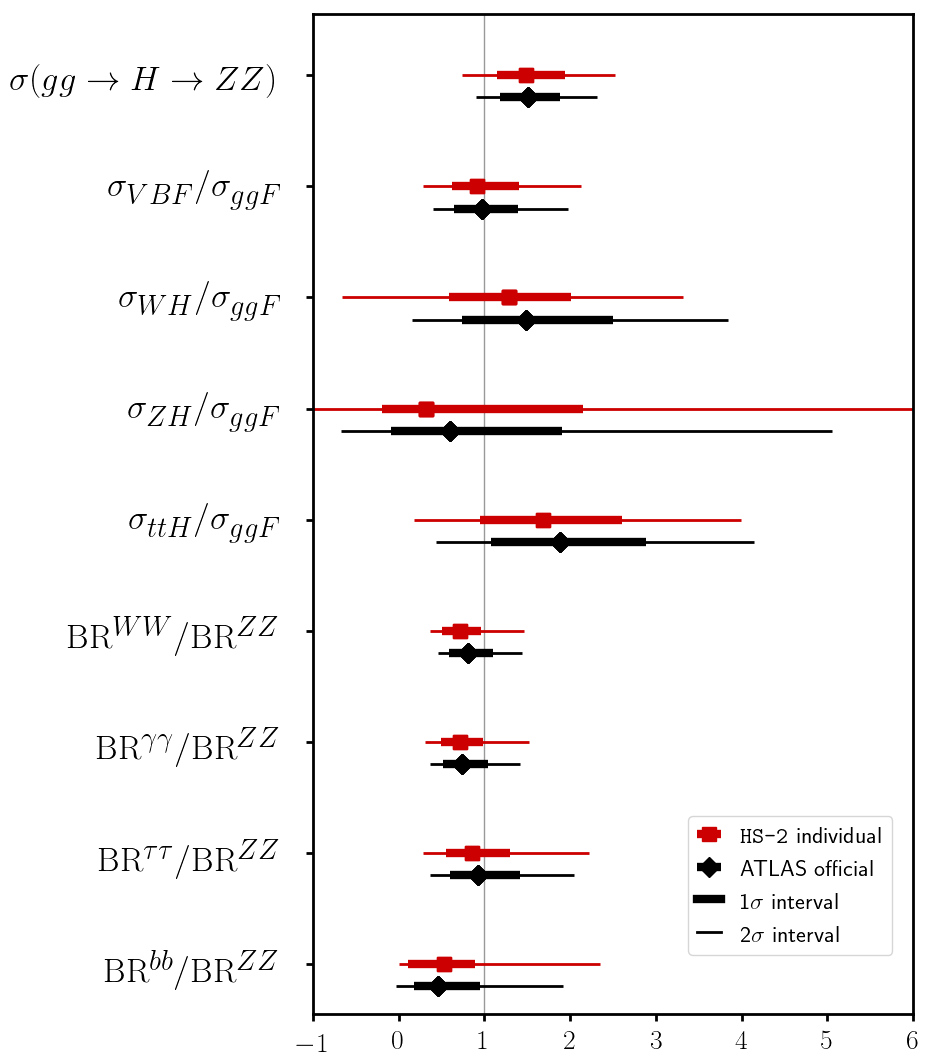}\hfill
    \includegraphics[height = 0.55\textwidth]{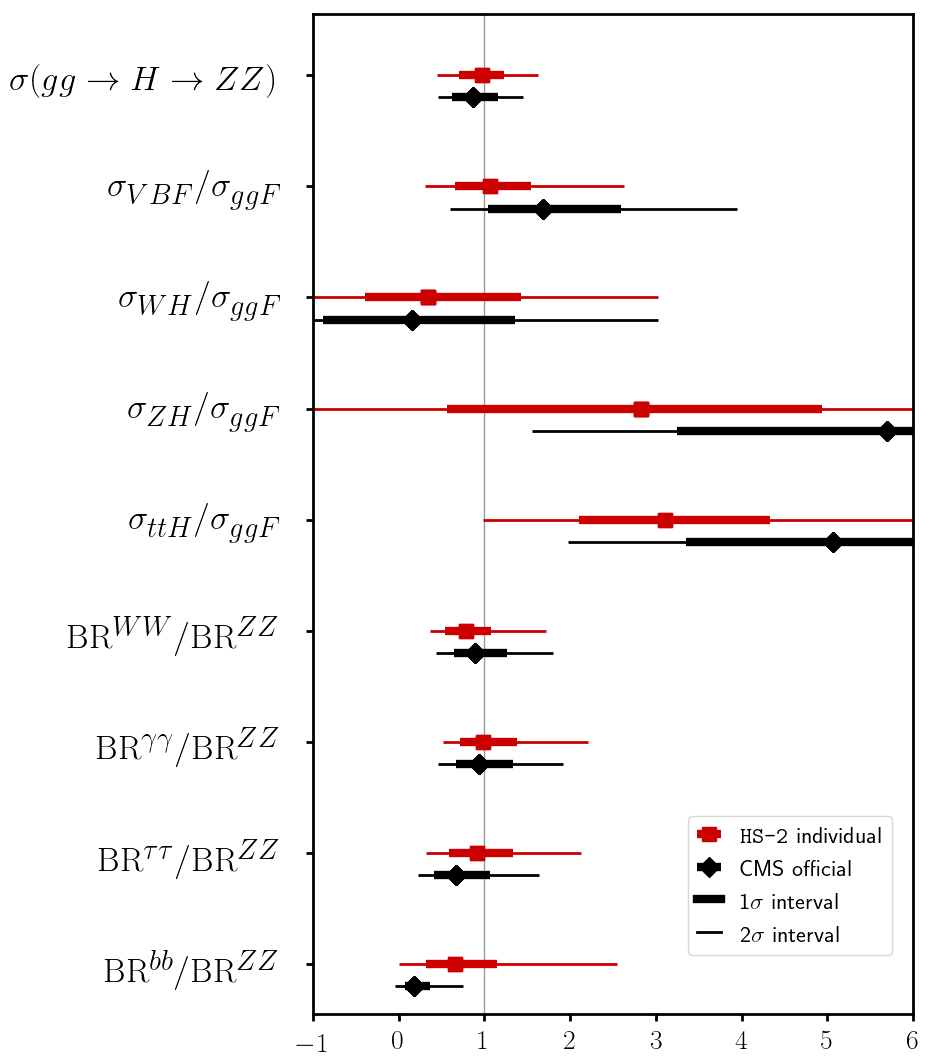}
    \caption{Results obtained with \HSv{2} in comparison with the official
        ATLAS-only (\emph{left panel}) and CMS-only (\emph{right panel}) fit
        results for the $\sigma (gg\to H\to ZZ)$ cross section and for
        ratios of cross sections and branching fractions. The \HSv{2} results
        and the official results are shown in red and black, respectively. The
        error bars indicate the $1\sigma$ (thick lines) and $2\sigma$ (thin
        lines) intervals. The results are normalised to the respective SM
        predictions.}%
    \label{fig:only}
\end{figure}

In summary, the performed comparisons in all three model parametrizations have
demonstrated very good agreement between the \HS\ implementation of the LHC
Run-1 measurements --- both using the individual and the combined experimental
input --- and the official ATLAS/CMS fit results. The agreement between the two
possible \HS implementations is on the one hand a successful closure test of the
\HS peak-centered $\chi^2$ method, and on the other hand motivates our choice of
using the LHC-Run-1 combined experimental measurements as default input for the
LHC Run-1 legacy $\chi^2$ evaluation in \HSv{2}, as described in
\cref{Sec:chi2LHCRun1}. Computationally, this implementation is much faster.
However, for very specific applications where the assumptions underlying the LHC
Run-1 combination are not fulfilled, the experimental input from the individual
Run-1 analyses is still available as the \Code{LHC7+8} observable set in the \HS
package.

\subsection{Examples for Run 2 Analyses in \HSv{2}} During Run 1 of the LHC,
Higgs rate measurements were mainly represented in terms of signal strengths,
$\mu=\sigma / \sigma_\mathrm{SM}$, and coupling modifiers, $\kappa_i$. For LHC
Run 2 the experimental collaborations increasingly made use of the STXS
framework to present their results (see \cref{sec:STXS}). In some analyses, both
STXS measurements and conventional signal strengths measurements in various
event categories were presented, along with the correlation matrices, which are
necessary to allow a comparison of the performance of the two experimental input
formats.

In this section we discuss the performance of \HSv{2} on the provided
experimental input for a selection of LHC Run-2 analyses. These examples are
primarily chosen to illustrate the level of agreement of the reconstructed \HS\
result with official results from ATLAS and CMS, and to highlight difficulties
in the usage of the experimental results, which are often related to incomplete
information in the public documentation of the experimental analysis.
With the increasing amount of data during Run-2 the statistical uncertainty
can be assumed to be Gaussian to very good approximation in most Higgs boson
search channels. However, a decreasing statistical uncertainty also entails
the fact that systematic uncertainties and their correlations among different
measurements become more relevant. Therefore, it has become common practice
for ATLAS and CMS to provide a correlation matrix of  the experimental
(statistical and systematic) uncertainties for the Run-2 measurements.

In the following we  discuss the performance of the two input types (signal
strength modifiers and STXS measurements) for a few selected Run-2 examples
that illustrate what experimental information is needed to enable a
successful application of the results to BSM models.
Unless otherwise noted, a Higgs boson mass of $m_H=\SI{125.09}{GeV}$ is assumed.
The chosen examples do not necessarily represent the latest measurements implemented in \HSv{2}. The
complete list of the ATLAS and CMS Run-2 Higgs signal rate measurements are
summarized in \cref{tab:expRun2ATLAS} and \cref{tab:expRun2CMS}, respectively.

\subsubsection{Input given in terms of signal strength modifiers}%
\label{sec:HS_with_muInput}

We discuss the CMS measurements in the $H\to W^+W^-$ channel at the LHC
Run-2 as an example for analyses implemented in terms of signal strength
modifiers. First Run-2 results were released based on 2015 and early 2016 data
with integrated luminosities of \SI{2.3}{\per\fb} and \SI{12.9}{\per\fb},
respectively~\cite{CMS:2017pzi}. The results of this early analysis
are given in terms of sub-channel signal strength modifiers, where the
different channels are tailored towards different Higgs productions modes (gluon
fusion, VBF, $ZH$ and $WH$). However, no signal efficiencies that would allow a
better estimate of the signal composition in the different channels were
provided.

\begin{figure}[t]
    \centering
    \includegraphics[scale=0.49]{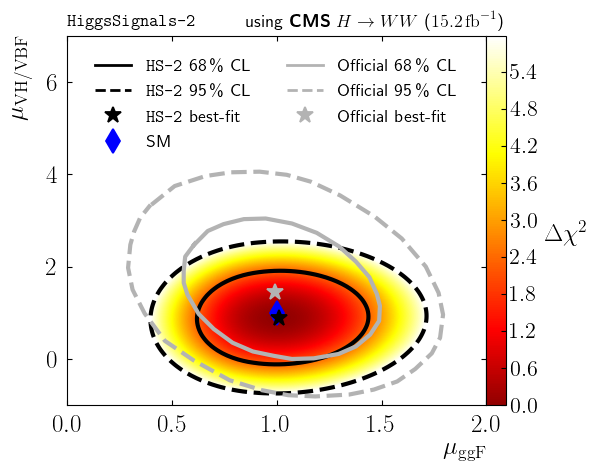}
    \includegraphics[scale=0.49]{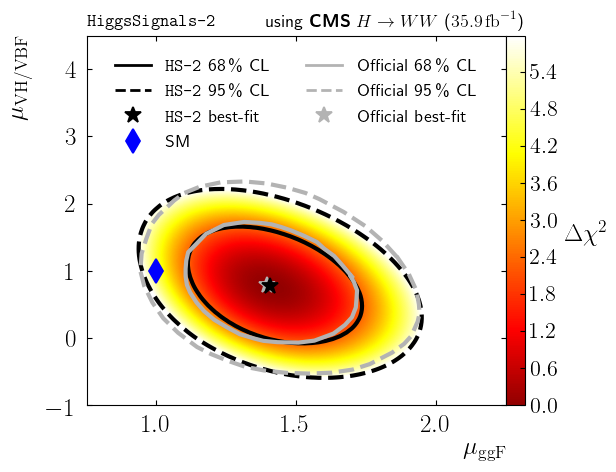}
    \caption{Comparison of the reconstructed fit results for the signal strength
            modifiers for fermionic ($\mu_{\mathrm{ggF}}$) and bosonic
            ($\mu_{VH/ \mathrm{VBF}}$) production modes with the official
            results for different CMS analyses in the $H\to WW$ decay channel at
            \SI{15.2}{\per\fb}~\cite{CMS:2017pzi} (\emph{left panel}) and
            $\SI{35.9}{\per\fb}$~\cite{Sirunyan:2018egh} (\emph{right panel}).
            The stars indicate the best fit points, and the solid (dashed)
            contours correspond to the \CL{68} (\CL{95}) regions. The \HS
            results are shown in dark and the official results in light shades.
            The diamonds indicate the SM prediction.}%
    \label{fig:CMS_HWW}
\end{figure}

The plot in the left panel of \cref{fig:CMS_HWW} illustrates the performance
achieved by \HS with the limited information available for the analysis of
Ref.~\cite{CMS:2017pzi}. The comparison is performed in the parametrization
of signal strength modifiers, with $\mu_{\text{ggF}}$ and $\mu_{VH/\text{VBF}}$
rescaling the SM cross section prediction for the fermionic and bosonic
production modes, respectively. The Higgs boson decay rates are fixed to their
SM prediction. The colormap represents the $\Delta\chi^2$ profiles reconstructed
by \HS, and the \SI{68}{\percent} (\SI{95}{\percent}) C.L.\ regions are
highlighted as black solid (dashed) lines. The corresponding official contours
published by CMS are overlaid as gray solid (dashed) lines. We find reasonable
agreement between the reconstructed and official intervals and the corresponding
best-fit point for $\mu_{\text{ggF}}$. However, the size of the allowed
$\mu_{VH/\text{VBF}}$ intervals and the observed anti-correlation between
$\mu_{\text{ggF}}$ and $\mu_{VH/\text{VBF}}$ is not reproduced. The reason for
this discrepancy is the lack of public information on signal efficiencies for
the  sub-channel rate measurements. The anti-correlation observed by CMS in the
left panel of \cref{fig:CMS_HWW} indicates that these sub-channels are
composed of signal contributions from both fermionic and bosonic production
modes.

Information on sub-channel signal efficiencies was made available in the CMS
$H\to W^+W^-$ analysis at \SI{35.9}{\per\fb}~\cite{Sirunyan:2018egh}. The right
panel of \cref{fig:CMS_HWW} shows the performance comparison for this analysis
using the sub-channel signal strength results and the corresponding
efficiencies. We find very good agreement between the reconstructed
$\Delta\chi^2$ and the official likelihood results. This demonstrates the
importance of publicly available detailed sub-channel information on signal
strengths and signal efficiencies. The analysis performed by CMS with
\SI{35.9}{\per\fb} also includes first results for $H\to W^+W^-$ in the stage-0
STXS framework. However no information on the correlations between the STXS bins
was provided for this analysis, which severely limits the usefulness of those
STXS results. The performance achieved using this partial STXS input (not shown
in the figure) is significantly worse than for the signal strength results in
the right panel of \cref{fig:CMS_HWW}.

A further update in the $H\to W^+W^-$ channel to
\SI{137}{\per\fb}~\cite{Sirunyan:2020tzo} has since been released by CMS and is
implemented in the current \HS datasets. The results of this analysis are given
in terms of $n$-jet differential cross sections in the STXS framework. Per-bin
signal efficiencies and inter-bin correlations are available as well. This
analysis provides the most complete input to date, and we expect its
implementation in \HS to be the best performing one. However, the analysis
presented in Ref.~\cite{Sirunyan:2020tzo} does not include any interpretations
that could be used for a performance comparison with \HS.

\subsubsection{Input given in terms of Simplified Template Cross Sections (STXS)}%
\label{sec:HS2_with_STXS}
We have seen that detailed information on sub-channel signal strengths and
signal efficiencies are important when Higgs coupling measurements are given in
terms of signal strengths. We now discuss some example applications for which
the input to \HSv{2} is given in the STXS framework instead.

\begin{figure}[tp]
    \centering
    \includegraphics[width = 0.49\textwidth]{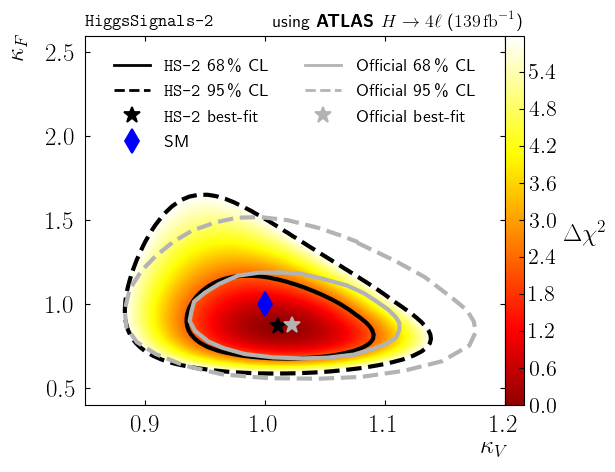}
    \includegraphics[width = 0.49\textwidth]{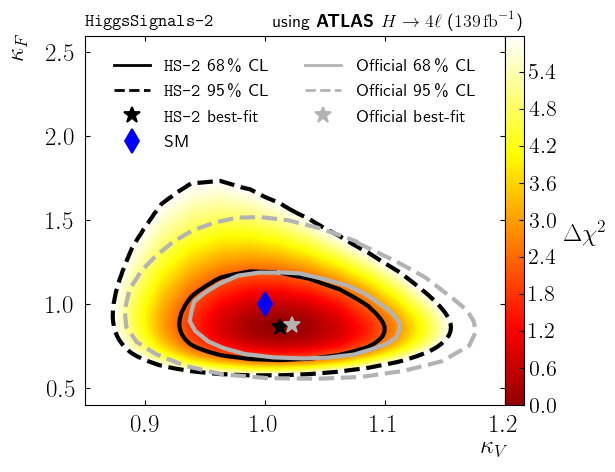}
    \caption{Performance test in the ($\kappa_V, \kappa_F$) parameter plane
        using the STXS measurements of the ATLAS $H \to ZZ$ analysis with
        \SI{139}{\per\fb}~\cite{Aad:2020mkp} as \HS\ input\@. Correlations of
        experimental uncertainties are included in both figures, while the
        correlations of theoretical uncertainties on the STXS bin predictions
        are neglected (\emph{left panel}) or taken into account (\emph{right
        panel}). The stars indicate the best fit points, and the solid (dashed)
        contours correspond to the \CL{68} (\CL{95}) regions. The \HS results
        are shown in dark and the official results in light shades. The diamonds
        indicate the SM prediction.}%
    \label{fig:ATL_HZZ_139}
\end{figure}

As a first example we study the \HS performance for ATLAS measurements in the
$H\to ZZ^*\to 4\ell$ channel with $\SI{139}{\per\fb}$ of data~\cite{Aad:2020mkp}
as STXS observables. These experimental results were given in 12 reduced
Stage-1.1 STXS bins along with the correlation matrix for the experimental
uncertainties (Fig.~10 of Ref.~\cite{Aad:2020mkp}). The \HS\ $\Delta\chi^2$
distribution in the ($\kappa_V$, $\kappa_F$) parameter plane based on this
input, neglecting correlations of theoretical uncertainties on the STXS bin
predictions, is shown in the left panel of \cref{fig:ATL_HZZ_139} in comparison
to the official ATLAS result (shown as gray contours). The agreement at lower
values of $\kappa_V$ and $\kappa_F$ with the ATLAS results is very good.
However, at larger values we find a small mismatch between the reproduced and
official confidence region contours. In these regions the agreement can be
improved if correlations of theoretical uncertainties on the gluon fusion STXS
bin predictions are included in the $\chi^2$ calculation, as shown in the right
panel of \cref{fig:ATL_HZZ_139}. These correlations were evaluated by the
ggF-subgroup of the LHC HXSWG and have been taken from Ref.~\cite{THUcorrmatrix}
(``2017 scheme''). The evaluation of similar correlations for the STXS bins of
other production modes is still in progress. As can be seen these correlations
lead to a flattening of the likelihood at large coupling scale factors, \ie in
the regions where the corresponding cross sections (and thus their
uncertainties) are larger than the SM prediction.

\begin{figure}[tp]
    \centering
    \includegraphics[width=0.49\textwidth]{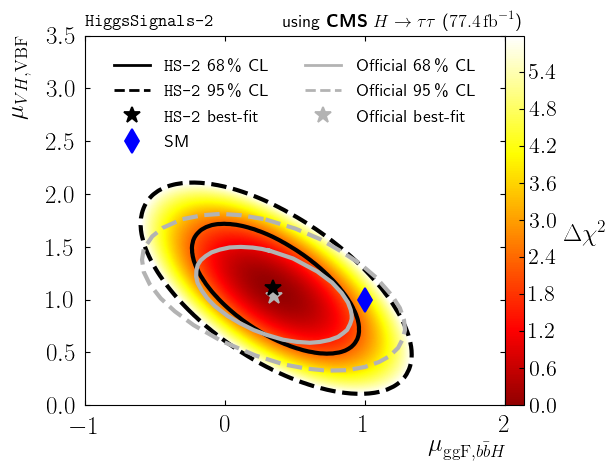}
    \includegraphics[width=0.49\textwidth]{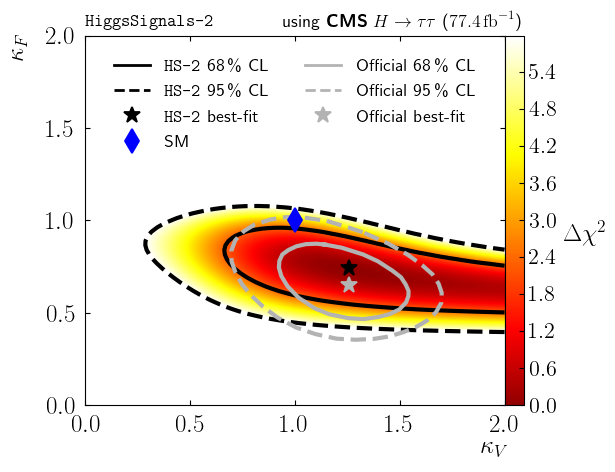}
    \caption{Performance test in the
        ($\mu_{\text{ggF},b\bar{b}H},\mu_{VH,\text{VBF}}$) (\emph{left panel})
        and ($\kappa_V, \kappa_F$) parameter plane (\emph{right panel}) using
        the STXS measurements of the CMS $H \to \tau\tau$ analysis with
        \SI{77}{\per\fb}~\cite{CMS:2019pyn} as \HS input. The stars indicate the
        best fit points, and the solid (dashed) contours correspond to the
        \CL{68} (\CL{95}) regions. The \HS results are shown in dark and the
        official results in light shades. The diamonds indicate the SM
        prediction.}%
    \label{fig:CMS_Htautau_77}
\end{figure}

The next analysis that we consider here is the CMS measurement in the
$H\to\tau\tau$ decay channel at \SI{77}{\per\fb}~\cite{CMS:2019pyn}. CMS provides cross
section measurements for nine different kinematic regions together with the
expected acceptances, the SM predictions and the correlations between the bins.
Figure~\ref{fig:CMS_Htautau_77} shows the comparison between the official and
the reproduced $\Delta\chi^2$ contours in the
($\mu_{\text{ggF},b\bar{b}H},\mu_{VH,\text{VBF}}$) (\emph{left panel}) and
($\kappa_V, \kappa_F$) (\emph{right panel}) parameter planes. We find reasonable
agreement in the former but a substantial disagreement in the latter. The source
of this discrepancy can be traced back to the fact that CMS included the
contribution from $H\to WW$ to the $e\mu$ final state to remove an unconstrained
direction along $\kappa_V$. As the $H\to WW$ contribution is not accounted for
in the presented STXS measurements it is not possible to properly implement it
in \HS. We therefore find this open direction. However, in a global picture
where $H\to \tau\tau$ and $H\to WW$ are simultaneously taken into account in
\HS, the flat direction is lifted. This is already the case when adding only the
measurements in the $e\mu$ final state \eg from the dedicated CMS $H\to WW$
analysis~\cite{Sirunyan:2018egh}.

As a final example in this context we discuss the ATLAS combination of Higgs
analyses in the $\gamma\gamma$, $ZZ^*$, $WW^*$, $\tau^+\tau^-$, $b\bar{b}$ and
$\mu^+\mu^-$ final states based on up to \SI{79.8}{\per\fb} of Run-2
data~\cite{Aad:2019mbh}. Within the STXS framework --- under the assumption that
the observed signals are associated with a single particle  --- the various
measurements are combined to determine the cross sections in various STXS bins.
These bins represent a production process in a specific kinematic regime,
e.g.~gluon fusion in association with one additional jet, and a Higgs boson
transverse momentum of $\SI{60}{\GeV} \le p_T^H \le  \SI{120}{\GeV}$, times the branching
fraction of the Higgs  boson to $Z$ bosons, $\BR{H\to ZZ^*}$.  In addition,
ratios of branching ratios are determined for the various final states, with
$\BR{H\to ZZ^*}$ taken in the denominator. Here we use these measurements (given
in Fig.~9 of Ref.~\cite{Aad:2019mbh}) and the corresponding correlation matrix
as experimental input for \HS.

\begin{figure}[tbp]
    \includegraphics[width=0.49\textwidth]{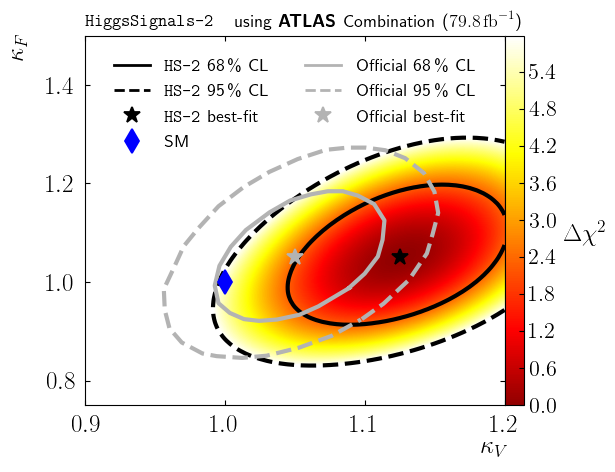}\hfill
    \includegraphics[width=0.49\textwidth]{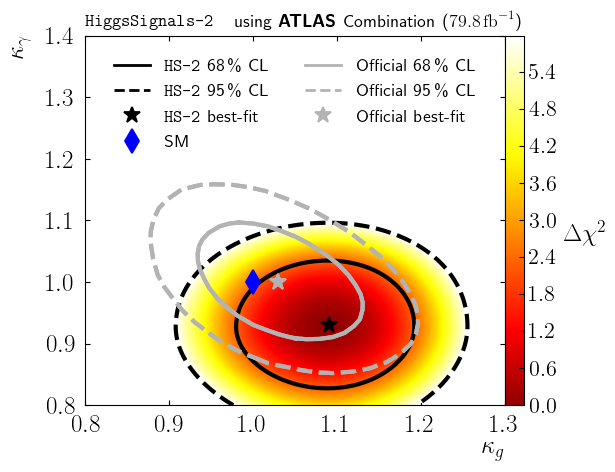}
    \includegraphics[width=0.49\textwidth]{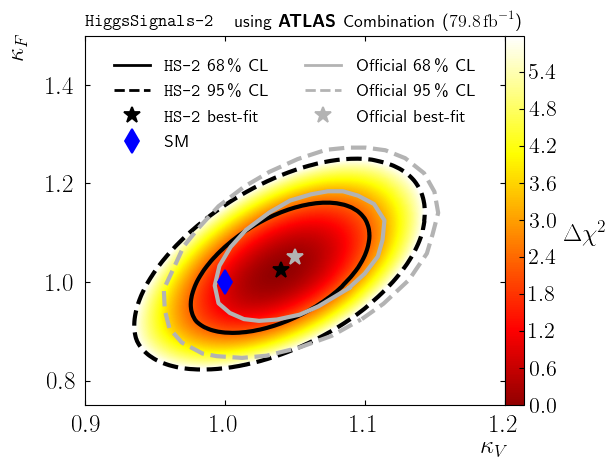}\hfill
    \includegraphics[width=0.49\textwidth]{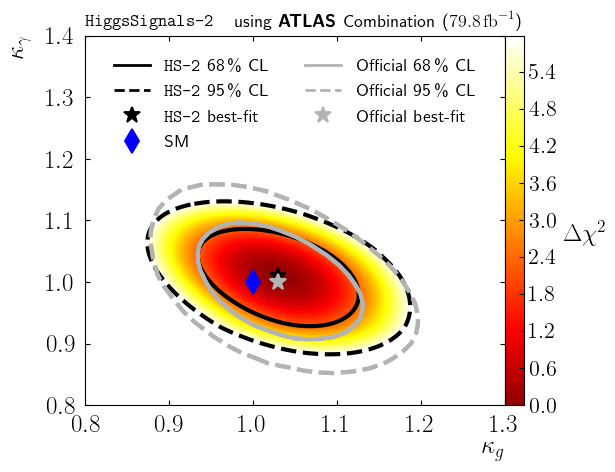}
    \includegraphics[width=0.49\textwidth]{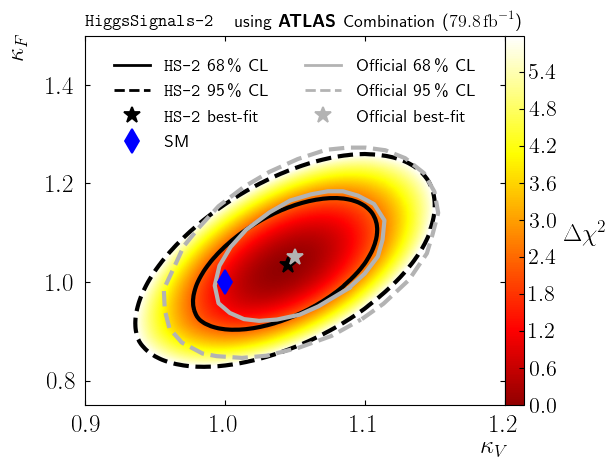}\hfill
    \includegraphics[width=0.49\textwidth]{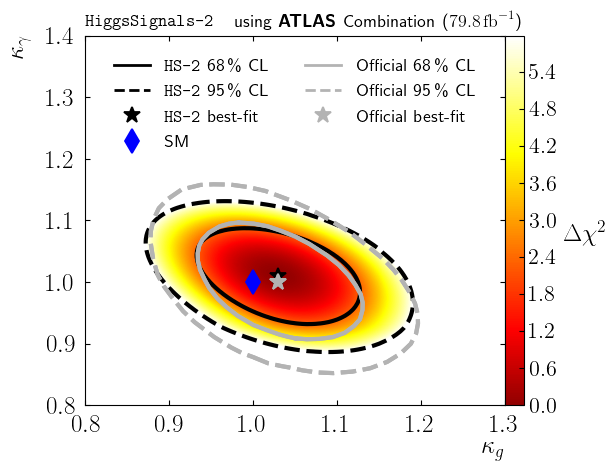}
    \caption{Performance test in the ($\kappa_F, \kappa_V$) plane (\emph{left
        panels}) and the ($\kappa_g,\kappa_\gamma$) plane (\emph{right panels})
        using the STXS measurements of the ATLAS Run-2 Higgs combination with
        \SI{80}{\per\fb} as \HS\ input. The contours and best fit points are
        indicated as in \cref{fig:CMS_HWW,fig:ATL_HZZ_139,fig:CMS_Htautau_77}.
        Correlations of experimental uncertainties are neglected in the top
        panels but included in the middle and bottom panels. Theoretical rate
        uncertainties for the $gg\to H$ process are treated as uncorrelated
        in the top and middle panels and as correlated in the bottom panels,
        see text for further details.}%
    \label{fig:ATLcomb}
\end{figure}

The performance of \HS for the results of the ATLAS Higgs combination is shown
in \cref{fig:ATLcomb}, where the left panels display the results in the
($\kappa_V$, $\kappa_F$) parameter plane and the right panels display the
results in the ($\kappa_g$, $\kappa_\gamma$) plane. For illustration, the top
panels show the resulting likelihood profile if correlations both between
experimental and theoretical uncertainties are neglected. We observe a clear
mismatch in size, shape and location of the allowed regions with respect to the
official ATLAS result (shown as gray contours). Once we include the correlations
of experimental uncertainties the agreement of the reproduced and official
confidence regions strongly improves, as shown in the middle panels of
\cref{fig:ATLcomb}. Finally, the bottom panels show the result when correlations
of theoretical uncertainties in the $gg\to H$ STXS bins~\cite{THUcorrmatrix} are
also included (see above for details). From the comparison of the middle and
bottom panels in \cref{fig:ATLcomb}, we find that these correlations have a
rather small impact, giving rise to a further slight improvement of the
agreement between the \HS result and the official result.

It should again be noted here that combination results with a separate
determination of production and (ratios of) decay rates rely on the assumption
that the signals are associated with a single Higgs boson. Thus, these
combined experimental results cannot be directly treated as peak or STXS
observables in \HS, where per default a signal can be composed of any
superposition of Higgs bosons in the model (see \cref{sec:overlappingHiggs}).
Therefore, in the officially provided observable sets we instead use the
individual (uncombined) measurements as default experimental input. If only one
Higgs boson is present in the model (or the other Higgs bosons have masses far
away from \SI{125}{\GeV}), the combined measurements can still be safely used as
experimental input (provided as observable set \Code{ATLAS_combination_Run2}
in the \HS\ package).

\subsection{Recommendations for the presentation of future Higgs signal rate measurements}

Based on our past experiences and the observations in the above performance
tests we summarize in this section the essential experimental information needed
to enable an accurate reproduction of the model interpretations of the data
(e.g.~in the $\kappa$ framework) provided by ATLAS and CMS and thus a reliable
application of the experimental measurements in BSM model tests. This concurs
with and partly extends recent recommendations on the presentation of LHC
measurements provided in a joint effort by the experimental and theory
community~\cite{Abdallah:2020pec}, as well as similar recommendations that we
recently provided in the context of BSM Higgs limits~\cite{Bechtle:2020pkv}. We
furthermore discuss the issue of the remaining model dependence in Higgs signal
rate measurements, and some ideas for a possible reduction of the model
dependence.

Early LHC results on the Higgs signal rates have been presented as signal
strength modifiers ($\mu$) in specific signal channels --- often binned in
various experimental categories --- targeted by the experimental analysis.
These measurements assume the kinematic properties of the SM Higgs boson,
relying on a SM MC simulation of the \emph{inclusive} signal process. The
result --- the signal strength modifier, $\mu$ --- is defined as the observed
signal rate normalized to the expected signal rate for a SM Higgs boson.

The complexity of calculating the expected signal rate in a BSM model depends on
whether the Higgs boson candidate has --- exactly or to a sufficiently accurate
approximation --- the same kinematic properties as the SM Higgs boson. If so,
the signal rate can be predicted within the BSM model easily without MC
simulation \emph{if} the signal efficiencies (or signal compositions) for the
corresponding experimental measurement are known. This is the approach pursued
per default in \HS, see \cref{Eq:mu_pred}. \emph{It is therefore essential that
the signal efficiencies (or signal compositions) expected for a SM Higgs boson
are provided for all $\mu$ measurements.} In contrast, if the Higgs boson
candidate has significantly different kinematic properties compared to the SM
Higgs boson, the signal needs to be processed through the full experimental
analysis, which includes a full MC simulation (including detector simulation) of
the signal and the application of analysis cuts. Given the high complexity of
most experimental analyses (using, for instance, machine learning techniques)
this is often impossible for theorists (who do not have access to full
simulations for the LHC detectors) and should preferentially be carried out by
the experimental collaborations.

In order to ameliorate this situation, \ie to allow a proper application of
measurements also to Higgs boson candidates with different kinematic properties
than in the SM, the STXS framework has been introduced. An STXS observable is
defined on a specific region in phase space at the MC particle level. The
experimental unfolding process only relies on the kinematic properties of the SM
Higgs boson within this specific phase space.\footnote{While this is true for
the central measurement, uncertainty correlations also depend on the SM Higgs
boson predictions (and thus the assumed SM Higgs boson kinematics) in other
phase space regions.} The model-dependence is therefore reduced, which, in turn,
allows a wider application of these measurements to BSM models. At the moment,
STXS observables are typically binned in jet multiplicity and/or Higgs boson
transverse momentum ($p_T^H$), depending on the Higgs boson production process.
As more LHC data is accumulated a finer STXS binning as well as binning in
additional kinematic variables (\eg angular observables) becomes possible, hence
various STXS stages have been defined (or are still being defined). In this way,
the model assumptions can be further reduced with increasing amounts of data.

As the STXS observables are defined for specific particle level topologies of
the Higgs boson production process, these can incorporate several production
processes that depend on different Higgs couplings. For instance, $gg\to H (+
\text{jets})$ STXS observables target the gluon fusion production mode including
$gg$-induced EW corrections. These are composed of virtual EW corrections to the
$gg\to H$ form factor as well as real EW corrections, corresponding to $gg\to
(Z\to q\bar{q})H$. In \HS, gluon fusion and $gg\to ZH$ are, however, treated as
separate processes, as their dependences on the Higgs coupling properties are
different. Another example are the STXS observables of the class ``EW qqH''
composed of the vector boson fusion (VBF) as  well as $qq \to (V\to qq) H$ (with
$V = W, Z$) processes, which all three are treated as separate processes in \HS{}.
While higher STXS stages aim at separating these subprocesses, the earlier
stages must be regarded as inclusive in these processes.\footnote{The general
claim for early STXS stage measurements is that, at the present level of
precision, these processes cannot be resolved. However, we want to remark that
this claim relies on the assumption of SM signal strengths for all involved
processes, and may not be true if a BSM model predicts a strong enhancement in
one or more of these processes.} For such STXS measurements it would be
beneficial to publicly release the signal efficiencies (or signal compositions)
for the involved processes, analogous to the case of $\mu$ measurements, see
above. Unfortunately, this has not been done by the experiments so far.

For both the inclusive ($\mu$) and STXS measurements, the correlations of
uncertainties --- preferably given separately for experimental and theoretical
sources --- need to be included in the $\chi^2$ test in order to achieve a good
performance. As the unfolding to pure production channels and/or STXS bins often
induces large correlations, experimental measurements of this kind are nowadays
always accompanied by the corresponding correlation matrix. However, even $\mu$
measurements in experimental categories (which are therefore composed of various
Higgs processes) have correlated experimental and theoretical uncertainties.
Therefore it would be useful if the experiments could provide correlation
matrices also for such cases, which so far has rarely happened.

We would furthermore like to encourage the experiments to accompany all signal
strength and STXS measurements with a reference value for the signal rate
expected for a SM Higgs boson. In case the measurement is quoted as a
normalized signal strength, this allows one to recalculate the absolute
observed signal rate. Furthermore, the signal strengths predicted in BSM
models can in many cases be approximated by a simple rescaling of the SM
reference value.

Lastly, we want to emphasize that the two-dimensional toy model interpretations
--- ($\kappa_V, \kappa_F$) and ($\kappa_g, \kappa_\gamma$) in the $\kappa$
framework, as well as the production rate rescaling models --- have proven
extremely valuable for validation checks. Unfortunately, we have recently
experienced that these interpretations are sometimes not performed when new
experimental results are released, see \eg
Ref.~\cite{ATLAS:2019jst,Sirunyan:2020tzo}.

\section{Summary}%
\label{sec:summary}

We have presented a new version of the public computer program \HS
for confronting the predictions of
arbitrary BSM models with the measured mass and rates of the
Higgs boson that has been detected at about \SI{125}{\GeV}. The
description of \HSv{2} provided in the present paper has focussed on the
improvements of the functionality and applicability of the program with
respect to version~\texttt{1.0} that was presented in
Ref.~\cite{Bechtle:2013xfa}. Besides the confrontation with the
properties of the detected Higgs signal, the phenomenological viability of
BSM Higgs sectors should also be tested against the limits that have been
obtained from the searches for additional Higgs bosons
--- a task that can be performed using the related public tool
\HB~\cite{Bechtle:2008jh,Bechtle:2011sb,Bechtle:2013wla,Bechtle:2015pma,Bechtle:2020pkv}.

On the basis of theoretical model input from the user --- in the
form of predicted Higgs-boson masses, production cross sections,
and decay rates in the \HB input format --- \HS evaluates a $\chi^2$ measure
to quantify the compatibility of the measured Higgs properties with the model
predictions as discussed in \cref{sec:newHS}. The peak-centered
$\chi^2$ method is used to evaluate the $\chi^2$ contribution from
signal strength ($\mu$) measurements, while STXS measurements give
rise to a $\chi^2$ contribution that is separately calculated in a
newly introduced module. In addition, the LHC Run-1 results give rise
to another $\chi^2$ contribution, employing the legacy ATLAS and CMS
combination of the Run-1 results. Lastly, the Higgs mass measurement(s)
lead to another contribution to the total $\chi^2$.
In total, the current version \HScurrent\ includes 86 Run-2 signal rate
measurements, 20 Run-1 measurements, and one (combined) Higgs mass
measurement as default experimental input.

The description in this paper has addressed the calculation of the various
$\chi^2$ parts, comprising as new feature the incorporation of experimental
results in the form of STXS measurements as well as further improvements that
have been highlighted. We have discussed possible features of BSM Higgs sectors
that can have an important impact in the comparison with the data. This includes
cases where the signal efficiencies differ significantly from the ones of a
SM-like Higgs boson, the treatment of theoretical uncertainties in the Higgs
mass predictions, and scenarios where the signal is composed of overlapping
contributions from several Higgs bosons.

In order to give some guidance on possible applications of \HSv{2} we
have thoroughly discussed the different interpretations of the $\chi^2$
output provided by \HS and spelled out the involved assumptions.
In particular, we have explained the application of
the obtained $\chi^2$ output in fits, for limit setting, and for
goodness-of-fit tests. We furthermore discussed possible attempts of using
\HS for ruling out the SM on the basis of possible deviations between the
measurements and the SM predictions, and stressed important caveats in this
context.

The experimental results used by \HSv{2} are either signal strength modifiers,
$\hat{\mu}$, in the various search channels and experimental categories, or
measured signal rates provided in the simplified template cross section (STXS)
framework. In both cases the results have to be supplemented by their
experimental uncertainties, $\Delta\hat{\mu}$ and $\Delta\hat{\sigma}$,
respectively. As performance tests of \HSv{2} we have presented several detailed
validations against published ATLAS and CMS results in \cref{sec:performance}.
These comparisons illustrate that the achievable agreement with the official
results published by ATLAS and CMS strongly depends on the information made
available by the experiments. As a first step, we compared to the ATLAS/CMS
combination of the full Run-1 data within the $\kappa$~framework for different
assumptions regarding the total width of the Higgs boson and found very good
agreement. Second, in our validations of Run-2 results against $\kappa$~fits
performed by ATLAS or CMS, we found the best agreement if the sub-channel signal
strength modifiers are given together with the corresponding signal
efficiencies. If signal strengths are given in terms of the targeted production
processes, the best-fit point is typically well reconstructed by \HSv{2} while
the correlations cannot be correctly reproduced. Therefore, we expect further
improvements from additional information about correlations of experimental
uncertainties. In case of STXS measurements we have seen that information about
experimental correlations is crucial in order to obtain reconstructed results
that are close to the ones published by ATLAS and CMS\@. If STXS measurements are
not given in terms of pure signal channels, for example by using intermediate
stage-1 bins, additional information about production processes could further
improve the reconstruction (\eg the $WH$ and $ZH$ composition in the leptonic
$VH$ bin for the considered ATLAS $H\to\gamma\gamma$ analysis). On the basis of
those findings we have formulated some recommendations for the presentation of
future Higgs signal rate measurements.

The \HS source code is available at
\begin{center}
    \url{https://gitlab.com/higgsbounds/higgssignals}
\end{center}
together with continuously updated technical documentation of
the program's subroutines, compilation
and interfacing procedures.

\section*{Acknowledgments}

We thank Henning Bahl, Thomas Biek{\"o}tter, Artur Gottmann, Sarah Heim, Bill
Murray, Frank Tackmann, Kerstin Tackmann and Roger Wolf for helpful discussions.
We are grateful for earlier contributions from the former \HS\ team members
Daniel Dercks and Oscar St{\aa}l. The work of S.H.\ was supported in part by the
MEINCOP (Spain) under contract FPA2016-78022-P and under contract
PID2019-110058GB-C21, in part by the Spanish Agencia Estatal de Investigaci\'on
(AEI), in part by the EU Fondo Europeo de Desarrollo Regional (FEDER) through
the project FPA2016-78645-P, in part by the ``Spanish Red Consolider MultiDark''
FPA2017-90566-REDC, and in part by the AEI through the grant IFT Centro de
Excelencia Severo Ochoa SEV-2016-0597. T.S.\ and G.W.\ acknowledge support by
the Deutsche Forschungsgemeinschaft (DFG, German Research Foundation) under
Germany's Excellence Strategy --- EXC 2121 ``Quantum Universe'' --- 390833306.
The work of J.W. is supported by the Swedish Research Council, contract number
2016-05996 and was in part funded by the European Research Council (ERC) under
the European Union's Horizon 2020 research and innovation programme, grant
agreement No 668679.

\clearpage
\appendix

\section{Implementation of STXS observables in \HS}\label{app:STXS} Each STXS
observable is defined by an individual file with the extension `\texttt{.stxs}'
located in the observable set folder. The observable set can contain signal
strength ($\mu$) observables and STXS observables simultaneously, however, the
user should avoid statistical overlap of the included measurements. The
structure of the \Code{.stxs} files is exemplified in \cref{Tab:STXSexample}.
The integer observable ID has to be unique among all STXS observables. As
mentioned earlier, the STXS framework also allows one to evaluate a $\chi^2$
contribution from potential Higgs mass measurements. The mass measurement must
be associated with one STXS observable and is specified on line 7 of the
\Code{.stxs} file. It is activated by setting the mass-observable flag on line 6
to 1. Independently of this mass measurement $\hat{m}$ and its quoted $1\sigma$
uncertainty $\Delta\hat{m}$, line 8 and 9 of the \Code{.stxs} file always specify a
reference mass, for which the rate measurement has been performed, as well as an
estimate of the mass resolution. The latter is decisive on whether a Higgs boson
in the model is assigned to the observable, \ie whether its signal rate is
assumed to contribute to the observable or not (see
\cref{sec:overlappingHiggs}). Line 10 of the \Code{.stxs} file specifies the number of
relevant signal channels, $N_c$, and another reference mass for the quoted
efficiencies (which usually coincides with the previous reference mass for the
measurement). Line 11 lists the $N_c$ channel IDs (separated by white spaces),
see \cref{Tab:pdidentifier}, and line 12 the corresponding relative channel
efficiencies in the SM, $\epsilon_{\text{SM},i}$. The last two lines (13, 14)
quote the observed and SM-predicted signal rate and the lower and upper
$1\sigma$ range, as quoted in the experimental result. If the rate is given in
SM-normalized form (SM-normalized-rate flag set to 1), the last line is ignored.

\begin{table}[t]
    \centering
    \small
    \begin{tabularx}{\textwidth}{c l >{\ttfamily}l}
        \toprule
        line & content                                                                        & example                           \\
        \midrule
             & lines starting with \texttt{\#} are comments                                   & \# A fictitious STXS measurement  \\
        1    & unique observable ID                                                           & 123456789                         \\
        2    & reference                                                                      & arXiv:1234.56789                  \\
        3    & collider, collaboration, experiment                                            & LHC, ATL, ATL                     \\
        4    & channel description                                                            & (pp)->h->gamma gamma (1j, pTH<60) \\
        5    & center-of-mass energy (TeV), $\mathcal{L}$, $\Delta\mathcal{L}$                & 13\quad 36.1\quad 0.032           \\
        6    & mass-observable flag, SM-normalized-rate flag                                  & 1\quad0                           \\
        7    & $\hat{m}$, $\Delta \hat{m}$ (\emph{line only exists for mass observable}) & 125.23\quad 0.56                  \\
        8    & reference mass                                                                 & 125.09                            \\
        9    & mass resolution                                                                & 2.5                               \\
        10   & $N_c$, reference mass for the efficiencies                                     & 1\quad125.09                      \\
        11   & channel IDs ($N_c$ entries)                                                    & 1.3                               \\
        12   & channel efficiencies      ($N_c$ entries)                                      & 1.0                               \\
        13   & observed rate (lower $1\sigma$, central, upper $1\sigma$)                      & 0.022\quad0.037\quad0.052         \\
        14   & SM predicted rate (lower $1\sigma$, central, upper $1\sigma$)                  & 0.047\quad0.063\quad0.080         \\
        \bottomrule
    \end{tabularx}
    \caption{Structure of an STXS observable implementation file for \HS. Note
        that line 7 is only present if the mass observable flag on line six is
        equal to \texttt{1}. $\mathcal{L}$ denotes the integrated luminosity
        with uncertainty $\Delta\mathcal{L}$, while $m$ is the measured mass
        value and $\Delta m$ the corresponding experimental $1\sigma$
        uncertainty.}%
    \label{Tab:STXSexample}
\end{table}

The STXS correlation matrix should be given in an additional file
(\texttt{.stxscorr}) in the same folder. The file contains three columns,
\begin{center}
    \texttt{observable ID 1} \hspace{1cm} \texttt{observable ID 2} \hspace{1cm} \texttt{correlation coefficient},
\end{center}
such that each line encodes the correlation between two observables.

\section{Experimental input}

Tables~\ref{tab:expRun2ATLAS} and~\ref{tab:expRun2CMS} list all LHC Run-2 Higgs signal measurements from ATLAS and CMS, respectively, which are implemented as the default observable set in \HScurrent.

\begin{table}
    \centering
    \footnotesize
    \renewcommand{\arraystretch}{1.05}
    \setlength{\tabcolsep}{5pt}
    \vspace{-1cm}
    \begin{tabularx}{\textwidth}{X c c c c}
        \toprule
        Channel                                                                             & Luminosity [$\text{fb}^{-1}$] & \multicolumn{2}{c}{Signal strength $\mu$}                   & Ref.                                                 \\
        \midrule
        VBF, $H\to b\bar{b}$                                                                & $30.6$                        & \multicolumn{2}{c}{$3.0\substack{+1.7                                                                                   \\-1.8}$ }     & \cite{Aaboud:2018gay}  \\
        $t\bar{t}H$, $H\to b\bar{b}$ ($1\ell$)                                              & $36.1$                        & \multicolumn{2}{c}{$0.67\substack{+0.71\\-0.69}$ } & \cite{Aaboud:2017rss}                                     \\
        $t\bar{t}H$, $H\to b\bar{b}$ ($2\ell$)                                              & $36.1$                        & \multicolumn{2}{c}{$0.11\substack{+1.36\\-1.41}$ } & \cite{Aaboud:2017rss}                                     \\
        $t\bar{t}H$, multilepton ($2\ell ss$)                                               & $79.9$                        & \multicolumn{2}{c}{$0.38\substack{+0.57                                                                                 \\-0.54}$} & \cite{ATLAS:2019nvo} \\
        $t\bar{t}H$, multilepton ($3\ell$)                                                  & $79.9$                        & \multicolumn{2}{c}{$0.93\substack{+0.58                                                                                 \\-0.52}$} & \cite{ATLAS:2019nvo} \\
        $t\bar{t}H$, multilepton ($4\ell$)                                                  & $79.9$                        & \multicolumn{2}{c}{$0.52\substack{+0.93                                                                                 \\-0.72}$} & \cite{ATLAS:2019nvo} \\
        $t\bar{t}H$, multilepton ($1\ell+2\tau_h$)                                          & $79.9$                        & \multicolumn{2}{c}{$0.30\substack{+1.01                                                                                 \\-0.90}$} & \cite{ATLAS:2019nvo} \\
        $t\bar{t}H$, multilepton ($2\ell+1\tau_h$)                                          & $79.9$                        & \multicolumn{2}{c}{$0.49\substack{+0.94                                                                                 \\-0.82}$} & \cite{ATLAS:2019nvo} \\
        $t\bar{t}H$, multilepton ($3\ell+1\tau_h$)                                          & $79.9$                        & \multicolumn{2}{c}{$0.43\substack{+1.10                                                                                 \\-0.85}$} & \cite{ATLAS:2019nvo} \\
        \midrule
                                                                                            &                               & $\sigma_\text{obs}$ [$\text{pb}$]                           & $\sigma_{\text{SM}}$ [$\text{pb}$] &                      \\
        \midrule
        $gg\to H$, $H\to W^+W^-$                                                            & $36.1$                        & $11.4\substack{+2.2                                                                                                     \\-2.1}$  &  $10.4\pm 0.6$ & \cite{Aaboud:2018jqu} \\
        VBF, $H\to W^+W^-$                                                                  & $36.1$                        & $0.50\substack{+0.29                                                                                                    \\-0.28}$  &  $0.81\pm 0.02$ & \cite{Aaboud:2018jqu} \\
        VBF, $H\to ZZ$ ($p_{T,H}$ high)                                                     & $139.0$                       & $0.0005\substack{+0.0079                                                                                                \\-0.0048}$                        & $0.00420 \pm 0.00018$              & \cite{Aad:2020mkp}  \\
        VBF, $H\to ZZ$ ($p_{T,H}$ low)                                                      & $139.0$                       & $0.15\substack{+0.064                                                                                                   \\-0.052}$                           & $0.1076\substack{+0.0024\\-0.0035}$                  & \cite{Aad:2020mkp}  \\
        $V(\text{had})H$, $H\to ZZ$                                                         & $139.0$                       & $0.021\pm 0.035$                                            & $0.0138\substack{+0.0004                                  \\-0.0006}$ & \cite{Aad:2020mkp} \\
        $V(\text{lep})H$, $H\to ZZ$                                                         & $139.0$                       & $0.022\substack{+0.028                                                                                                  \\-0.018}$ & $0.0164\pm0.0004$ &      \cite{Aad:2020mkp}                        \\
        $gg\to H$, $H\to ZZ$ ($p_{T,H}$ high)                                               & $139.0$                       & $0.038\substack{+0.021                                                                                                  \\-0.016}$                          & $0.015\pm 0.004$                   & \cite{Aad:2020mkp}  \\
        $gg\to H$, $H\to ZZ$ ($0j,~p_{T,H}$ high)                                           & $139.0$                       & $0.630\pm 0.110$                                            & $0.55\pm 0.04$                     & \cite{Aad:2020mkp}   \\
        $gg\to H$, $H\to ZZ$ ($0j,~p_{T,H}$ low)                                            & $139.0$                       & $0.17\pm 0.055$                                             & $0.176\pm 0.025$                   & \cite{Aad:2020mkp}   \\
        $gg\to H$, $H\to ZZ$ ($1j,~p_{T,H}$ high)                                           & $139.0$                       & $0.009\substack{+0.016                                                                                                  \\ -0.012}$ & $0.020\pm 0.004$ &  \cite{Aad:2020mkp} \\
        $gg\to H$, $H\to ZZ$ ($1j,~p_{T,H}$ low)                                            & $139.0$                       & $0.05\pm 0.08$                                              & $0.172\pm 0.025$                   & \cite{Aad:2020mkp}   \\
        $gg\to H$, $H\to ZZ$ ($1j,~p_{T,H}$ med.)                                           & $139.0$                       & $0.17\pm 0.05$                                              & $0.119\pm 0.018$                   & \cite{Aad:2020mkp}   \\
        $gg\to H$, $H\to ZZ$ ($2j$)                                                         & $139.0$                       & $0.040\pm 0.075$                                            & $0.127\pm 0.027$                   & \cite{Aad:2020mkp}   \\
        $t\bar{t}H$, $H\to ZZ$                                                              & $139.0$                       & $0.025\substack{+0.022                                                                                                  \\ -0.013}$ & $0.0154\substack{+0.0010\\-0.0013}$ &  \cite{Aad:2020mkp} \\
        $gg\to H$, $H\to \gamma\gamma$ ($0j$)                                               & $139.0$                       & $0.039\pm 0.006$                                            & $0.0382 \substack{+0.0019                                 \\ -0.0018}$ & \cite{ATLAS:2019jst} \\
        $gg\to H$, $H\to \gamma\gamma$ ($1j$)                                               & $139.0$                       & $0.0162\substack{+0.0031                                                                                                \\-0.0022}$ & $0.0194 \substack{+0.0018 \\ -0.0019}$ & \cite{ATLAS:2019jst} \\
        $gg\to H$, $H\to \gamma\gamma$ ($2j$, $\Delta\Phi_{jj} \in [-\pi,-\tfrac{\pi}{2}]$) & $139.0$                       & $0.0023\pm 0.0007$                                          & $0.0024 \pm 0.0002$                & \cite{ATLAS:2019jst} \\
        $gg\to H$, $H\to \gamma\gamma$ ($2j$, $\Delta\Phi_{jj} \in [-\tfrac{\pi}{2},0]$)    & $139.0$                       & $0.0011\pm 0.0004$                                          & $0.0020 \pm 0.0002$                & \cite{ATLAS:2019jst} \\
        $gg\to H$, $H\to \gamma\gamma$ ($2j$, $\Delta\Phi_{jj} \in [0, \tfrac{\pi}{2}]$)    & $139.0$                       & $0.0014\pm 0.0004$                                          & $0.0020 \pm 0.0002$                & \cite{ATLAS:2019jst} \\
        $gg\to H$, $H\to \gamma\gamma$ ($2j$, $\Delta\Phi_{jj} \in [\tfrac{\pi}{2},\pi]$)   & $139.0$                       & $0.0021\pm 0.0007$                                          & $0.0024 \pm 0.0002$                & \cite{ATLAS:2019jst} \\
        $t\bar{t}H$, $H\to \gamma\gamma$                                                    & $139.0$                       & $1.59\substack{+0.43                                                                                                    \\ -0.39}$ & $1.15 \substack{+0.09 \\ -0.12}$ & \cite{ATLAS:2019jst} \\
        VBF, $H\to \tau^+\tau^-$                                                            & $36.1$                        & $0.28\substack{+0.14                                                                                                    \\ -0.13}$ & $0.237\pm 0.006$ & \cite{Aaboud:2018pen} \\
        $gg\to H$, $H\to \tau^+\tau^-$                                                      & $36.1$                        & $3.10\substack{+1.90                                                                                                    \\ -1.60}$ & $3.05\pm 0.13$ & \cite{Aaboud:2018pen} \\
        $WH$, $H\to W^+W^-$                                                                 & $36.1$                        & $0.67\substack{+0.36                                                                                                    \\ -0.30}$ & $0.293\pm 0.007$ & \cite{Aad:2019lpq} \\
        $ZH$, $H\to W^+W^-$                                                                 & $36.1$                        & $0.54\substack{+0.34                                                                                                    \\ -0.25}$ & $0.189\pm 0.007$ & \cite{Aad:2019lpq} \\
        $WH$, $H\to b\bar{b}$ ($p_{T,V} \in [150, 250]~\mathrm{GeV}$)                       & $139.0$                       & $0.0190\pm 0.0121$                                          & $0.0240\pm 0.0011$                 & \cite{Aad:2020jym}   \\
        $WH$, $H\to b\bar{b}$ ($p_{T,V} \ge 250~\mathrm{GeV}$)                              & $139.0$                       & $0.0072\pm 0.0022$                                          & $0.0071\pm 0.0030$                 & \cite{Aad:2020jym}   \\
        $ZH$, $H\to b\bar{b}$ ($p_{T,V} \in [75, 150]~\mathrm{GeV}$)                        & $139.0$                       & $0.0425\pm 0.0359$                                          & $0.0506\pm 0.0041$                 & \cite{Aad:2020jym}   \\
        $ZH$, $H\to b\bar{b}$ ($p_{T,V} \in [150, 250]~\mathrm{GeV}$)                       & $139.0$                       & $0.0205\pm 0.0062$                                          & $0.0188\pm 0.0024$                 & \cite{Aad:2020jym}   \\
        $ZH$, $H\to b\bar{b}$ ($p_{T,V} \ge 250~\mathrm{GeV}$)                              & $139.0$                       & $0.0054\pm 0.0017$                                          & $0.0049\pm 0.0005$                 & \cite{Aad:2020jym}   \\
        \bottomrule
    \end{tabularx}
    \caption{ATLAS Higgs rate measurements from LHC Run-2 included in the default observable set \Code{LHC13_Apr2020} in \HScurrent.}%
    \label{tab:expRun2ATLAS}
\end{table}

\begin{table}
    \centering
    \footnotesize
    \renewcommand{\arraystretch}{1.05}
    \setlength{\tabcolsep}{5pt}
    \vspace{-1cm}
    \begin{tabularx}{\textwidth+5mm}{X c c c c}
        \toprule
        Channel                                                     & Luminosity [$\text{fb}^{-1}$] & \multicolumn{2}{c}{Signal strength $\mu$} & Ref.                                               \\
        \midrule
        $pp\to H$, $H\to \mu^+\mu^-$                                & $35.9$                        & \multicolumn{2}{c}{$1.0\substack{+1.1                                                               \\-1.1}$ }     & \cite{Sirunyan:2018hbu}\\
        $WH$, $H\to b\bar{b}$                                       & $35.9$                        & \multicolumn{2}{c}{$1.7\substack{+0.7                                                               \\-0.7}$ }     & \cite{Sirunyan:2017elk}\\
        $ZH$, $H\to b\bar{b}$                                       & $35.9$                        & \multicolumn{2}{c}{$0.9\substack{+0.5                                                               \\-0.5}$ }     & \cite{Sirunyan:2017elk}\\
        $pp\to H$ (boosted), $H\to b\bar{b}$                        & $35.9$                        & \multicolumn{2}{c}{$2.3\substack{+1.8                                                               \\-1.6}$}   & \cite{Sirunyan:2017dgc}\\
        $t\bar{t}H$, $H\to b\bar{b}$ ($1\ell$)                      & $35.9 \oplus 41.5$            & \multicolumn{2}{c}{$0.84\substack{+0.52                                                             \\-0.50}\oplus 1.84\substack{+0.62\\-0.56}$}   & \cite{Sirunyan:2018mvw,CMS:2019lcn}\\
        $t\bar{t}H$, $H\to b\bar{b}$ ($2\ell$)                      & $35.9 \oplus 41.5$            & \multicolumn{2}{c}{$-0.24\substack{+1.21                                                            \\-1.12}\oplus 1.62\substack{+0.90\\-0.85}$}   & \cite{Sirunyan:2018mvw,CMS:2019lcn}\\
        $t\bar{t}H$, $H\to b\bar{b}$  (hadronic)                    & $41.5$                        & \multicolumn{2}{c}{$-1.69\substack{+1.43                                                            \\-1.47}$}   & \cite{Sirunyan:2018mvw}\\
        $t\bar{t}H$, multilepton ($1\ell+2\tau_h$)                  & $35.9 \oplus 41.5$            & \multicolumn{2}{c}{$-1.52\substack{+1.76                                                            \\-1.72} \oplus 1.4\substack{+1.24\\-1.14}$} & \cite{Sirunyan:2018shy,CMS:2018dmv}\\
        $t\bar{t}H$, multilepton ($2\ell ss+1\tau_h$)               & $35.9 \oplus 41.5$            & \multicolumn{2}{c}{$0.94\substack{+0.80                                                             \\-0.67} \oplus 1.13\substack{+1.03\\-1.11}$} & \cite{Sirunyan:2018shy,CMS:2018dmv}\\
        $t\bar{t}H$, multilepton ($2\ell ss$)                       & $35.9 \oplus 41.5$            & \multicolumn{2}{c}{$1.61\substack{+0.58                                                             \\-0.51} \oplus 0.87\substack{+0.62\\-0.55}$} & \cite{Sirunyan:2018shy,CMS:2018dmv}\\
        $t\bar{t}H$, multilepton ($3\ell+1\tau_h$)                  & $35.9 \oplus 41.5$            & \multicolumn{2}{c}{$1.34\substack{+1.42                                                             \\-1.07} \oplus -0.96\substack{+1.96\\-1.33}$} & \cite{Sirunyan:2018shy,CMS:2018dmv}\\
        $t\bar{t}H$, multilepton ($3\ell$)                          & $35.9 \oplus 41.5$            & \multicolumn{2}{c}{$0.82\substack{+0.77                                                             \\-0.71} \oplus 0.29\substack{+0.82\\-0.62}$} & \cite{Sirunyan:2018shy,CMS:2018dmv}\\
        $t\bar{t}H$, multilepton ($4\ell$)                          & $35.9 \oplus 41.5$            & \multicolumn{2}{c}{$0.57\substack{+2.29                                                             \\-1.57} \oplus 0.99\substack{+3.31\\-1.69}$} & \cite{Sirunyan:2018shy,CMS:2018dmv}\\
        \midrule
                                                                    &                               & $\sigma_\text{obs}$ [$\text{pb}$]         & $\sigma_{\text{SM}}$ [$\text{pb}$] &                    \\
        \midrule
        $gg\to H$, $H\to W^+W^-$ ($0j$)                             & $137.0$                       & $0.0423\substack{+0.0063                                                                            \\ -0.0059}$ & $0.0457\substack{+0.0029\\ -0.0018}$ & \cite{Sirunyan:2020tzo} \\
        $gg\to H$, $H\to W^+W^-$ ($1j$)                             & $137.0$                       & $0.0240\substack{+0.0057                                                                            \\ -0.0051}$ & $0.0217\substack{+0.0023\\ -0.0022}$ & \cite{Sirunyan:2020tzo} \\
        $gg\to H$, $H\to W^+W^-$ ($2j$)                             & $137.0$                       & $0.0151\substack{+0.0051                                                                            \\ -0.0046}$ & $0.0100\substack{+0.0020\\ -0.0011}$ & \cite{Sirunyan:2020tzo} \\
        $gg\to H$, $H\to W^+W^-$ ($3j$)                             & $137.0$                       & $0.0050\substack{+0.0045                                                                            \\ -0.0042}$ & $0.0033\substack{+0.0002\\ -0.0004}$ & \cite{Sirunyan:2020tzo} \\
        $gg\to H$, $H\to W^+W^-$ ($4j$)                             & $137.0$                       & $0.0064\substack{+0.0039                                                                            \\ -0.0034}$ & $0.0018\substack{+0.0001\\ -0.0002}$ & \cite{Sirunyan:2020tzo} \\
        VBF, $H\to ZZ$                                              & $137.1$                       & $0.279\substack{+0.211                                                                              \\-0.162}$     & $0.450\pm 0.010$ & \cite{CMS:2019chr}\\
        $gg/b\bar{b}\to H$, $H\to ZZ$                               & $137.1$                       & $5.328\pm 0.611$                          & $5.550\substack{+0.600                                  \\-0.650}$ & \cite{CMS:2019chr}\\
        $VH$, $H\to ZZ$                                             & $137.1$                       & $0.305\substack{+0.243                                                                              \\-0.194}$     & $0.270\pm 0.010$ & \cite{CMS:2019chr}\\
        $t\bar{t}H,tH$, $H\to ZZ$                                   & $137.1$                       & $0.0078\pm 0.0552$                        & $0.060\substack{+0.011                                  \\-0.012}$ & \cite{CMS:2019chr}\\
        $gg\to H$, $H\to \gamma\gamma$ ($0j$)                       & $77.4$                        & $0.072 \pm0.0122$                         & $0.0610\substack{+0.0037                                \\-0.0031}$ & \cite{CMS:1900lgv} \\
        $gg\to H$, $H\to \gamma\gamma$ ($1j,~p_{T,H}$ high)         & $77.4$                        & $0.0029\substack{+0.0017                                                                            \\-0.0012}$ & $0.0017\pm 0.0002$ &\cite{CMS:1900lgv} \\
        $gg\to H$, $H\to \gamma\gamma$ ($1j,~p_{T,H}$ low)          & $77.4$                        & $0.021\substack{+0.0090                                                                             \\-0.0075}$ & $0.015\pm 0.0015$& \cite{CMS:1900lgv} \\
        $gg\to H$, $H\to \gamma\gamma$ ($1j,~p_{T,H}$ med.)         & $77.4$                        & $0.0076\pm 0.0040$                        & $0.010\pm 0.001$                   & \cite{CMS:1900lgv} \\
        $gg\to H$, $H\to \gamma\gamma$ ($2j$)                       & $77.4$                        & $0.0084\substack{+0.0066                                                                            \\ -0.0055}$ & $0.011\pm 0.002$& \cite{CMS:1900lgv} \\
        $gg\to H$, $H\to \gamma\gamma$ (BSM)                        & $77.4$                        & $0.0029\pm 0.00104$                       & $0.0013\pm 0.0003$                 & \cite{CMS:1900lgv} \\
        VBF, $H\to \gamma\gamma$                                    & $77.4$                        & $0.0091\substack{+0.0044                                                                            \\-0.0033}$ & $0.0011\pm 0.002$ &\cite{CMS:1900lgv} \\
        $t\bar{t}H$, $H\to \gamma\gamma$                            & $137.0$                       & $0.00156\substack{+0.00034                                                                          \\-0.00032}$&  $0.0013\substack{+0.00008\\-0.00011}$   & \cite{Sirunyan:2020sum} \\
        $V(\text{had})H$, $H\to \tau^+\tau^-$                       & $77.4$                        & $-0.0433\substack{+0.057                                                                            \\-0.054}$ & $0.037\pm 0.001$ &\cite{CMS:2019pyn} \\
        VBF, $H\to \tau^+\tau^-$                                    & $77.4$                        & $0.114\substack{+0.034                                                                              \\-0.033}$ & $0.114\pm 0.009$ &\cite{CMS:2019pyn} \\
        $gg\to H$, $H\to \tau^+\tau^-$ ($0j$)                       & $77.4$                        & $-0.680\substack{+1.292                                                                             \\-1.275}$ & $1.70\pm 0.10$ &\cite{CMS:2019pyn} \\
        $gg\to H$, $H\to \tau^+\tau^-$ ($1j$, $p_{T,H}$ high)         & $77.4$                        & $0.108\substack{+0.071                                                                              \\-0.061}$ & $0.060\pm 0.010$ &\cite{CMS:2019pyn} \\
        $gg\to H$, $H\to \tau^+\tau^-$ ($1j$, $p_{T,H}$ low)          & $77.4$                        & $-0.139\substack{+0.562                                                                             \\-0.570}$ & $0.410\pm 0.060$ &\cite{CMS:2019pyn} \\
        $gg\to H$, $H\to \tau^+\tau^-$ ($1j$, $p_{T,H}$ med.)         & $77.4$                        & $0.353\substack{+0.437                                                                              \\-0.420}$ & $0.280\pm 0.040$ &\cite{CMS:2019pyn} \\
        $gg\to H$, $H\to \tau^+\tau^-$ ($2j$)                       & $77.4$                        & $0.0987\substack{+0.1911                                                                            \\-0.1806}$ & $0.210\pm 0.050$& \cite{CMS:2019pyn} \\
        $gg\to H$, $H\to \tau^+\tau^-$ ($1j$, $p_{T}^{j_1}> \SI{200}{\GeV}$) & $77.4$                        & $0.0199\substack{+0.0145                                                                            \\-0.0148}$ & $0.0141\pm 0.0004$ &\cite{CMS:2019pyn} \\
        $gg\to H$, $H\to \tau^+\tau^-$ (Rest)                       & $77.4$                        & $-0.195\substack{+0.506                                                                             \\-0.491}$ & $0.184\pm 0.005$ &\cite{CMS:2019pyn} \\
        \bottomrule
    \end{tabularx}
    \caption{CMS Higgs rate measurements from LHC Run-2 included in the default observable set \Code{LHC13_Apr2020} in \HScurrent.}
    \label{tab:expRun2CMS}
\end{table}

\clearpage
\begin{multicols}{2}[\printbibheading]
    \renewcommand*{\bibfont}{\small}
    \printbibliography[heading=none]
\end{multicols}
\end{document}